\documentclass[sn-mathphys-num]{sn-jnl}

\usepackage{graphicx}%
\usepackage{multirow}%
\usepackage{amsmath,amssymb,amsfonts}%
\usepackage{amsthm}%
\usepackage{mathrsfs}%
\usepackage[title]{appendix}%
\usepackage{xcolor}%
\usepackage{textcomp}%
\usepackage{manyfoot}%
\usepackage{booktabs}%
\usepackage{algorithm}%
\usepackage{algpseudocode}%
\usepackage{listings}%
\usepackage[nice]{nicefrac}
\usepackage{todonotes}
\usepackage{multicol}
\usepackage{placeins}
\usepackage{enumerate}
\usepackage{array}

\usepackage{xcolor}
\usepackage{amsmath}
\usepackage{subfig}
\usepackage[permil]{overpic}
\usepackage{anyfontsize}
\graphicspath{{Figures/}}

\newcommand{\ve}[1]{\text{\boldmath${#1}$}} 
\newcommand{\te}[1]{\text{\boldmath$\mathrm{#1}$}} 
\newcommand{\grads}{{\nabla^\mathrm{S}}} 
\renewcommand{\bar}[1]{{\overline{#1}}} 

\def\stmdocstextcolor#1{}

\newcommand{\fbf}{{\ve{f}}}
\newcommand{\Xbf}{{\ve{X}}}
\newcommand{\fDeltaX}{{\Delta \ve{{X}}}}
\newcommand{\fDeltaXhat}{\Delta{\ve{\hat{X}}}}
\newcommand{\p}[2]{{p}(#1\mid #2)}

\def\allDraft{false} 
\def\draftSim{\allDraft}
\def\draftMeshExamples{\allDraft}
\def\draftVascularModelling{\allDraft}
\def\draftUQresults{\allDraft}

\begin{document}

\title[Article Title]{Geometric Uncertainty of Patient-Specific Blood Vessels and its Impact on Aortic Hemodynamics}
\author*[1]{\fnm{Domagoj} \sur{Bo\v{s}njak}}\email{bosnjak@tugraz.at}
\author[2]{\fnm{Richard} \sur{Schussnig}\email{richard.schussnig@ruhr-uni-bochum.de}}
\author[3]{\fnm{Sascha} \sur{Ranftl}\email{ranftl@tugraz.at}}
\author[4,5]{\fnm{Gerhard A.} \sur{Holzapfel}\email{holzapfel@tugraz.at}}
\author[1]{\fnm{Thomas-Peter} \sur{Fries}\email{fries@tugraz.at}}

\affil[1]{\orgdiv{Institute of Structural Analysis}, \orgname{Graz University of Technology}, \orgaddress{\street{Lessingstrasse 25}, \city{ 8010 Graz},  \country{Austria}}}
\affil[2]{\orgdiv{Faculty of Mathematics}, \orgname{Ruhr University Bochum}, \orgaddress{\street{Universitätsstraße 150}, \city{44795 Bochum}, \country{Germany}}}
\affil[3]{\orgdiv{Institute of Theoretical and Computational Physics}, \orgname{Graz University of Technology}, \orgaddress{\street{Petersgasse 16/II}, \city{8010 Graz},  \country{Austria}}}
\affil[4]{\orgdiv{Institute of Biomechanics}, \orgname{Graz University of Technology}, \orgaddress{\street{Stremayrgasse 16/II}, \city{8010 Graz}, \country{Austria}}}
\affil[5]{\orgdiv{Department of Structural Engineering}, \orgname{Norwegian University of Science and Technology}, \orgaddress{\street{7491 Trondheim}, \country{Norway}}}

\abstract{In the context of numerical simulations of the vascular system, local geometric uncertainties have not yet been examined in sufficient detail due to model complexity and the associated large numerical effort. Such uncertainties are related to geometric modeling errors resulting from computed tomography imaging, segmentation and meshing. This work presents a methodology to systematically induce local modifications and perform a sufficient number of blood flow simulations to draw statistically relevant conclusions on the most commonly employed quantities of interest, such as flow rates or wall shear stress. The surface of a structured hexahedral mesh of a patient-specific aorta is perturbed by displacement maps defined via Gaussian random fields to stochastically model the local uncertainty of the boundary. 
Three different cases are studied, with the mean perturbation magnitude of $0.25$, $0.5$ and $1.0$~mm. Valid, locally perturbed meshes are constructed via an elasticity operator that extends surface perturbations into the interior. Otherwise, identical incompressible flow problems are solved on these meshes, taking physiological boundary conditions and Carreau fluid parameters into account. Roughly $300\,000$ three-dimensional non-stationary blood flow simulations are performed for the three different perturbation cases to estimate the probability distributions of the quantities of interest. Convergence studies justify the spatial resolution of the employed meshes. Overall, the results suggest that moderate geometric perturbations result in reasonable engineering accuracy (relative errors in single-digit percentage range) of the quantities of interest, with higher sensitivity for gradient-related measures, noting that the observed errors are not negligible.
}


\maketitle

\section{Introduction}

\textit{In silico} modeling of the vascular system in health and disease has become increasingly relevant due to tremendous achievements in computational biomechanics in recent years~\cite{Menon2024, RolfPissarczyk2024}. Target applications in clinical scenarios include the comparison of alternative treatment options, parameter studies on virtual cohorts and detailed analyses of pathologies and related interventions. 
Numerous \textit{in silico} models have been devised to capture the underlying physics and reliably predict outcomes in conditions such as~\cite{Avril2021a, Fonken2021, Lan2022}, aortic dissection~\cite{Baeumler2020, Schussnig2024, Baeumler2024_submitted, Zimmermann2021}, thrombosis~\cite{Menichini2016a, Jafarinia2022a, Armour2022b, Schussnig2022b}, and more. Reproducing patient-specific flow conditions enables the analysis of otherwise only invasively obtainable or completely inaccessible data such as wall shear stress (WSS) and related biomarkers. For various conditions, such biomarkers have been associated with adverse outcomes, allowing configurations to be classified (see, e.g., \cite{Moretti2023a, Kimura2022a, Li2022a} in the context of aortic dissection). Such biomarkers allow comparing available treatment options \textit{in silico}, indicating likely outcomes. Therefore, these approaches possess great potential, given the computational models are capable of reproducing physiological conditions despite the potentially high inter-patient variability. 

Depending on the scenario, vessel wall compliance needs to be considered. Neglecting it translates to flow simulations on fixed meshes, which has been shown to yield excellent agreement with \textit{in vivo} and \textit{in vitro} data for certain cases~\cite{Zorilla2020a, Cheng2014a, Zimmermann2023a}. For other scenarios, however, assuming a fixed lumen introduces big modeling errors, such that fluid--structure interaction has to be accounted for~\cite{Alimohammadi2015a, Baeumler2020, Khannous2020a}. Here, differences in the results and derived biomarkers can be drastic, potentially leading to different conclusions. The physical coupling of the blood flow and vessel wall deformation (or thoughtful neglection thereof) has received great attention, but has to be decided on a case-by-case basis.

Similarly, it is unclear, \textit{how accurate} the geometric representation needs to be to achieve engineering accuracy. Engineering accuracy refers to a practically relevant error threshold in the target quantities of interest (QoI) typical for biomedical engineering. To apply \textit{in silico} models in clinical support, patient-specific data, boundary conditions, \textit{and} the geometry have to be approximated. 
Typically, convergence studies are employed to determine the required spatio-temporal resolution, but a discussion of the medical image data and the absolute limit in the achievable error in the target quantity is often completely ignored. The geometric approximation, uncertainties therein and their influence on the results have not been comprehensively investigated.

In this context, we interpret the error introduced by geometric approximation differently, namely, as the difference between two almost identical configurations or individuals. Even if the remaining model parameters remain unchanged, does (mild) geometric variation \textit{alone} already introduce significant alterations in the model output? Despite its seemingly academic character, this question is indeed of high practical relevance, as patient-specific data are usually severely limited in practical scenarios, such that patient-specific datasets are often augmented by additional data obtained elsewhere. Moreover, the geometric approximation inherently admits errors: on the one hand, the spatial resolution of CT scanners has a clearly quantifiable error of up to \(0.5\)~mm, with modern improvements aiming to reduce this to around $0.15$-$0.2$~mm~\cite{Lin_2009a, Kwan_2020a}. On the other hand, segmentation errors cannot be  clearly measured because manual segmentation~\cite{Taha_2015a} is often considered ground truth.ures and WSS) under geometric variations. Gralton~et~al.~\cite{Gralton_2024a} adopt a similar methodology to investigate a linear elastic abdominal aortic aneurysm, and fluid flow in an idealized cylinder with \(100\) examples. However, to draw statistically relevant conclusions, a \emph{large} number of simulations need to be executed. Naturally, performing a significant amount of simulations in a realistic time frame requires, among other things, a geometry description of suitable size with high mesh quality and representation accuracy. Note that the focus of the present paper is on standard three-dimensional finite element meshes. For non-trivial geometries such as blood vessels, the usual approach is unstructured tetrahedral meshing~\cite{Owen_2000a}. 
While it offers robustness and flexibility, several mesh properties are hard to control, namely mesh size and local structure. Thus, we turn to structured hexahedral meshing~\cite{Pietroni_2022a}, where additional effort and reduced robustness are traded for complete control over the above aspects of the mesh. Structured meshing has been well studied in biomedical applications, as blood vessels, to a degree, still enable the imposition of a structure~\cite{Zhang_2007a, Livesu_2016a, Ghaffari_2017a, Decroocq_2023a, Bosnjak_2024b}. Herein, we make use of a skeleton-based block-structured meshing approach~\cite{Bosnjak_2023a, Bosnjak_2023b}. The main benefits include straightforward control of boundary layers, and the ability to produce extremely coarse meshes ($\sim 300$ elements for an aorta with subvessels) of any desired polynomial order. While such a coarse mesh may not be directly useful, it can be utilised to construct a nested collection of meshes with the finest mesh being the final simulation mesh, thereby enabling the use of geometric multigrid methods~\cite{Wesseling_2001a}. The block-structured approach is therefore used to generate a hierarchy of meshes of a patient-specific aorta, which is subsequently used as input to the simulation pipeline.

\subsubsection*{Geometric uncertainty}
Uncertainty quantification has long been regarded a necessary and insightful aspect of biomechanics \cite{MILLER2013} and cardiovascular modeling \cite{Eck2016}. Several major types of uncertainty arise in cardiovascular models, such as material parameters \cite{ranftl2022stochastic,Ranftl2023}, boundary conditions \cite{schiavazzi2017patient} and geometry. 
The uncertainty and influence of geometry has been investigated in several studies \cite{SANKARAN2015167, sankaran2015fast, Antiga2003, maher2021geometric, du2022reducing}. Antiga~et~al.~\cite{Antiga2003} introduced patient-specific
reconstruction and meshing of blood vessels, e.g., abdominal aortas and carotid bifurcations. Sankaran et al. 
\cite{sankaran2015fast, SANKARAN2015167, sankaran2014real} studied the sensitivity of fractional flow reserve in the coronary arteries to lumen segmentation uncertainty with a sectional Gaussian perturbation model associated to the centerline. Sankaran et al. \cite{Sankaran2016} then also studied the interaction of so-defined geometric uncertainty with uncertainty of (Newtonian) blood viscosity and boundary resistances in sensitivity of fractional flow reserve. Cilla et al. \cite{cilla2020parametric} studied the impact of geometric features of the aortic arc. Du and Wang \cite{du2022reducing} modeled geometric uncertainty via variation of global shape descriptors, such as segmental lumen radii, of the abdominal aorta.
Hoeijmakers et al. \cite{Hoeijmakers2021} investigated the shape uncertainty of the aortic valve and found that uncertainty of only voxel size may already lead to substantial uncertainty in transvalvular pressure drop.

\subsubsection*{Global shape uncertainty and local boundary uncertainty}
We observe that several studies have investigated \textit{shape} uncertainty \cite{Liang2017a}, i.e., \textit{global} geometric uncertainty. 
In contrast, \textit{local} features of geometric uncertainty remain mostly unclear. Of the few available studies, we put our work briefly in context with two notable ones: in their study, Biehler~et~al.~\cite{biehler2015towards} modeled the geometric parameter of aortic wall thickness as a random field. Aortic wall thickness however cannot typically be measured from CT, but, e.g., from \textit{ex vivo} samples. In contrast, in this contribution we aim to study \textit{local} geometric uncertainty of the boundary itself instead of the wall thickness, which can be typically measured from CT, in the context of CT instrument resolution and segmentation uncertainty. We will model local variations of the boundary as a random field similarly to~\cite{biehler2015towards}. Maher~et~al.~\cite{maher2021geometric} considered local geometric uncertainty implicitly combined with global geometric uncertainty (shape variability) via volume image inputs for several QoIs for pathological aortas using a machine learning model. With $\sim 10^2$ samples, aortic radii uncertainty assumed well below image resolution, and the mixing of local and global geometric uncertainty in their model without control of local geometric uncertainty, it is difficult to differentiate their results to understand and isolate the effect of local geometric uncertainty. Note that in our study, we do not make any additional machine learning-based surrogates or approximations.

This contribution is structured as follows: Sec.~\ref{sec:Methods} covers the required methodologies, namely modeling the boundary as a random field to vary the local geometric features, accounting for these geometric features on the mesh level, vascular modeling and QoIs, and finally the propagation of the uncertainty through the model. Sec.~\ref{sec:Results} discusses the details of the numerical method, the convergence studies, and shows the results of the simulations and the uncertainty propagation. It is followed by the discussion and interpretation of said results in Sec.~\ref{sec:Discussion}. The conclusions in Sec.~\ref{sec:Conclusion} outline the most important findings and sketch future work.

\section{Methods}\label{sec:Methods}
\begin{figure}[!h]
    \centering
    \subfloat
    {
        \includegraphics[width=0.16\linewidth,draft=\draftMeshExamples]{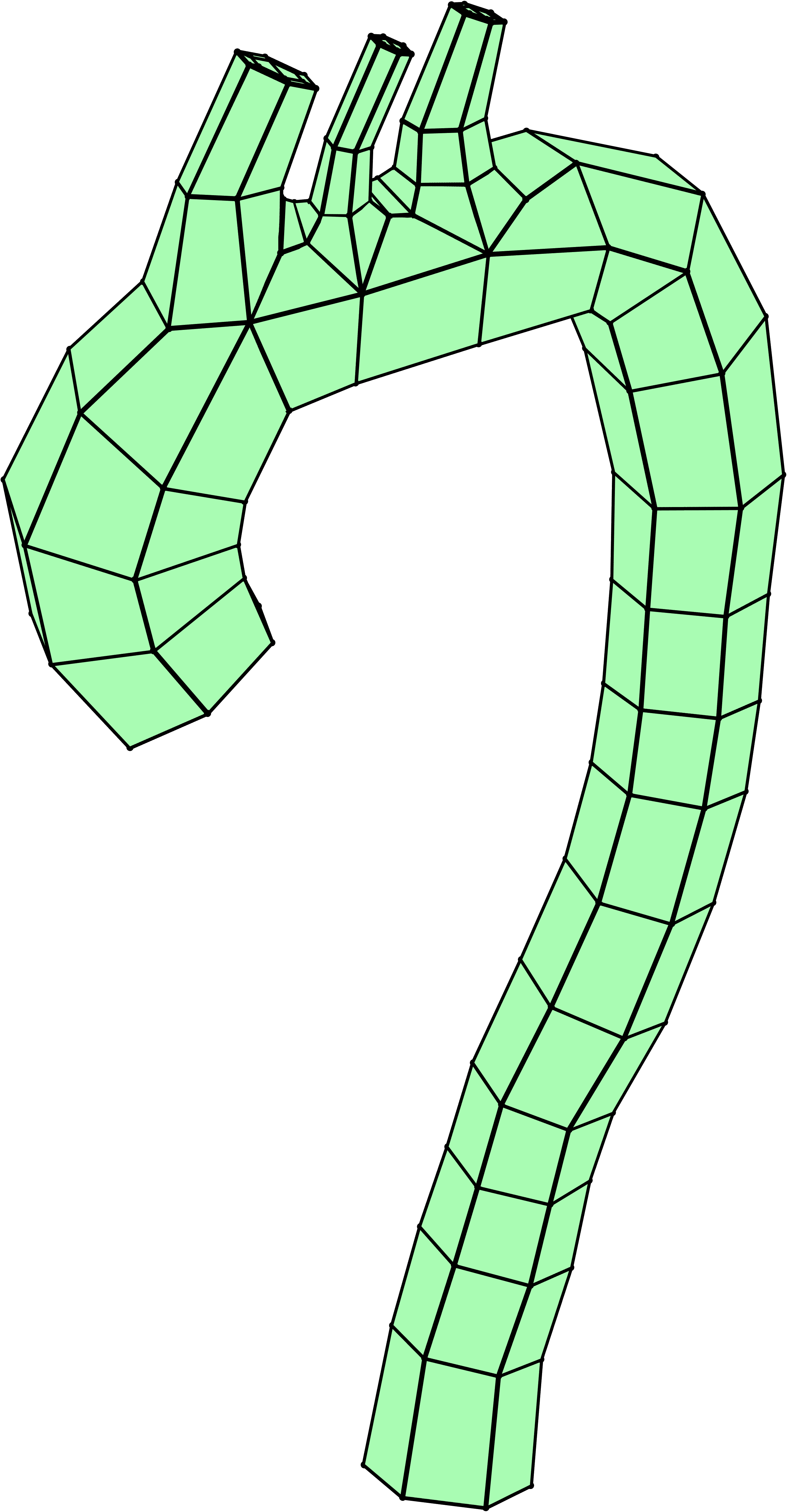}
    }
    \hspace{0.02\linewidth}
    \subfloat
    {
        \includegraphics[width=0.16\linewidth,draft=\draftMeshExamples]{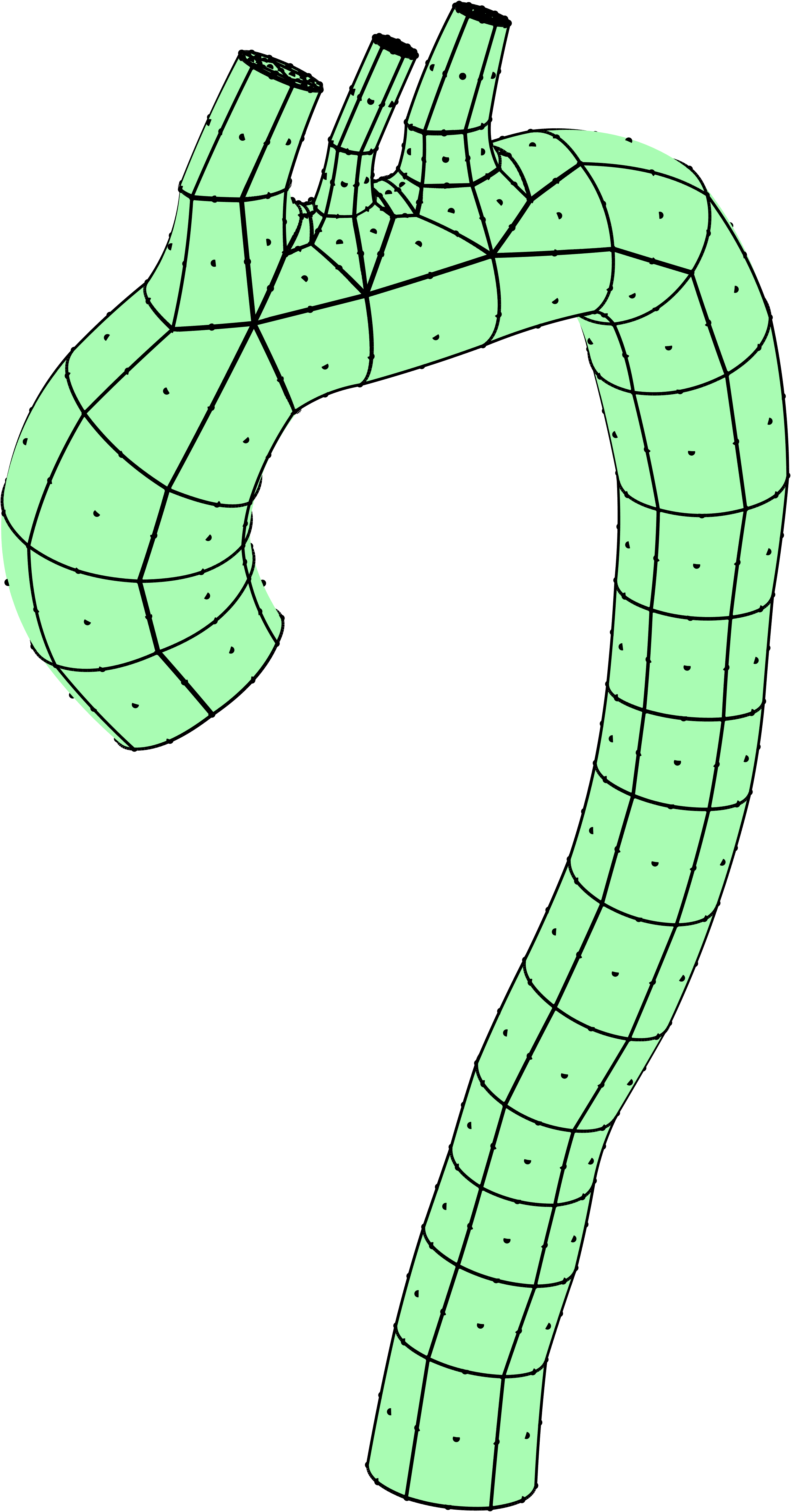}
    }
    \hspace{0.02\linewidth}
    \subfloat
    {
        \includegraphics[width=0.16\linewidth,draft=\draftMeshExamples]{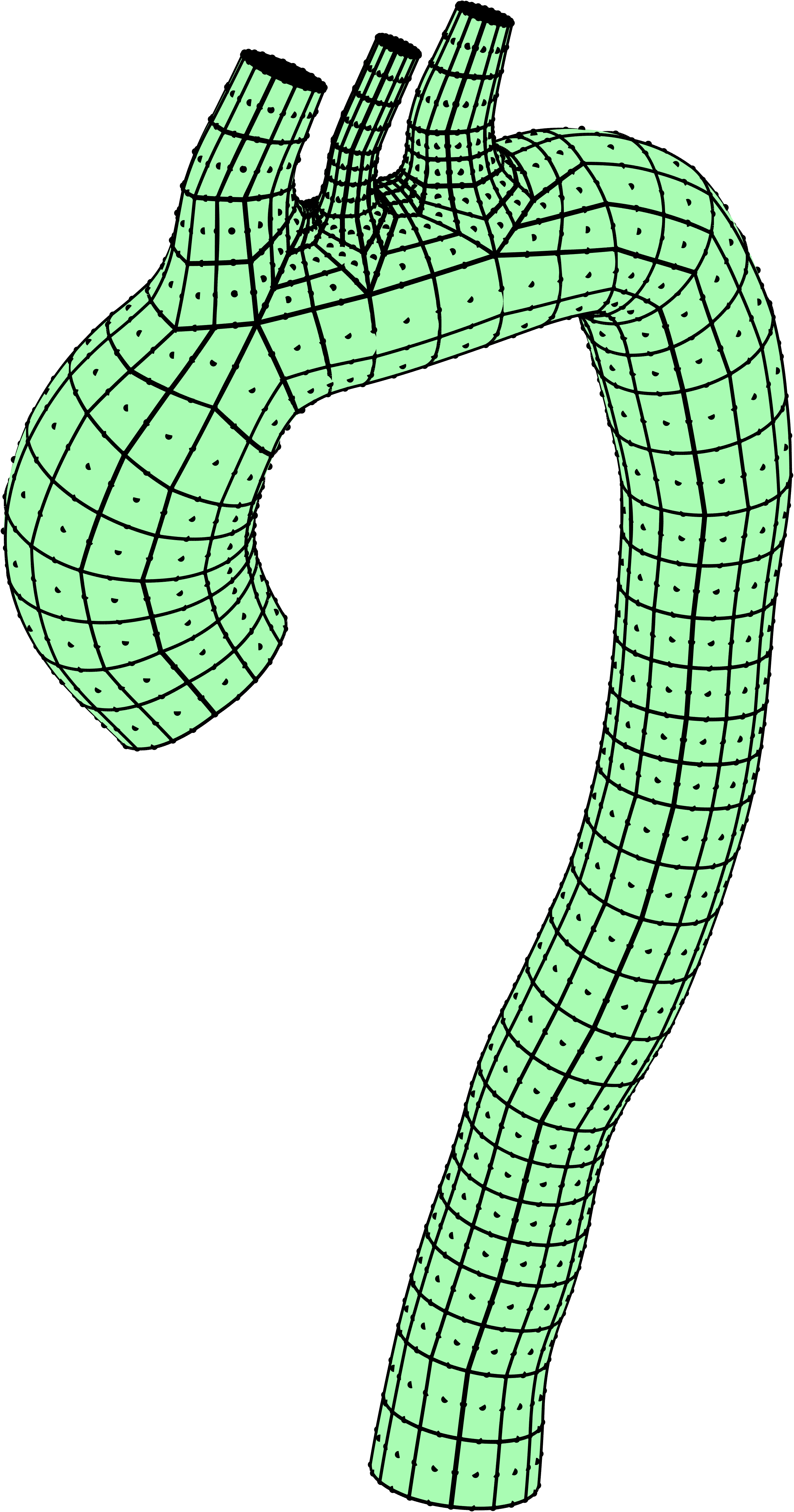}
    }
    \hspace{0.02\linewidth}
    \subfloat
    {
        \includegraphics[width=0.16\linewidth,draft=\draftMeshExamples]{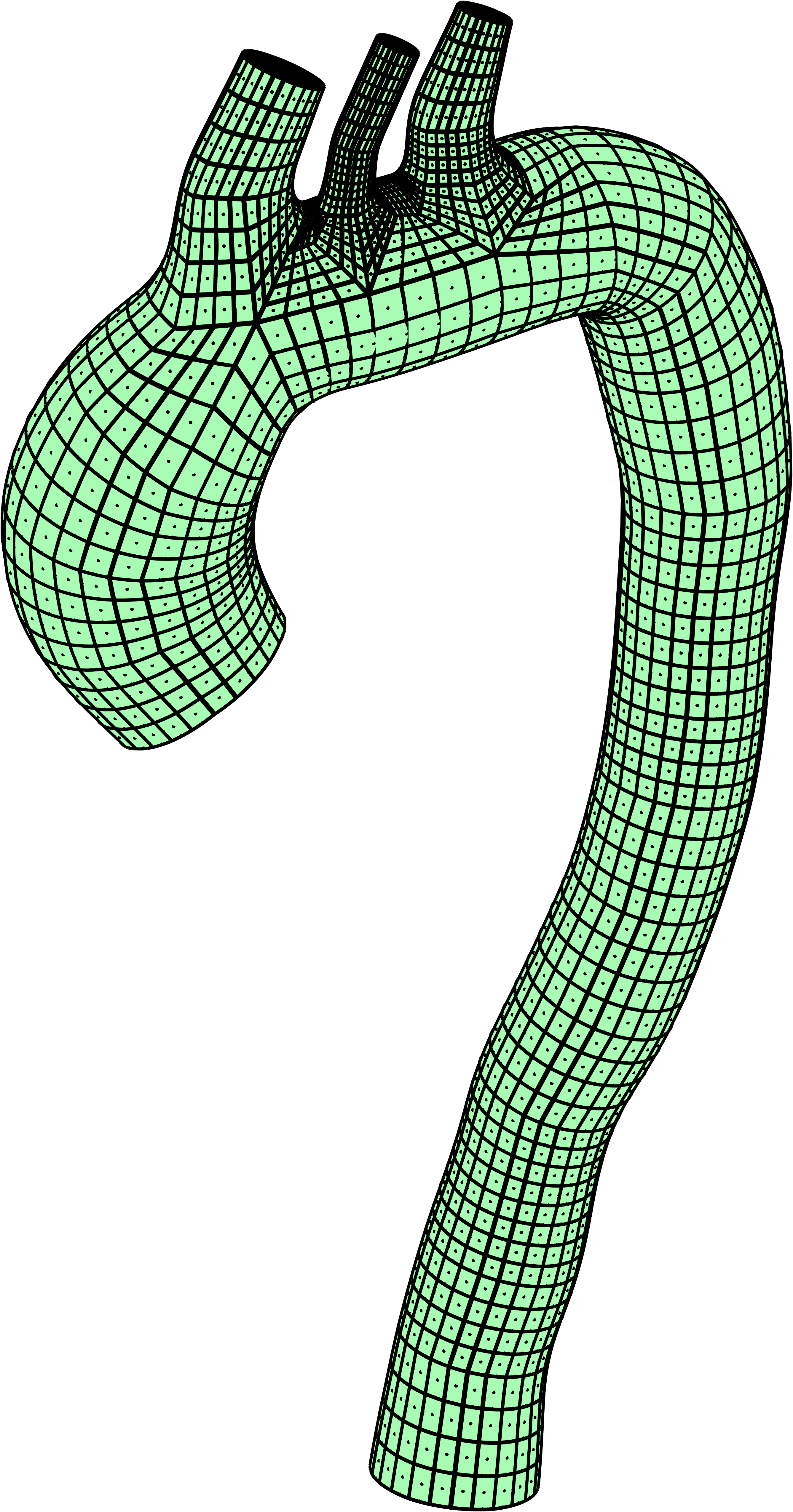}
    }
    \caption{The employed hierarchy of nested, hexahedral meshes: (a) linear mesh with \(224\) elements, followed by quadratic meshes with (b) \(224\) elements, (c) \( 1\, 792\) elements, and (d) \(14\, 336\) elements, respectively. }
    \label{fig:MeshPerturbation_Input}
\end{figure}

To realize this analysis, we focus on a single patient-specific aorta meshed with a block-structured approach~\cite{Bosnjak_2023a, Bosnjak_2023b}. The main features of the said approach which we capitalize on are the capabilities to generate coarse meshes, as well as higher-order meshes. Figure~\ref{fig:MeshPerturbation_Input} shows a sample nested hierarchy of structured meshes using linear (on the coarsest mesh) and quadratic finite elements. The fact that the meshes can be so coarse, while still leading to reasonable errors and good domain representation, enables the execution of this study. Namely, a mesh that is too large would significantly impair the capability to perform a substantial amount of simulations in a realistic time window. 

This section describes the methods employed to perturb the geometry to consider for local geometric variations with mean values of $0.25~\text{mm}$, $0.5~\text{mm}$ and $1.0~\text{mm}$, which is intended to reproduce small uncertainties on the boundary as introduced by medical imaging, segmentation and meshing. This requires automated and systematic construction of perturbation fields, applying these fields to deform the initial geometry, solving flow problems  reproducing physiological flow on the perturbed meshes, computing the QoIs, and performing the statistical analysis of the obtained results.

\subsection{The boundary as a random field}\label{subsec:BoundaryRandomField}

We model local perturbations to the boundary, or to the mesh, respectively, as a Gaussian random field. A random field is, colloquially, the generalization of random variables to field variables: the observable is a collection of distinct random variables, one random variable at each point in the field. In our case, each point in space is assigned a random variable, and all of these random variables have a defined correlation. The simplest example is a Gaussian random field, and the definition reads that from the collection of random variables, every finite subset follows a (multi-variate) Gaussian distribution. In our case, this corresponds to perturbations entering the domain discretization. In other words, all mesh boundary nodes together, or rather the perturbations of the nodes' spatial coordinates, follow a joint multi-variate Gaussian distribution. Formally, a random field $\mathcal{F}(\Omega_{0})$ is a function that assumes random values at every point in the reference domain $\Omega_0$, see Fig.~\ref{fig:MeshPerturbation_Input}. We denote the reference domain's vessel wall boundary as $\Gamma_{0,\mathrm{wall}}$ and evaluate the random field at $\Xbf \in \Gamma_{0,\mathrm{wall}}, \Xbf = (X_1, X_2, X_3)^{\rm T}$. A Gaussian random field is defined as a multi-input Gaussian process, i.e.,
\begin{equation} \label{eq:def:gp}
    \mathcal{F} \sim {\cal{GP}}\big( {{\mu}(\Xbf}), \mathrm{k}\big(\Xbf,\Xbf^\prime ) \big) \;,
\end{equation}
with a mean function ${{\mu}(\Xbf)}$ and a positive semi-definite covariance function $\mathrm{k}(\Xbf, \Xbf^\prime)$ between any two points $\Xbf \in \Gamma_{0,\mathrm{wall}}$ and $\Xbf^\prime \in \Gamma_{0,\mathrm{wall}}$. In this study, we will assume a zero-mean prior ($\mu \equiv 0$), and an isotropic, non-stationary correlation function $\mathrm{k}(\Xbf, \Xbf^\prime) = A \exp{
[ - (\Xbf - \Xbf^\prime)^2/2\ell^2]\varsigma^2(X_3)} $.

The factor $A \varsigma^2$ controls the magnitude of perturbations, while the correlation length $\ell$ controls the length scale of spatial variability. We will relate this magnitude to \textit{combined} CT instrument \textit{and} segmentation uncertainty. The length scale is, as the name suggests, as scale variable, and thus naturally bounded by instrument resolution and spatial scale of the aorta.
The factor $\varsigma^2$ is often chosen to be homogeneous, i.e., constant in space. In our study however, the supraaortic 
arteries  demand considerable regional mesh refinement. Hence large perturbations are (i) not physiological and (ii) can result in invalid meshes. Hence the factor $\varsigma^2$ was amended to reduce the perturbation magnitude at the corresponding region as compared to the abdominal or ascending aorta as follows: 
$\varsigma^2(X_3) = \{ 1 + \mathrm{tanh}[ (X_{3,0} - X_3)/2] \} / 2 + 0.05$, where $X_{3,0} = 333 ~\mathrm{mm}$ is the highest point immediately below the subclavians. The values for the parameters  $A$ and $\ell$ were set to distinct values for three distinct experiments in Sec.~\ref{sec:numerical_results-UQ-analysis}. 

The correlation length was $\ell = 5$~mm and isotropic in all cases. Three test cases are analyzed, with the mean magnitude of perturbations of \(A = 0.25,\, 0.5 \text{ and }1\)~mm, respectively. The cases reflect both a lower bound on the segmentation uncertainty of the boundary due to typical instrumental precision of 1 voxel in contemporary CT, and an upper bound where perturbations are not physiological (e.g., the typical radius of a healthy aorta is in the range of $\sim 15$~mm, and an uncertainty of \(10\%\) of that is already unrealistically high).

We define $\hat{\Xbf} = \{\Xbf^{(n_{\rm x})} \}_{n_{\rm x}=1}^{N_{\rm x}}$, where $\Xbf^{(n_{\rm x})} = (X_1^{(n_{\rm x})}, X_2^{(n_{\rm x})}, X_3^{(n_{\rm x})})^\top$, as the set of mesh coordinates in the discretization of the reference configuration's vessel wall boundary $\Gamma_{0,\mathrm{wall}}$. That is, $N_{\rm x}$ is the total number of nodes in the boundary mesh. We will later identify the vector of values $\fbf = (f^{(1)},\ldots,f^{(n_{\rm x})})^\top$ of the random field at each node $\Xbf^{(n_{\rm x})}$ with the vector of mesh perturbations $\fDeltaX^{(N_{\rm x})}$ at each node on 
$\Gamma_{0,\mathrm{wall}}$.
Then, according to the definition of a Gaussian random field, the joint probability density function (PDF) $p$ for the random field values $\fbf$, or mesh perturbations $\fDeltaXhat = (\fDeltaX^{(1)}, \ldots,\fDeltaX^{(N_{\rm x})})^\top$,
at all $\Xbf^{(n_{\rm x})} \in \hat{\Xbf}$ is simply a multi-variate Gaussian distribution, i.e.,
\begin{equation}\label{eq:def-discrete-gauss-random-field}
    \p{\fDeltaXhat}{\hat{\Xbf}} = {\mathcal{N}}(0; \mathbf{K}) 
    \qquad \mbox{with} \qquad 
    [\mathbf{K}]_{uv} = \mathrm{k}(\Xbf^{(u)}, \Xbf^{\prime(v)}),
\end{equation}
where ${\mathcal{N}}$ represents the multivariate normal distribution with the mean $\mu=0$ and the covariance matrix 
$[\mathbf{K}]$ with the components $u$ and $v$ of size $N_{\rm x}$ given by the number of nodes in the boundary mesh. The components of $\mathbf{K}$ are determined by the chosen covariance function $\mathrm{k}(\Xbf, \Xbf^\prime)$. We then draw samples from this distribution, that is, we generate realizations of random fields which are consistent with so defined statistics.

\subsection{Sampling Gaussian random fields} \label{sec:sampling-gaussian-fields}
There are several ways of sampling from multi-variate Gaussian distributions \cite{Liu2019a_randomfields}. The simplest method is the Cholesky decomposition. In this study we use a simple Karhunen-Loéve approximation \cite{spanos1989stochastic,Ghanem1991}. For the sake of convenient computation, we exploit the fact that the eigenvalue spectra of such covariance matrices typically fall off exponentially. Based on an initial covariance matrix eigenvalue decomposition of only the first \(\sim 100\) eigenvalues, we fitted an exponential model to estimate the truncation index of the Karhunen-Loéve expansion such that the eigenvalue spectral energy of the Karhunen-Loéve approximation retains $99.9\%$ of the exact decomposition obtained with a Lanczos scheme, i.e., the deviations from the Gaussian function are negligible. 
Each sample then realizes a random mesh perturbation. A simple addition of the random perturbations to the mesh as 
$\hat{\Xbf} + \fDeltaXhat$ very often yields an invalid mesh. Thus, this step requires more attention, as explained in the following subsection.

\subsection{Mesh perturbation}
\label{sec:methods:mesh-perturbation}
To induce the perturbations in the mesh, its surface nodes are moved in the direction of their respective normal vectors. The aforementioned random field determines the magnitudes of the movement, as well as the signs, which specify whether the movement is oriented towards the outside or the inside of the mesh. The nodal normal vectors are interpolated from the mesh, and normalized so as not to affect the magnitude. Applying surface perturbations without extra precautions leads to elements becoming invalid, rendering the mesh unusable. Therefore, a more detailed consideration is necessary. Namely, the mesh perturbations are treated as solid body deformations, and a linear pseudo-elasticity problem is set up accordingly~\cite{Stein_2003a}. Thus, 
\begin{eqnarray}
    \label{eqn:linelast}
\nabla \cdot 
\left[ 
    \lambda_\mathrm{m} (\mathrm{tr}\ve{\epsilon})\te{I} + 2 \mu_\mathrm{m} \, \ve{\epsilon}
\right]  = \ve{0},
\quad
\mathrm{in~}
\Omega_0,
\quad
\text{with}
\quad
\ve{\epsilon} = \nabla^S\ve{d},
\end{eqnarray}
where the symmetric gradient is defined as $\nabla^S(\bullet) := \nicefrac{1}{2}\left[\nabla (\bullet) + \nabla^\top (\bullet)\right]$ and $\ve{d}$ is a continuous displacement field, describing the invertible map from the reference configuration $\Omega_0$ to $\Omega$, the deformed domain considering for the geometric perturbations. The surface node perturbations following from~\eqref{eq:def-discrete-gauss-random-field} enter as Dirichlet boundary conditions in the discretized form of \eqref{eqn:linelast}, while the inlet and the outlet faces are kept fixed. 
The mesh nodes' positions \(\ve{X}^{(1)},\dots,\ve{X}^{(N_x)}\), as denoted in Sec.~\ref{subsec:BoundaryRandomField}, are perturbed via \( \{ \fDeltaXhat^{(1)}, \dots , \fDeltaXhat^{(N_\mathrm{x})}\}\), i.e.,
\begin{eqnarray}
    \fDeltaXhat^{(n_{\mathrm{x}})}
    =
    \ve{d}^{(n_\mathrm{x})} , \quad n_x = 1,\dots,N_x. 
\end{eqnarray}
Thus, the perturbations are enforced exactly, and all of the non-perturbed nodes are updated accordingly. The Lam{\'e} material parameters are computed from the Young's modulus \(E\) and Poisson's ratio \(\nu\) as \(\mu_\mathrm{m} = E
/ (1 + 2\nu)\) and \(\lambda_\mathrm{m} = E\nu / [(1+\nu)(1-2\nu)]\). The Poisson's ratio is set to \(0\). The Young's modulus is computed locally based on the volume of the corresponding finite element \(E = \left(V_{\text{Elem}}\right)^k\), with \(k\geq 0\). Varying \(k\) produced no difference in mesh validity, thus it was set to \(k=1\). Such approaches are well-known in the fluid-structure interaction community, employed to tackle mesh movement~\cite{Stein_2003a}.

Each perturbation case therefore implies a partial differential equation to be solved, e.g., via the finite element method. Hence, each case would require the assembly of a matrix and a right-hand side vector, and solving the associated linear system of equations. However, we observe that it is only the boundary conditions that vary, not the system matrix, and the number of nodes/unknowns is usually rather small due to the use of structured meshes. It is therefore assembled once, at the beginning of the algorithm. Additionally, we apply the LU decomposition with full pivoting immediately, thereby avoiding the need for advanced system-solving schemes. Then, for each case, the right-hand side vector is updated according to the Dirichlet boundary conditions, i.e., the surface perturbations, and the system is solved with a simple forward/backward substitution. Since the different perturbation cases are independent of each other, and the system matrix is pre-assembled and pre-decomposed, the algorithm is completely straightforward to parallelize. 

\begin{figure}[!h]
    \centering
    \subfloat
    {
        \includegraphics[width=0.16\linewidth,draft=\draftMeshExamples]{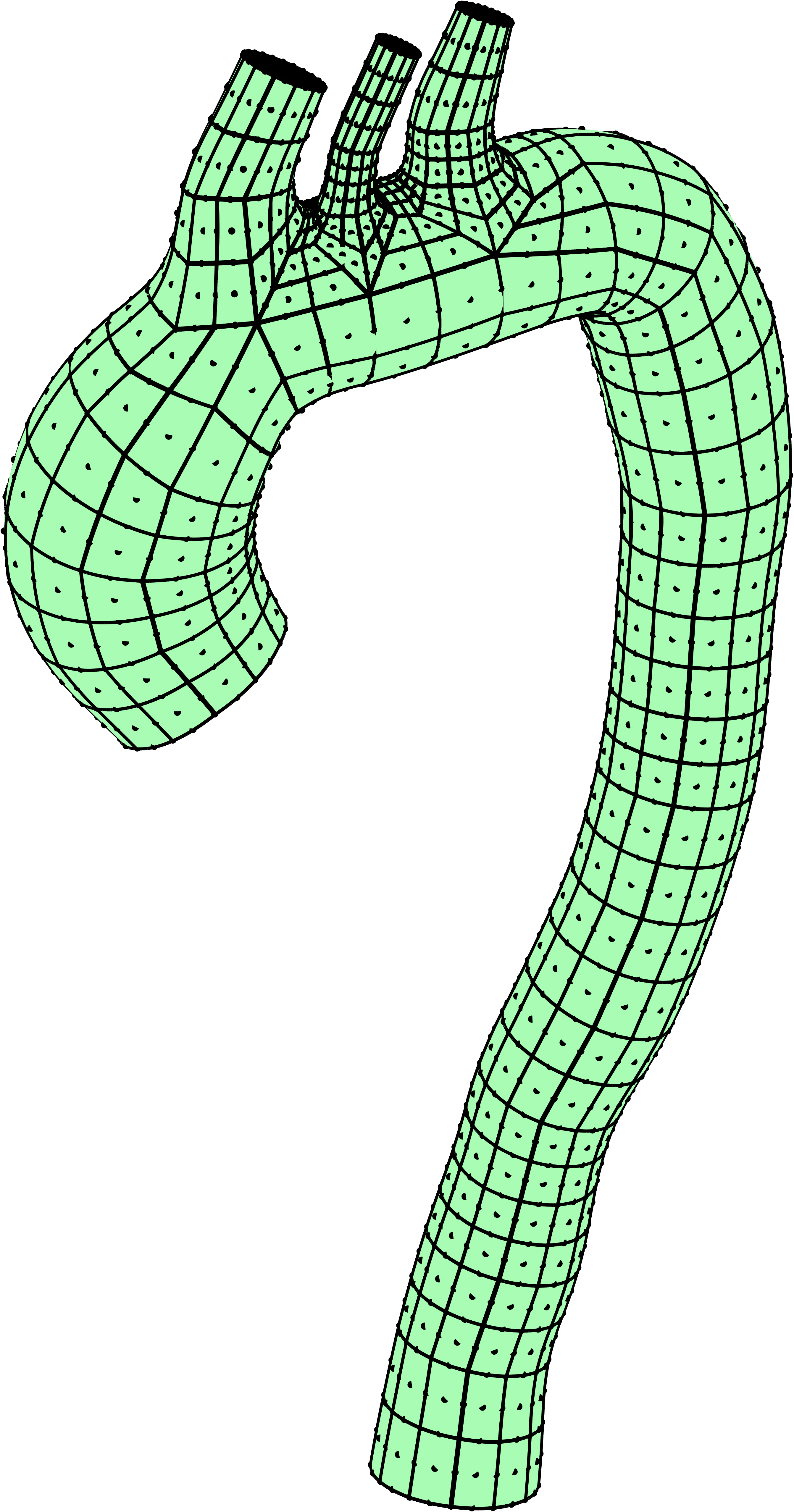}
    }
    \hspace{0.02\linewidth}
    \subfloat
    {
        \includegraphics[width=0.16\linewidth,draft=\draftMeshExamples]{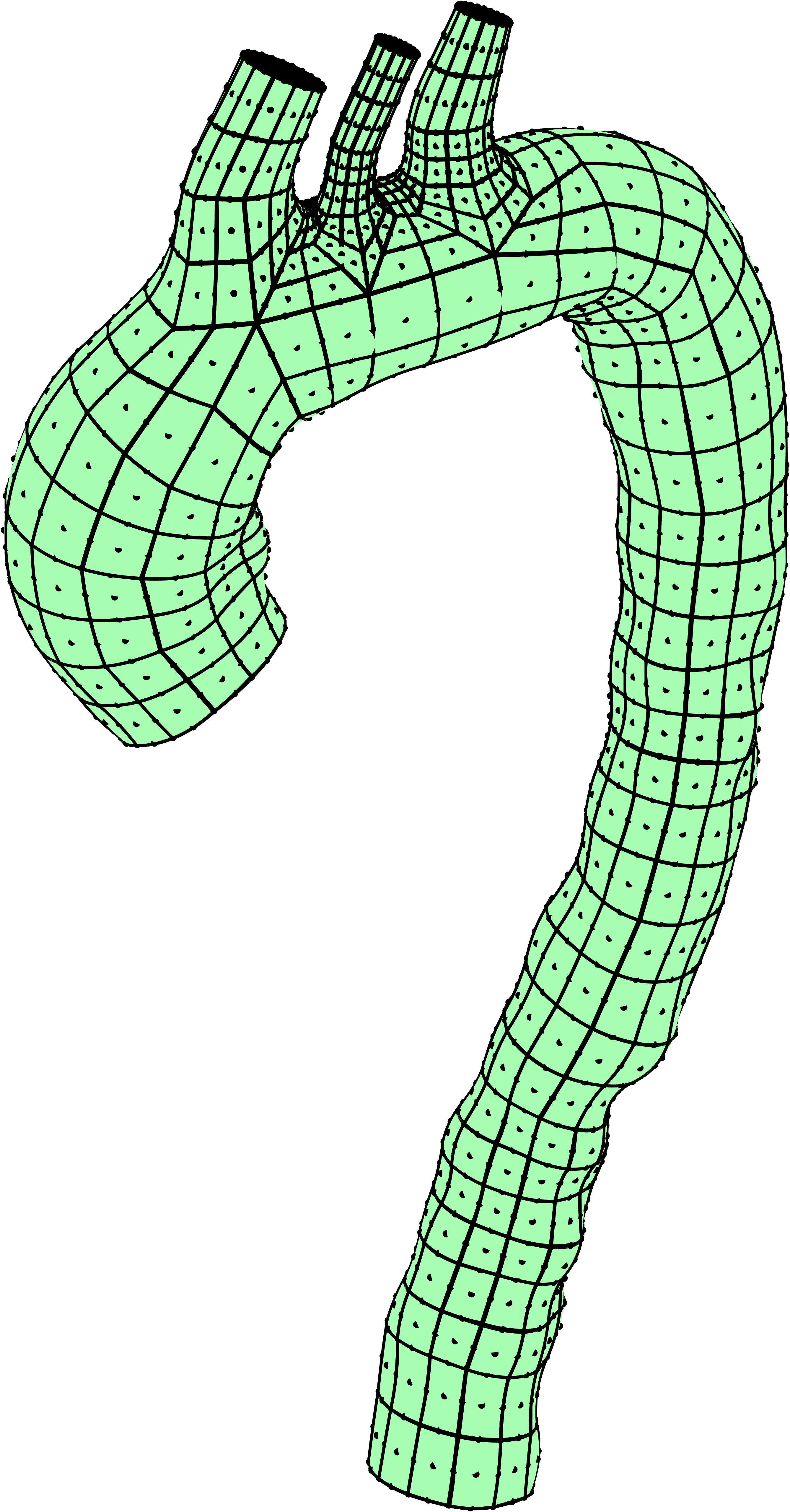}
    }
    \hspace{0.02\linewidth}
    \subfloat
    {
        \includegraphics[width=0.16\linewidth,draft=\draftMeshExamples]{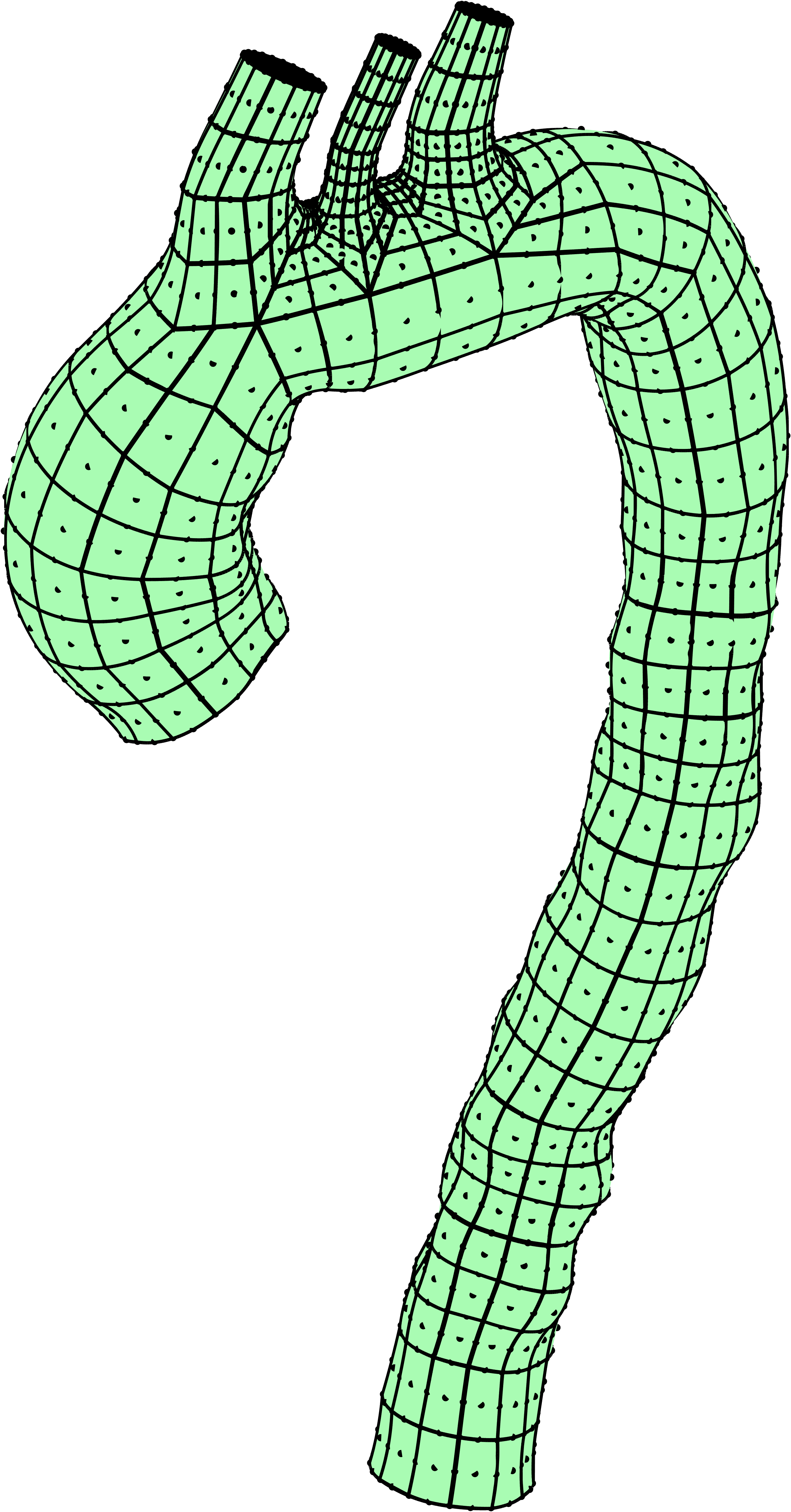}
    }
    \hspace{0.02\linewidth}
    \subfloat
    {
        \includegraphics[width=0.16\linewidth,draft=\draftMeshExamples]{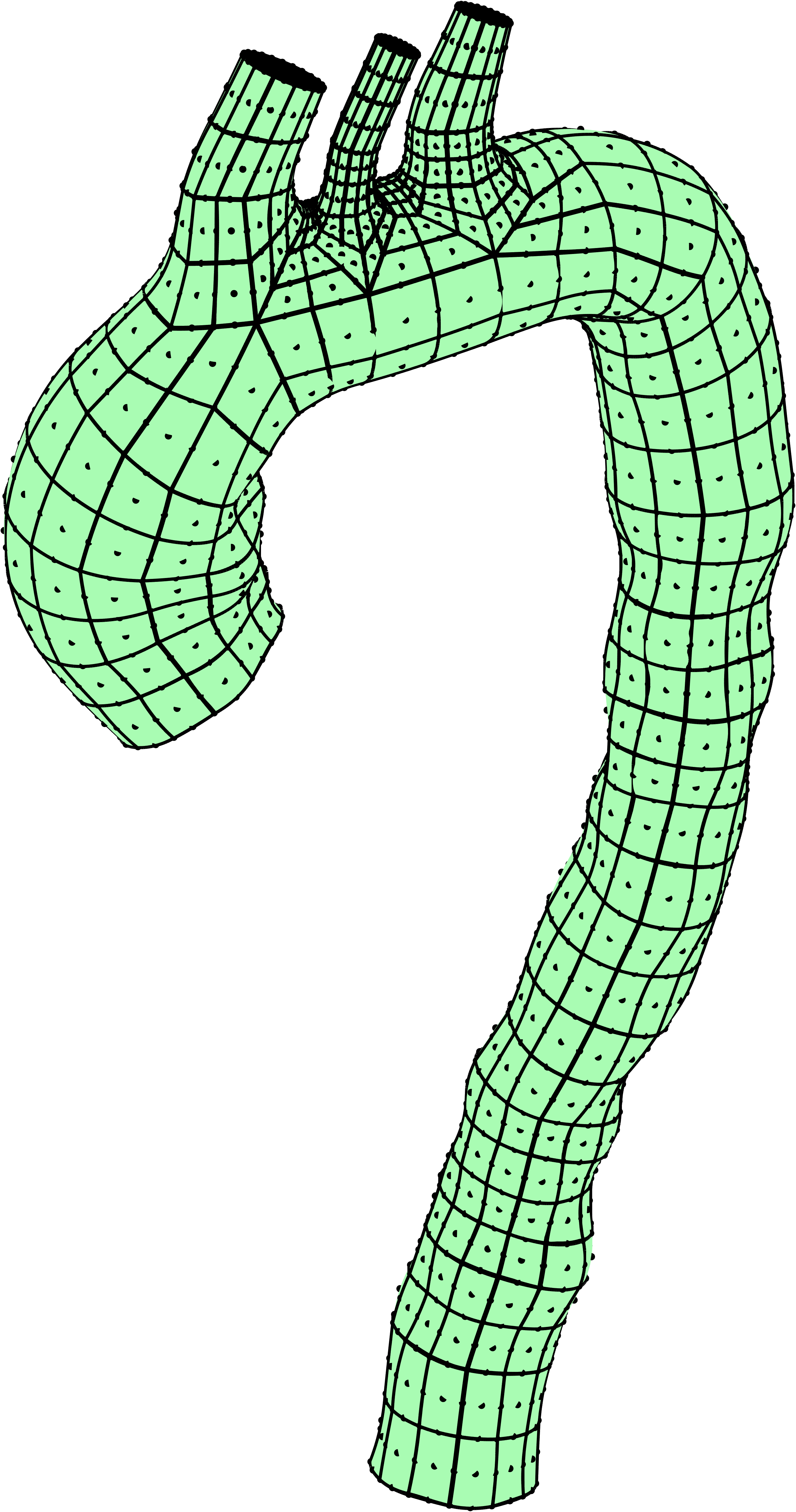}
    }
    \caption{Geometrical changes caused by the surface perturbation process: (a) original unperturbed mesh, and (b)-(d) perturbed meshes, with the same perturbation magnitude.}
    \label{fig:MeshPerturbation_Example}
\end{figure}

\begin{figure}[!h]
    \centering
    \subfloat
    {
        \centering
        \begin{overpic}[width=0.15\textwidth]{Figures/MeshPerturbation/Mesh_Unpert.png}
        \put(-85,45){\frame{\includegraphics[width=0.09\linewidth]{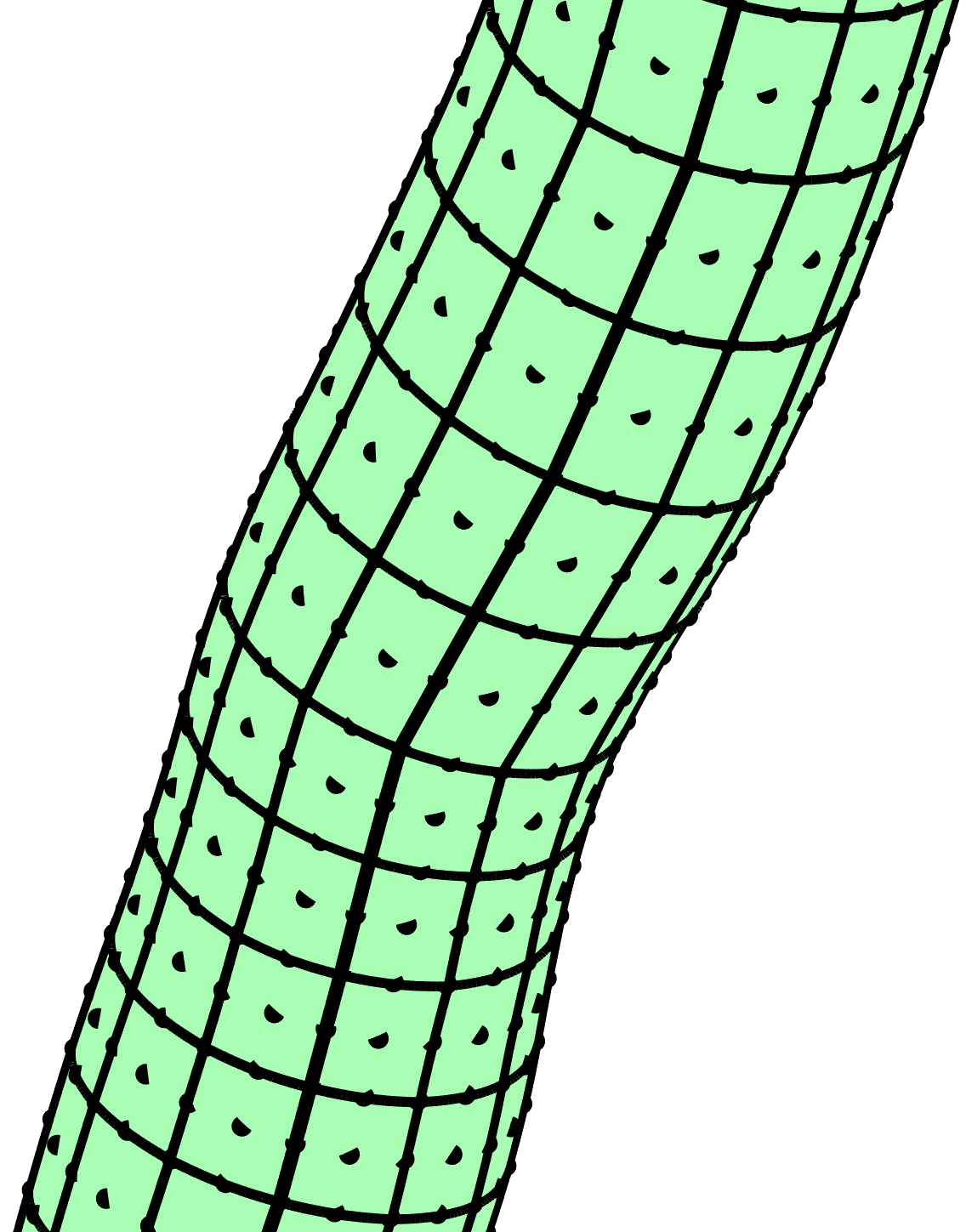}}}
        \end{overpic}
    }
    \hspace{0.02\linewidth}
    \subfloat
    {
        \begin{overpic}[width=0.15\textwidth]{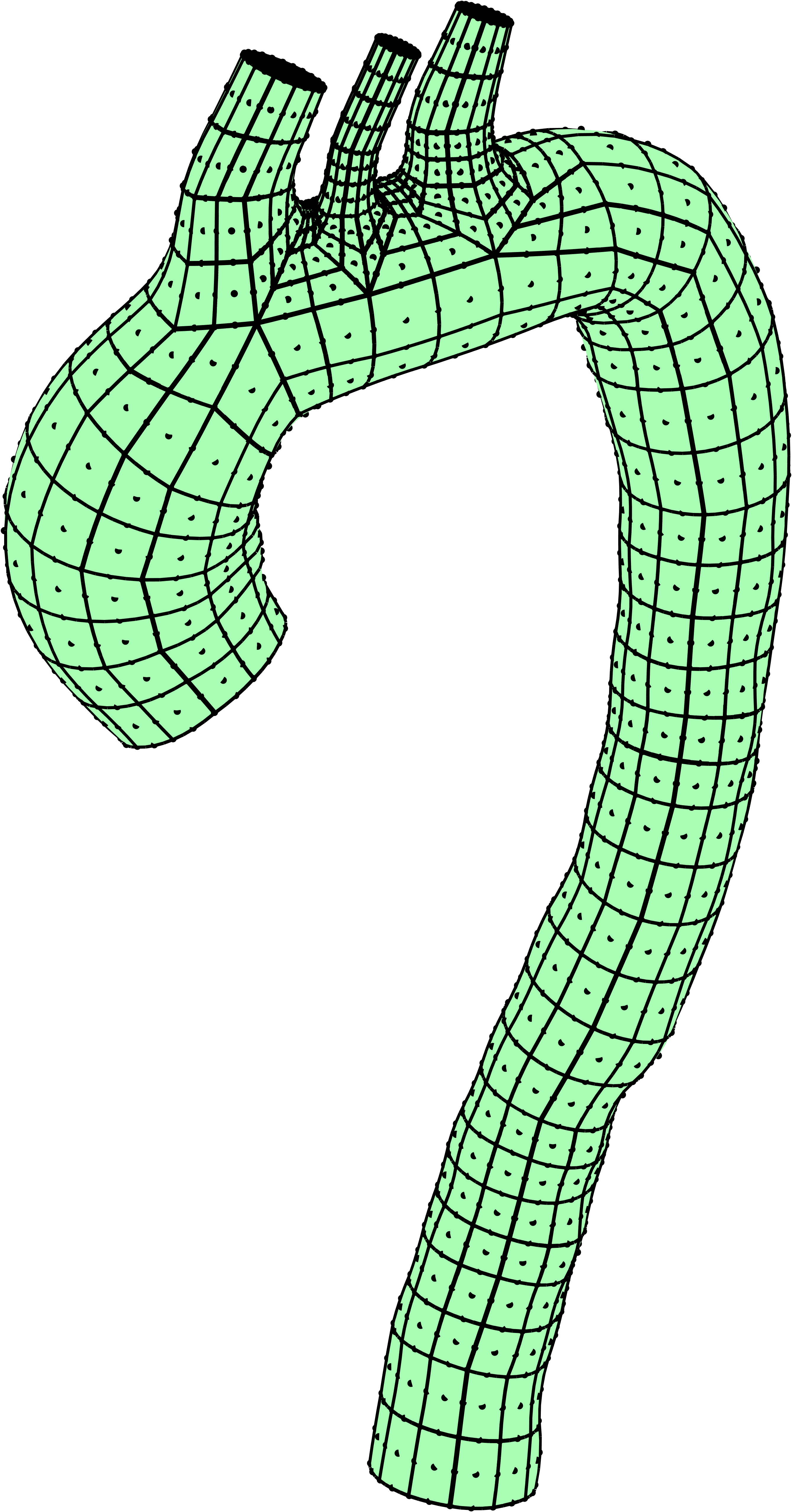}
        \put(-85,45){\frame{\includegraphics[width=0.09\linewidth]{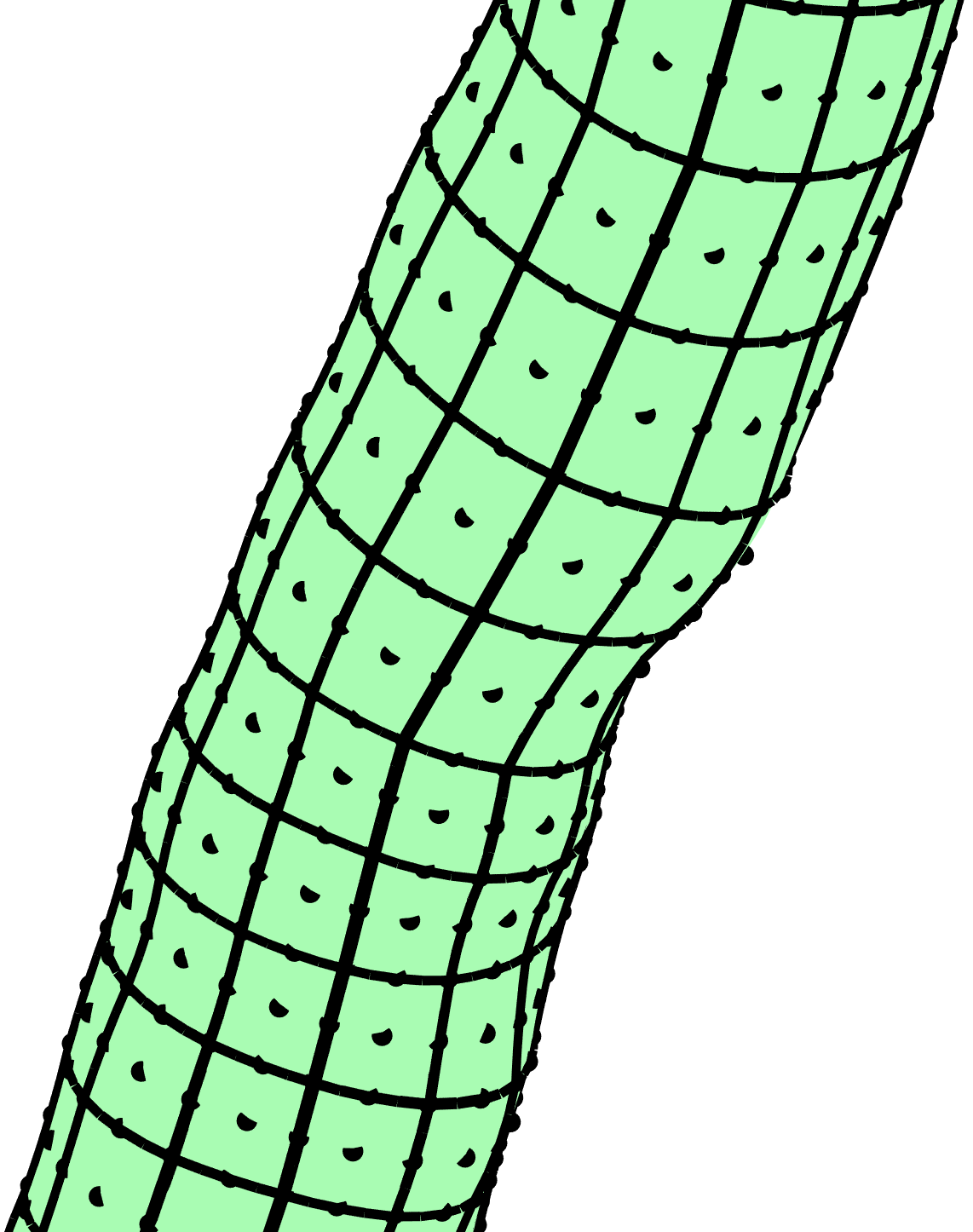}}}
        \end{overpic}
    }
    \hspace{0.02\linewidth}
    \subfloat
    {
        \begin{overpic}[width=0.15\textwidth]{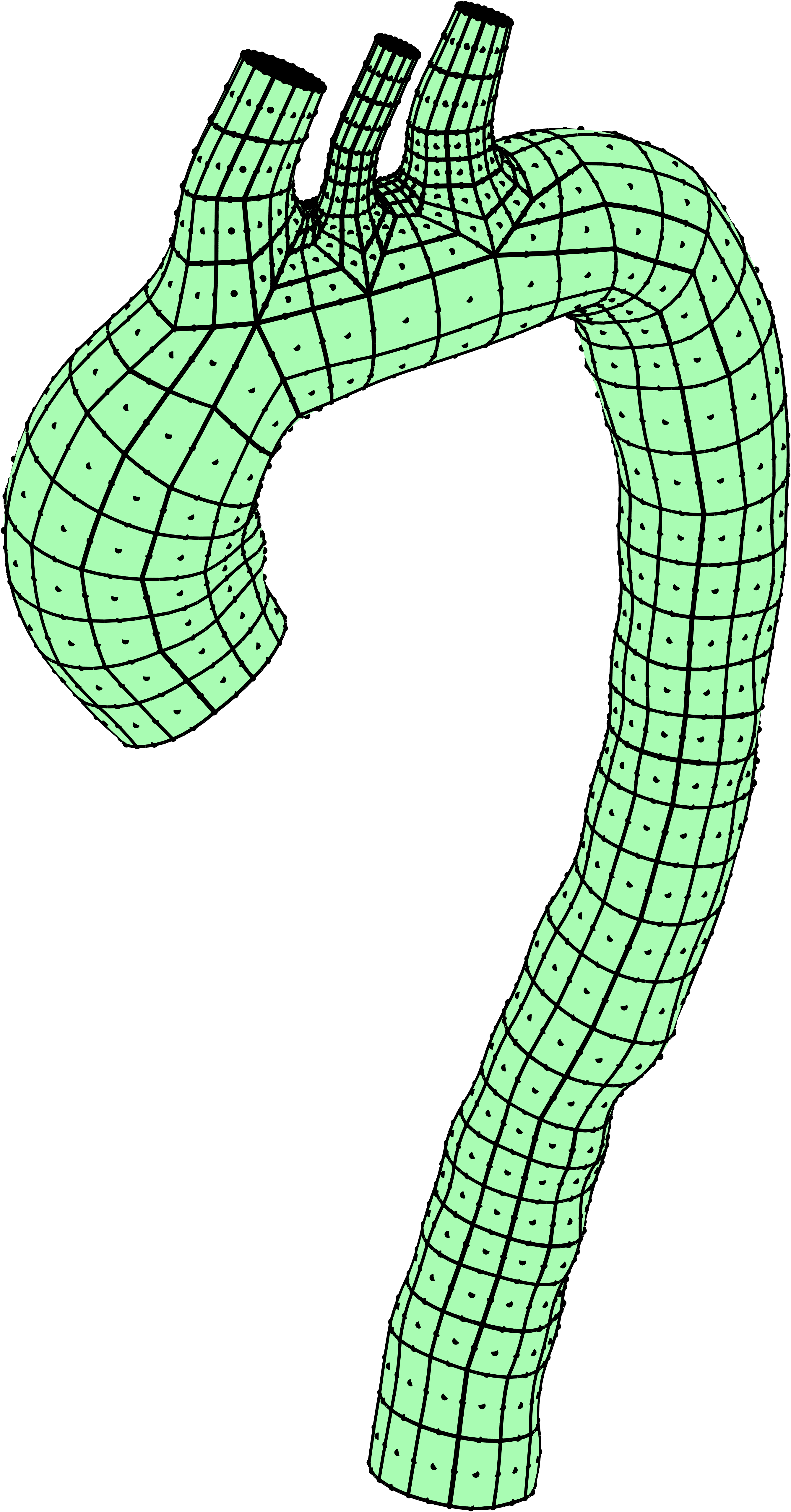}
        \put(-85,45){\frame{\includegraphics[width=0.09\linewidth]{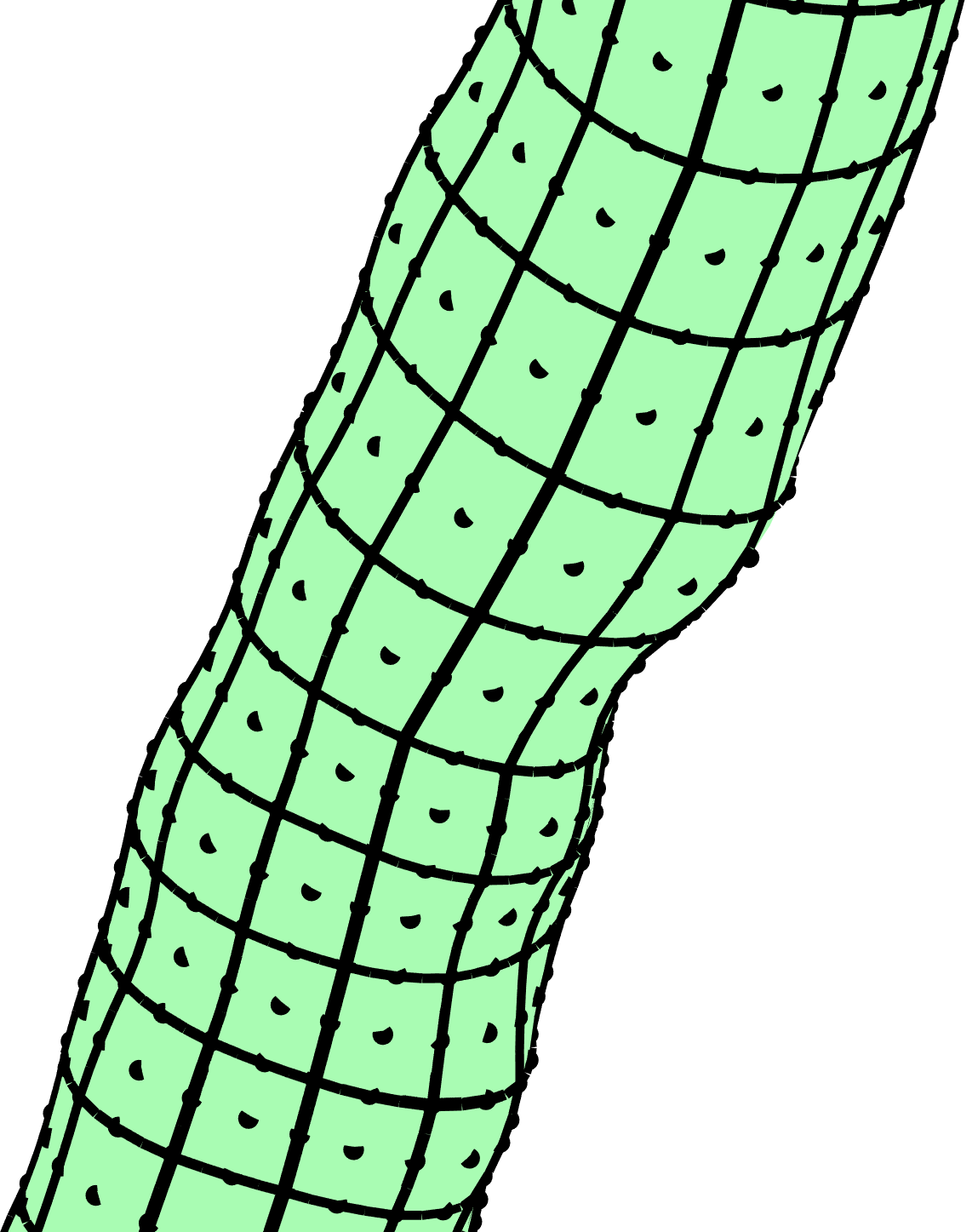}}}
        \end{overpic}
    }
    \hspace{0.02\linewidth}
    \subfloat
    {
        \begin{overpic}[width=0.15\textwidth]{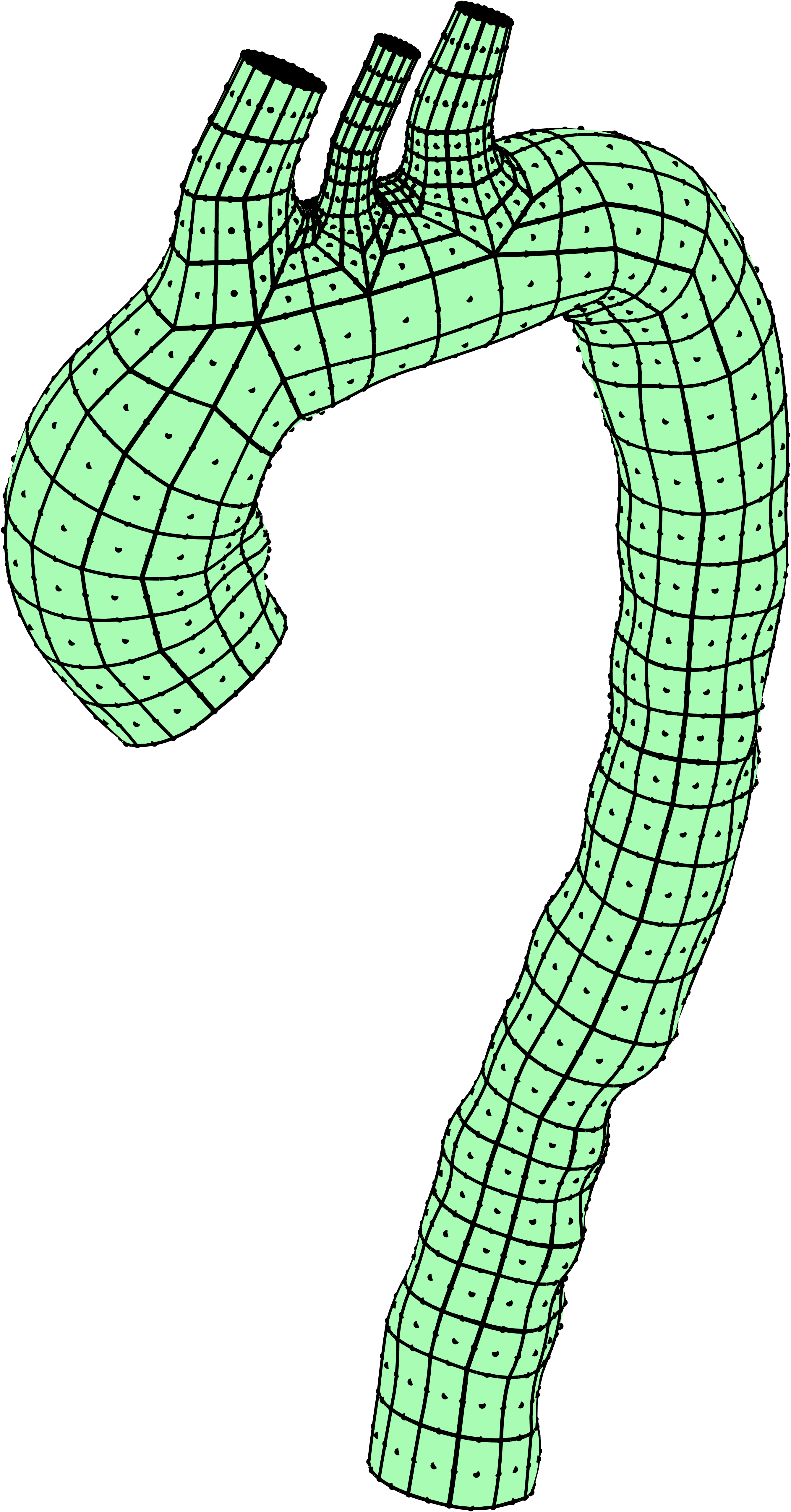}
        \put(-85,45){\frame{\includegraphics[width=0.09\linewidth]{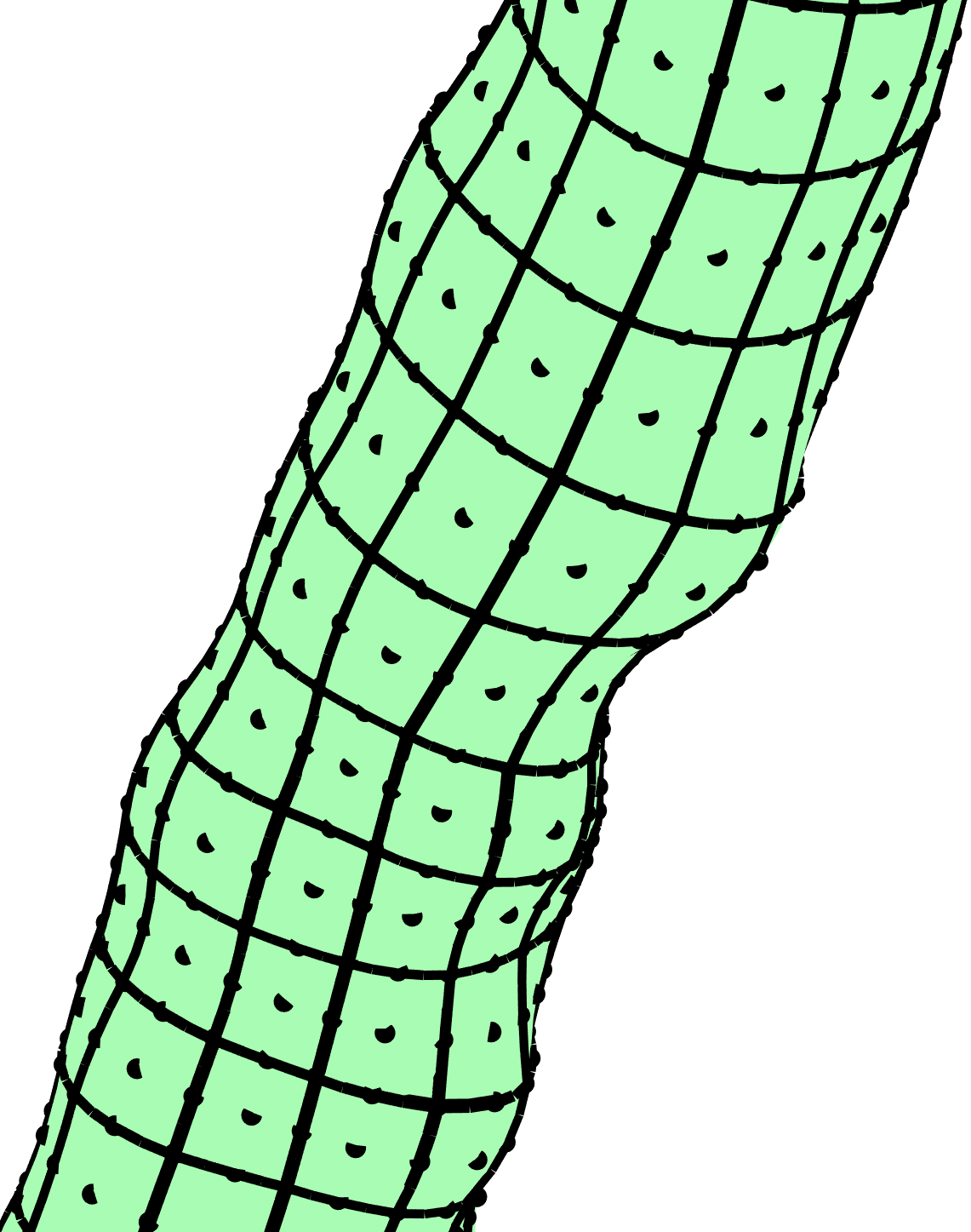}}}
        \end{overpic}
    }
    \caption{Comparison between different perturbation magnitudes for the same case: (a) original unperturbed mesh, (b)-(e) perturbed meshes 
    ($A = 0.25$, $0.5$, $1.0$~mm).}
    \label{fig:MeshPerturbation_Comparison}
\end{figure}

Meshes perturbed in this way only rarely (less than \(0.66\%\) of the cases {even for $A=1.0~\text{mm}$}) suffer from validity issues. The approach is viable for higher-order meshes as well, which are in general even more sensitive to perturbations compared to low-order (i.e., linear) meshes. Altogether, this yields a very efficient algorithm to produce simulation meshes from surface perturbations of a fixed input mesh. Figure~\ref{fig:MeshPerturbation_Example} illustrates several examples of random field perturbations, with visible geometrical changes in all of the cases. For comparison purposes, Fig.~\ref{fig:MeshPerturbation_Comparison} illustrates the geometrical differences between varying perturbation magnitudes, for the same perturbation case. Note here also that this algorithm is not limited to \textit{local} perturbation as used herein, but may also be adopted to generate \textit{globally} perturbed meshes, where tailored constitutive models generalizing the underlying elasticity problem~\eqref{eqn:linelast} might allow for even larger perturbations.



\subsection{Vascular modeling}\label{sec:vascular-modeling}
Reproducing realistic flow rates and pressures in the aorta is of critical importance here, since we aim to demonstrate the clinical relevance of geometric alterations and their effect on quantities of direct interest \textit{under physiological conditions}. Note, however, that we do not present a patient-specific case in the sense of a full dataset obtained from a single person, but rather combine available data in a meaningful way. Such an approach is common practice in the field of patient-specific modeling due to severely limited data availability. The following section discusses the employed models and related parameter choices. For the details regarding numerical treatment adopted herein, we refer the interested reader to~\cite{Schussnig2022c}, covering the more general case of fluid--structure interaction. Within this contribution, however, we limit ourselves to flow problems on a stationary domain denoted as $\Omega$, neglecting vessel motion completely. Under this assumption, blood flow is governed by the Navier--Stokes equations for incompressible flow
\begin{align}
    \rho \frac{\partial}{\partial t}\ve{u} 
    + 
    \rho \ve{u} \cdot \nabla \ve{u}
    - 
    2 \rho \nabla \cdot \left(\eta \nabla^\mathrm{S}\ve{u} \right)
    + 
    \nabla p
    &= 
    \ve{0}
    & \text{in }\Omega
    ,
    \label{eqn:momentum_balance}
    \\
    \nabla \cdot \ve{u} 
    &= 
    0
    & \text{in }\Omega
    \label{eqn:continuity_equation}
    .
\end{align}
The fluid velocity $\ve{u}$ and pressure $p$ are governed by the linear momentum balance~\eqref{eqn:momentum_balance} and continuity~\eqref{eqn:continuity_equation} equations, with density $\rho = 1060~\text{kg/m}^3$ in the acceleration term 
$\rho \partial \ve{u}/\partial t$ and nonlinear convective term $\rho \ve{u}\cdot \nabla \ve{u}$, and kinematic viscosity $\eta$. The latter shear stress term is considered to be nonlinear in viscosity to capture the complex shear-thinning behavior of the blood, using a so-called quasi-Newtonian Carreau model, i.e.,
\begin{equation}
    \eta (\dot\gamma) = \eta_\infty + (\eta_0 - \eta_\infty) \left[ 1.0 + (\lambda \dot\gamma)^2 \right]^{\frac{n-1}{2}}
    ,
    \quad
    \text{with} 
    \quad
    \dot{\gamma} := \sqrt{2 \nabla^\mathrm{S} \mathbf{u} : \nabla^\mathrm{S} \mathbf{u}}
    .
    \label{eqn:carreau}
\end{equation}
In~\eqref{eqn:carreau}, physiological parameters according to~\citet{Ranftl2023} are chosen, that is 
$\eta_\infty\approx3.04\times10^{-6}$~m$^2$/s, 
$\eta_0\approx25.67$~m$^2$/s, 
$\lambda=1.556$~1/s 
and $n=0.462$, 
corresponding to 50\% hematocrit. Note here that generalized Newtonian models such as~\eqref{eqn:carreau} are among the accepted choices according to the current state of the art for blood flow modeling. 
Despite their simplicity, they are able to capture some of the intricate rheological behaviors while being computationally cheaper (and easier to implement) than multi-phase approaches or viscoelastic models. As \cite{Ranftl2023} shows, enriching the constitutive model compared to a purely Newtonian approach with constant viscosity is indeed favorable, \textit{particularly} when only population mean parameters are available.

\begin{figure}[!ht]
    \centering
    \begin{overpic}[height=4cm,draft=\draftVascularModelling]{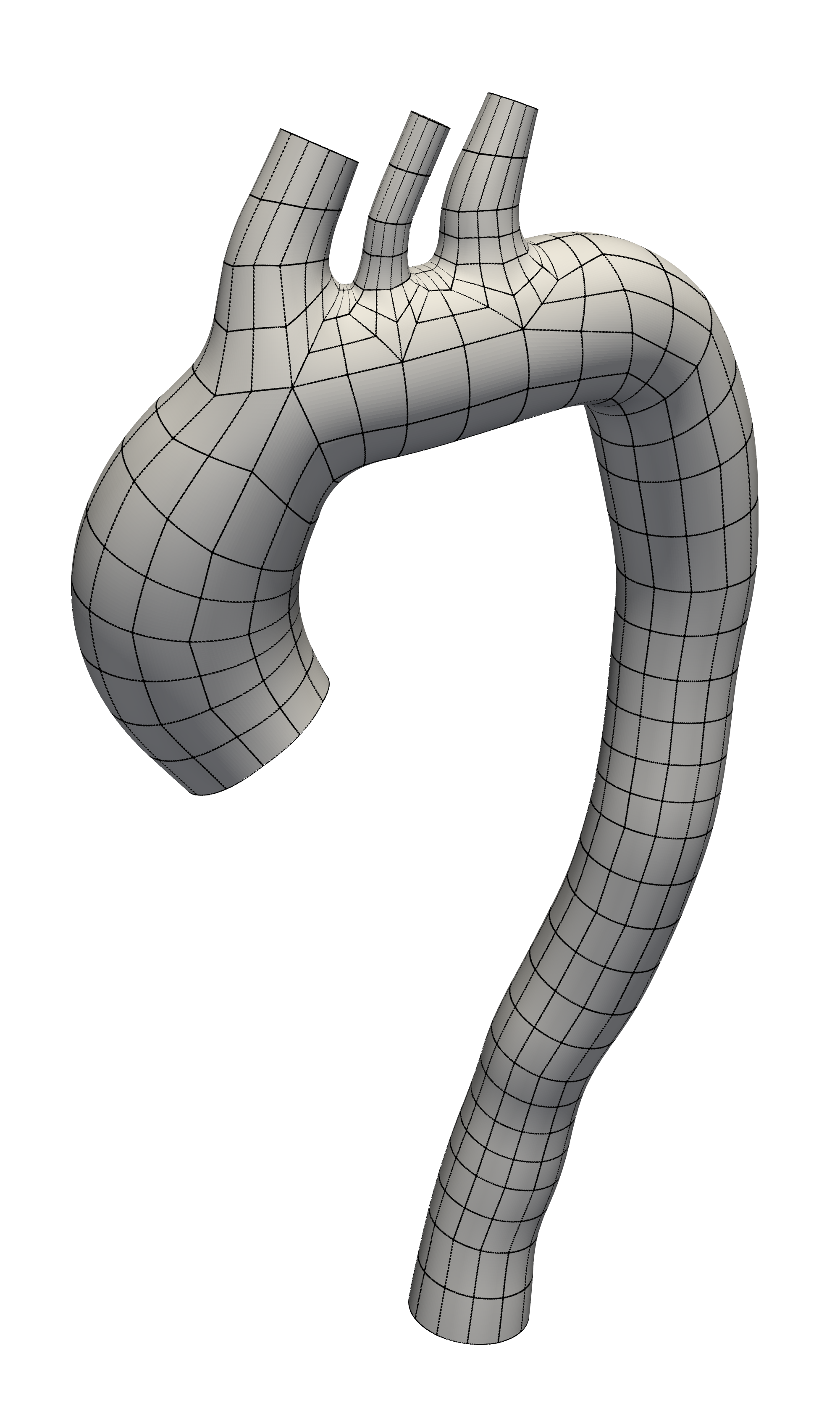}%
        \put(60,360){$\Gamma_0$}
        \put(150,930){$\Gamma_1$}
        \put(250,960){$\Gamma_2$}
        \put(350,990){$\Gamma_3$}
        \put(160,100){$\Gamma_4$}
    \end{overpic}
    \hspace{0.02\linewidth}
    {\includegraphics[height=4cm,draft=\draftVascularModelling]{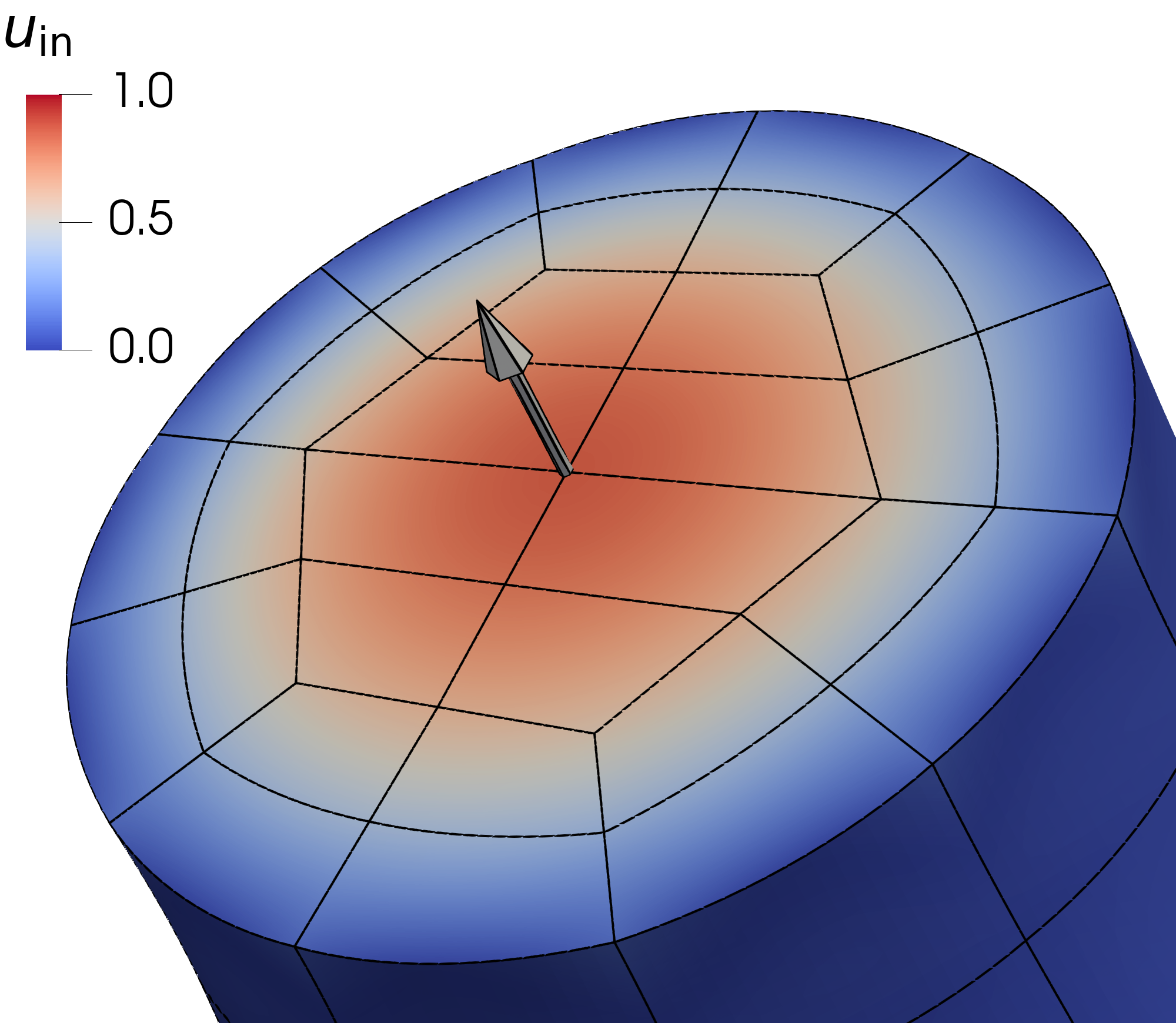}}
    \hspace{0.02\linewidth}
    {\includegraphics[height=4cm,draft=\draftVascularModelling]{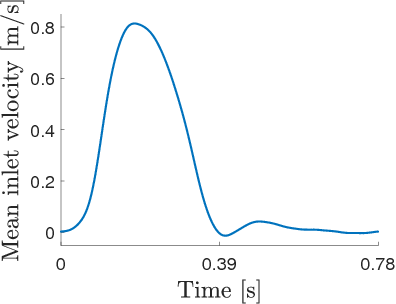}}
    \caption{The aortic geometry considered with inlet $\Gamma_0$ at the aortic root, outlets at the supra-aortic vessels: brachiocephalic trunk $\Gamma_1$, left common carotid artery $\Gamma_2$, left subclavian artery $\Gamma_3$, and main outlet $\Gamma_4$ positioned in the abdominal aorta (left). At $\Gamma_0$, the inflow is prescribed taking data from \citet{Baeumler2020}, enforcing a quasi-parabolic profile (middle) and a temporal scale fit to clinical measurements (right).}
    \label{fig:boundary_conditions_aorta}
\end{figure}

The governing equations including the constitutive model~\eqref{eqn:momentum_balance}--\eqref{eqn:carreau} are supplied with suitable boundary conditions to ensure physiological flow splits and pressure levels. The inflow rate at the aortic root is enforced by means of a Dirichlet boundary condition, scaling physiological data from~\cite{Baeumler2020} (from a different patient) to achieve an identical ratio of inlet area to volumetric flow rate, preserving the Reynold's number that characterizes the flow field. Based on the target flow rate, a velocity profile on the non-circular inlet is constructed adopting a technique by~\cite{Takizawa2010}, see Fig.~\ref{fig:boundary_conditions_aorta} for the temporal scale and the resulting quasi-parabolic scale in space.

Since the vessel wall is stationary and no-slip conditions are assumed on the aortic walls, the only remaining complication lies in estimating suitable pressure at truncated vessels. Here, we consider Windkessel models, simultaneously targeting the flow splits reported in~\cite{Baeumler2020}, at the same time obtaining realistic absolute pressure values at the abdominal aorta of $120~\text{mmHg}$ in systole and $75~\text{mmHg}$ in diastole~\cite{Mills1970}. In summary, we achieve physiological flow conditions by starting from the fitting scheme by~\cite{Grinberg2008}, and tuning Windkessel parameters until the set listed in Table~\ref{tab:windkessel_parameters} meets the described requirements. Note that this parameter set is not unique, but the result of manual tuning. This, however, suffices for the purpose of the present contribution, since we (i) do not have patient-specific flow data available, and (ii) merely aim for physiological flow conditions, not exact reproduction of a flow field within an actual patient at some time.

\begin{table}
	\centering
	\caption{Obtained flow splits $q_i:= \int_{t} Q_{\Gamma_i} ~\mathrm{d}t$ fit to clinical data~\cite{Baeumler2020} using the scheme by \citet{Grinberg2008}, where the outlet in the abdominal aorta serves as a reference Windkessel with a distal resistance of $R_4 = 2.5\times10^7$~Pa~s/m$^3$. The distal pressure is $p_\mathrm{d} = 10^3$~Pa at all outlets, and the capacitance $C_i$ is set per outlet to obtain stable results. This leads to $120~\text{mmHg}$ in systole and $75~\text{mmHg}$ in diastole at the abdominal aorta, which lies in the physiological range~\cite{Mills1970}.}
	\vspace{2mm}
	\label{tab:windkessel_parameters}
    \begin{tabular}{||l|>{\centering\arraybackslash}p{0.15\linewidth}|>{\centering\arraybackslash}p{0.15\linewidth} |>{\centering\arraybackslash}p{0.15\linewidth}||}
        \hline
		&&&\\[-1.6ex]
        & $q_i$~[\%]
		& $\nicefrac{R_i}{R_4}$~[-]
        & $C_i$~[$10^{-10}$~m$^3$/Pa]
        \\
		\hline\hline
		&&&\\[-1ex]
		  Brachiocephalic trunk, $\Gamma_1$ & 20.5 & 3.1 & \phantom{0}4  \\
        Left common carotid, $\Gamma_2$& \phantom{0}6.7  & 8.6 & \phantom{0}2  \\
        Left subclavian artery, $\Gamma_3$& 14.5 & 3.5 & \phantom{0}4  \\
        Abdominal aorta, $\Gamma_4$& 58.3 & 1.0 & 12
		\\[1.4ex]
        \hline
	\end{tabular}
\end{table}

\subsection{Target quantities and relevant biomarkers}
\label{sec:QoIs}
Having the physiological flow field in terms of velocity, pressure and viscosity available as volumetric time-dependent data allows for detailed investigations of the hemodynamics. Changing the present configuration in terms of boundary conditions, geometry or other input parameters could be evaluated for their effects on the flow field and derived measures. Such quantities of interest are, e.g., flow rates, pressure/pressure drops, vortical structures or disturbances in the flow field, and elevated shear stress. While the former are directly related to the primary variables, for example flow rate $Q_i$ or spatial mean pressure $\bar{p}_i$ over some inlet or outlet $\Gamma_{i}$,
\begin{equation}
    Q_{\Gamma_i} := \int_{\Gamma_i} \ve{u}\cdot \ve{n} ~\mathrm{d}\Gamma_i
    \,
    ,
    \quad
    \bar{p}_{\Gamma_i}
    :=
    \frac{\int_{\Gamma_i} p ~\mathrm{d}\Gamma_i}{\int_{\Gamma_i} 1 ~\mathrm{d}\Gamma_i}  
    \,,
    \label{eqn:definitions_Q_pmean}
\end{equation}
the latter are functions of the wall shear stress vector $\te{\tau}_\mathrm{wall}\in\mathbb{R}^3$ defined as
\begin{equation}
    \label{eqn:shear_stress_vector}
    \ve{\tau}_\mathrm{wall} 
    :=
    2 \rho\eta\, \ve{n} \cdot \grads \ve{u} 
    - 
    2 \rho\eta
    \left[ \ve{n} \cdot \left(\ve{n} \cdot \grads \ve{u} \right) \right]\ve{n}
    \,.
\end{equation}
Here, one can easily retrace the direct impact of the surface approximation quality on the unit outward normal on the vessel wall boundary $\Gamma_\mathrm{wall}:= \partial\Omega \setminus \bigcup_{i=0}^4 \Gamma_i$, denoted as $\ve{n}$ in~\eqref{eqn:definitions_Q_pmean}--\eqref{eqn:shear_stress_vector}, and thus its influence on the biomarkers derived from the shear stress. This relation renders accurate surface representations particularly important in the present context, i.e., when shear stresses are to be investigated. Naturally, higher-order, smooth surface representations enable lower spatial resolution compared to piecewise linear surfaces, to be discussed in Sec.~\ref{sec:numerical_results}.

Based on the wall shear stress vector~\eqref{eqn:shear_stress_vector}, the time-averaged wall shear stress (TAWSS) and related biomarkers such as the oscillatory shear index (OSI), highly oscillatory, low magnitude shear (HOLMES)~\cite{Alimohammadi2016} and endothelial cell activation potential (ECAP)~\cite{DiAchille2014} are defined as
\begin{align}
    \mathrm{TAWSS} 
    &:= 
    \nicefrac{1}{T_p} \int_{(i-1)T_p}^{i T_p} |\ve{\tau}_\mathrm{wall}| ~\mathrm{d}t
    \,,
   &
   \mathrm{OSI}
   &:=
   \nicefrac{1}{2} - \frac{\left\rvert \nicefrac{1}{T_p} \int_{(i-1)T_p}^{i T_p} \ve{\tau}_\mathrm{wall} ~\mathrm{d}t \right\rvert}{2\,\mathrm{TAWSS}}
    \,,
    \label{eqn:TAWSS_OSI}
    \\
    \mathrm{HOLMES}
    &:=
    \mathrm{TAWSS} \left( \nicefrac{1}{2} - \mathrm{OSI} \right)
    \,,
    &
    \mathrm{ECAP}
    &:=
    \frac{\mathrm{OSI}}{\mathrm{TAWSS}}
    \,,
    \label{eqn:HOLMES_ECAP}
\end{align}
where $i\geq 1$ refers to the $i$-th cardiac cycle of length $T_p$ equal to $0.78~\text{s}$ in the present scenario.
Luminal wall under high TAWSS for an extended period is at greater risk of rupture~\cite{Alimohammadi2016, Alimohammadi2018}, whereas low TAWSS promotes endothelial cell degeneration, and is additionally adopted as model parameter in some thrombus growth models~\cite{Menichini2017a,Chong2022a,Jafarinia2022a}. The OSI indicates the orientation of the wall shear stress. The OSI vanishes in regions with unidirectional shear, while high OSI indicates regions with heavily varying or even reversing wall shear stress. In combination, high OSI and low TAWSS correlate with rupture and remodeling~\cite{Alimohammadi2016,Qiao2019a}, giving rise to the HOLMES index. And lastly, ECAP relates OSI and TAWSS similar to HOLMES, leading to an index indicating regions of endothelial vulnerability~\cite{DiAchille2014}. 

\begin{figure}[!ht]
    \centering
    {\includegraphics[width=0.3\linewidth,draft=\draftVascularModelling]{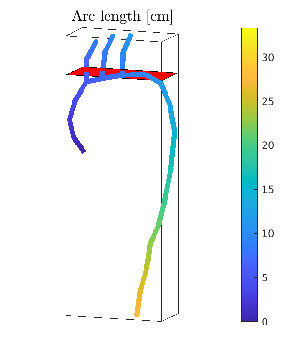}}\hspace{0.02\linewidth}%
    {\includegraphics[width=0.3\linewidth,draft=\draftVascularModelling]{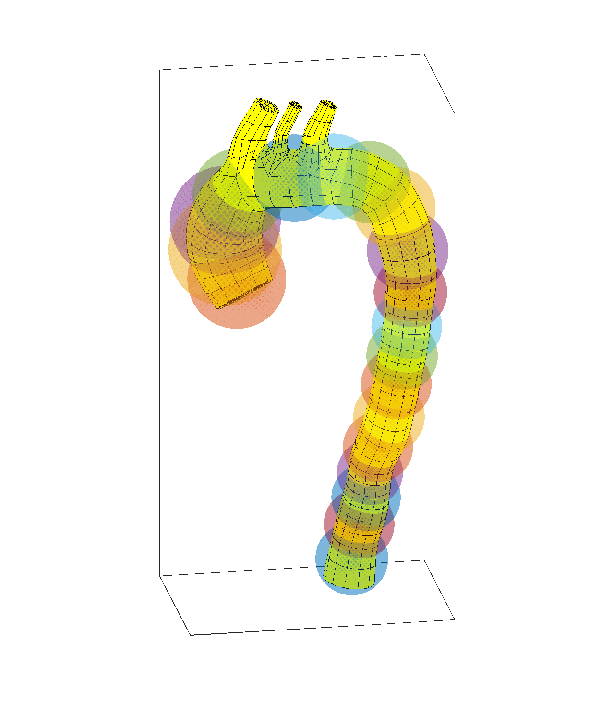}}\hspace{0.02\linewidth}%
    {\includegraphics[width=0.3\linewidth,draft=\draftVascularModelling]{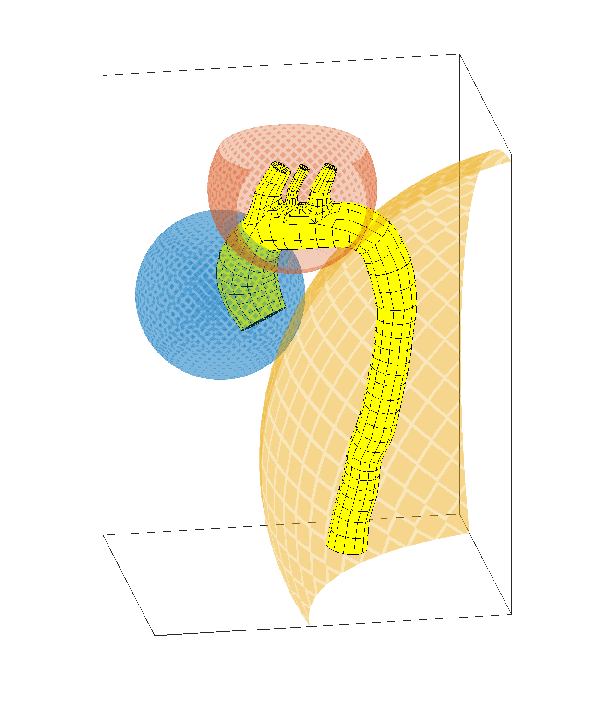}}
    \caption{Arc length along the centerline of the vessel with red plane indicating $X_{3,0} = 333~$mm (left), where we place spheres to define the regions for the integral mean of the wall shear stress~\eqref{eqn:wss_integral_avg}. Finer-grained results are achieved with smaller spheres of 1.5$\times$ the vessel radius at the respective points, while larger regions for the ascending aorta, aortic arc, and abdominal aorta regions are defined via manual definitions (see Table~\ref{tab:spheres}).}
    \label{fig:wss_spheres}
\end{figure}

The biomarkers in Eqn.~\eqref{eqn:TAWSS_OSI}--\eqref{eqn:HOLMES_ECAP} are all based on the wall shear stress vector $\ve{\tau}_\mathrm{wall}$. Hence, to simplify matters, we will focus the investigations herein on the integral mean of the wall shear stress norm over the vessel wall $\Gamma_\mathrm{wall}\subset\partial \Omega$, denoted herein as $w_\Gamma$,
\begin{align}
    w_\Gamma 
    :=
    \frac{
        \int_\Gamma ||\ve{\tau}_\mathrm{wall}||
        ~\mathrm{d}\Gamma
    }{
        \int_\Gamma 1
        ~\mathrm{d}\Gamma
    }
    \,,
    \label{eqn:wss_integral_avg}
\end{align}
which is of interest on the luminal wall (excluding inlets and outlets), or subsections thereof. This allows us to derive a single scalar value (per defined sub-part of the boundary, $\Gamma$) at each time $t$, which corresponds to the wall shear stress acting on the vessel wall or a portion of it. The latter approach is meaningful in the sense that one can capture region-dependent effects of perturbations in the boundary-value problem, that is, geometric variations in the present scenario. We define regions $\Gamma_{\mathrm{wall},i}$ by specifying spheres with midpoints lying on the centerline and radius depending on the vessel radius at the specific point or, alternatively, by region: ascending aorta, aortic arc, and abdominal aorta. The spheres are illustrated in Fig.~\ref{fig:wss_spheres} along with the arc length $s$ along the centerline (and listed in Table~\ref{tab:spheres} for completeness). The integration domain is then defined as the sum of all finite element faces $\partial\Omega_e$ on the exterior boundary, with the cell center lying within the respective sphere. Note here, that such a measure considered with $\Gamma_\mathrm{wall}$, that is, the vessel wall boundary, is better suited for a convergence study. On the other hand, comparisons employing the same principal finite element mesh, but under different deformation maps (the geometric variations imposed) give insight into regional differences of practical interest, which we aim to capture and compare with the smaller integration domains defined via the defined spheres. With such a measure, we are in the position to identify region-dependent effects without considering for highly complex integration domains emerging from intersecting the vessel surface with the actual spheres. Additionally, point-wise comparisons of QoIs are prone to show strongly localized effects, which are not of immediate interest, while integrated QoIs reduce this effect by averaging contributions, delivering a clearer picture of the greater trends.

\subsection{Propagating the uncertainty from geometry to QoI} 
Since we utilize extremely efficient implementations on high performance computing clusters, we do not need a surrogate model or machine learning; we can propagate the uncertainty with simple Monte Carlo estimates benefiting from trivial parallelization. Note that in similar studies based on surrogate models or machine learning, the Monte Carlo estimate is typically used as the \lq true\rq~ reference value \cite{biehler2015towards}.

\begin{figure}
    \centering
    \hspace{5mm}
    {
        \begin{overpic}[width=0.30\linewidth,draft=\draftSim]{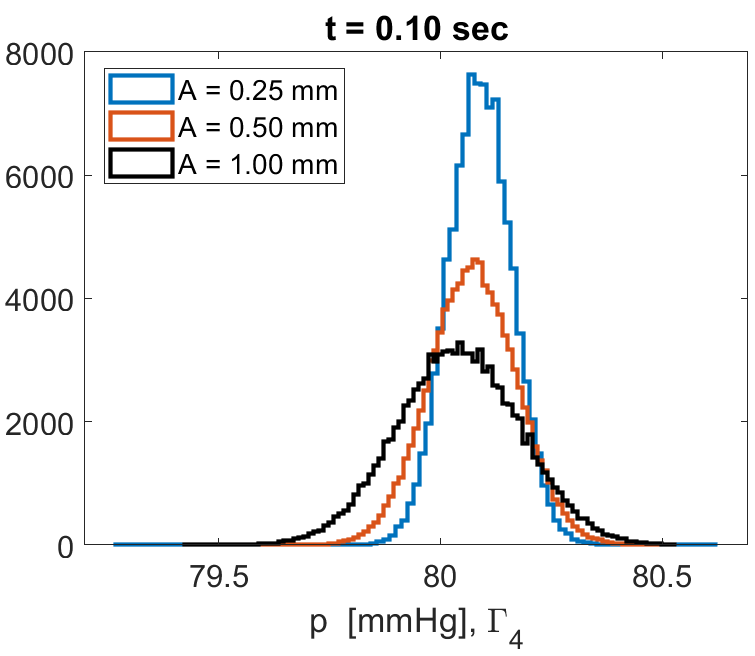}
        \put(-100,500){\scriptsize\rotatebox[origin=c]{90}{Frequency [1]}}
        \put(410,25){\scriptsize\colorbox{white}{\rotatebox[origin=c]{0}{$\bar{p}_{\Gamma_4}$ [mmHg]}}}
        \end{overpic}
    }
    {
        \begin{overpic}[width=0.30\linewidth,draft=\draftSim]{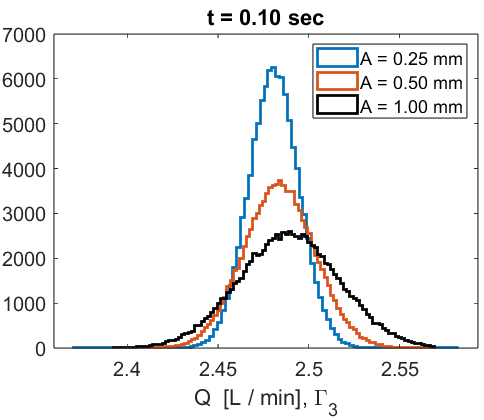}
        \put(345,0){\scriptsize\colorbox{white}{\rotatebox[origin=c]{0}{$Q_{\Gamma_3}$ [lit/min]}}}
        \end{overpic}
    }
    {
        \begin{overpic}[width=0.30\linewidth,draft=\draftSim]{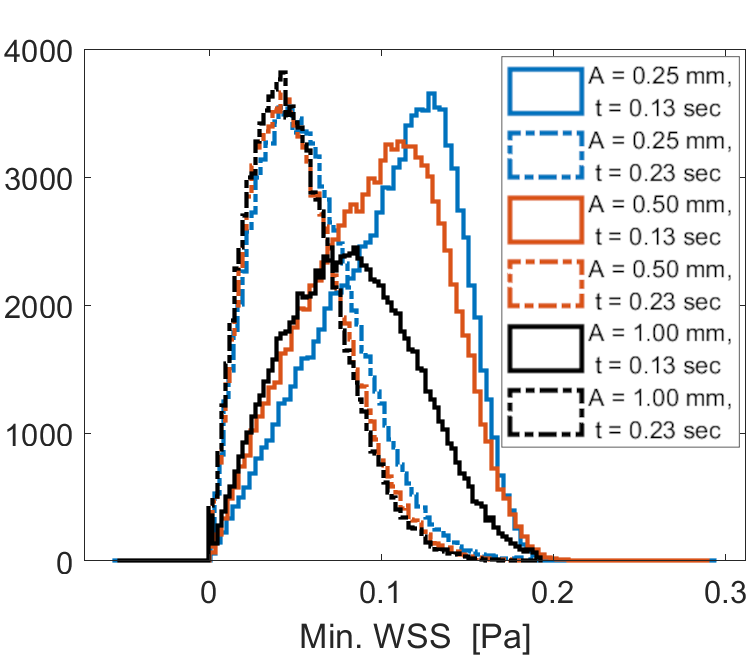}
        \put(380,0){\scriptsize\colorbox{white}{\rotatebox[origin=c]{0}{Min. WSS [Pa]}}}
        \end{overpic}
    }
    \caption{Example histograms of the spatial mean pressure $\bar{p}_{\Gamma_4}$ at the outlet $\Gamma_4$ (left), flow rate $Q$ at the outlet $\Gamma_3$ (middle), and minimum WSS (right). Note that the width of the histogram typically increases with the perturbation magnitude $A$. Note also that the distributions are uni-modal and not generally Gaussian, but are occasionally skewed, especially for strictly non-negative QoIs at small values. With this in mind, the notation of \lq mean $\pm$ uncertainty' remains meaningful, but must be interpreted with caution.}
    \label{fig:pedagogical_histograms}
\end{figure}

We estimate the probability density function (PDF) for the QoI $y$ given data $D$, formally $\p{y}{D}$, via Monte Carlo sampling by a simple procedure:
\begin{enumerate}
    \item[(i)] Sample $N = 10^5$ Gaussian random field realizations according to Sec.~\ref{subsec:BoundaryRandomField}--\ref{sec:sampling-gaussian-fields} and compute corresponding mesh perturbations according to \eqref{eqn:linelast}.
    \item[(ii)] Solve the model equations defined above (\eqref{eqn:momentum_balance}--\eqref{eqn:carreau}, Sec.~\ref{sec:vascular-modeling}) for each perturbed mesh, yielding a dataset $D$ of pairs of inputs $\fDeltaXhat_{(n)}$ and outputs $y_{(n)}$, $D = \{\fDeltaXhat_{(n)}, y_{(n)}\}_{n=1}^N$.
    \item[(iii)] Aggregate all simulation results $D$ in histograms in order to approximate the PDF $\p{y}{D}$. Pedagogical examples of such histograms are shown in Fig.~\ref{fig:pedagogical_histograms}.
\end{enumerate}
We define the \lq uncertainty\rq~ $\Delta_y$ as the standard deviation of $y$ derived from this PDF.

As it turns out, when running \(300\, 000\) simulations with \(300\, 000\) distinct meshes automatically, a small number of simulations are outliers. These outliers need to be identified and separated as otherwise the estimates for mean and uncertainty are strongly distorted. The median and confidence intervals are measures that are more robust to outliers and offer a compelling alternative to mean and variance. However, the median is defined via the monotonic ordering of QoI values. This order depends on the QoI of choice. Since we have several QoIs, this order is neither unambiguous nor generally consistent across QoIs. Hence we opt for summarizing the statistics through mean and variance in combination with outlier removal as follows: (i) datasets are removed judged by physical sensibility if outlet flow rates and pressures do not lie within a \(\pm50\)\% range of the mean at systole and diastole or if the end time was not reached due to divergence. Note that the latter aspect \emph{naturally} competes with a minimal time to solution through (non)linear solver and adaptive time step choices/settings and is hence expected; 
(ii) remaining outliers were removed if any QoI, at any time or any location, was outside the $10\sigma$-interval symmetrically around the corresponding mean. This heuristic is sometimes referred to as \lq z-test'. Under the assumption of a Gaussian distribution, the probability for a sample outside the $10\sigma$-interval is $\mathcal{O}(10^{-40})$. Although our probability distributions are not generally Gaussian and values can generally differ by several orders of magnitude, we take this pragmatic judgement to identify extreme events with confidence.
Mesh quality could be assessed at mesh generation time and hence was not a criterion for outliers at this stage. The total number of outliers so identified was $0.16 \%$, $0.11 \%$ and $0.59 \%$ for $A=0.25$~mm, $A=0.50$~mm and $A=1.00$~mm, respectively, of the total of $N=100\, 000$ samples each. It is sensible that more outliers occur at high mesh perturbation magnitude. A summary of the performed outlier removal per perturbation level \(A\) is given in 
Table~\ref{tab:diverged_runs_overview}.
\begin{table}[ht!]
	\centering
	\caption{Removal of simulations per mean perturbation level \(A\) due to insufficient mesh quality, divergence or unphysical results, and outlier detection. Stages are passed top to bottom, where percentages are given relative to the number of simulations passing the previous stage.}
	\vspace{2mm}
	\label{tab:diverged_runs_overview}
    \begin{tabular}{||r|>{\centering\arraybackslash}p{0.175\linewidth}|>{\centering\arraybackslash}p{0.175\linewidth}|>{\centering\arraybackslash}p{0.175\linewidth}||}
        \hline
		&&&\\[-1.6ex]
        Perturbation level \(A\)~[mm] & 0.25 & 0.50 & 1.00
        \\
		\hline\hline
		&&&\\[-1ex]
		  Mesh quality insufficient~[\%] 
        & \(0.000\) 
        & \(0.001\) 
        & \(0.645\) 
        \\
        End time not reached~[\%] 
        & \(0.004\) 
        & \(0.003\) 
        & \(0.003\) 
        \\
        Unphysical~[\%] 
        & \(0.079\) 
        & \(0.079\) 
        & \(0.078\) 
        \\
        Outlier~[\%] & 0.16 & 0.11 & 0.59
        \\[1.4ex]
        \hline
	\end{tabular}
\end{table}
\section{Results}
\label{sec:Results}
This section describes the numerical tools employed to achieve robust and accurate hemodynamics simulations in the present scenario. First, we perform convergence studies to determine the required spatiotemporal resolution required to achieve engineering accuracy in the QoIs considering the three levels of perturbations. After identifying an appropriate finite element mesh and temporal resolution, we select numerical settings and perform simulations with \(300\, 000\) perturbed geometries. The QoIs determined for all cases are analyzed for the individual perturbation level and compared to identify trends with increasing mean perturbation.

\subsection{Discretization and numerical setup}
\label{sec:numerical_results}

The solution of the Navier--Stokes euqations~\eqref{eqn:momentum_balance}--\eqref{eqn:continuity_equation} governing incompressible flow is a challenging task. Tailored numerical solvers taking the generalized Newtonian behavior into account involve time-splitting schemes~\cite{Pacheco2021c}, coupled approaches~\cite{Schussnig2021b} or, as considered within this work, projection methods~\cite{Karniadakis1991}. We consider a second-order backward differentiation formula with adaptive timestepping ($\mathrm{CFL} = 0.5$) and inf-sup stable finite-element pairs in an $L^2$-conforming discontinuous Galerkin method~\cite{Fehn2017}, which enforces mass conservation and energy stability via stabilisation terms and is hence suited for underresolved turbulent flow simulations~\cite{Fehn2019}. 
The simulations are performed using \href{https://github.com/exadg/exadg}{\texttt{ExaDG}}~\cite{ExaDGgithub} (see \cite{Fehn2021b} for an overview and \cite{Arndt2020}), which is based on \texttt{deal.II}~\cite{dealII95}, and its matrix-free infrastructure~\cite{KronbichlerKormann2012, KronbichlerWall2018}, and supports $hp$-multigrid preconditioners based on~\cite{KronbichlerWall2018, Fehn2020, Fehn2021b}.

\begin{figure}
    \begin{center} 
    {
        \begin{overpic}[height=4.5cm,draft=\draftSim, trim=0 0 0 0]{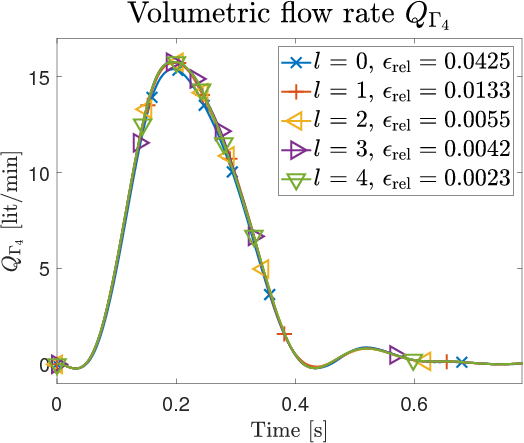}
        \end{overpic}
    }
    \hspace{0.02\linewidth}
    {
        \begin{overpic}[height=4.5cm,draft=\draftSim, trim=0 0 0 0]{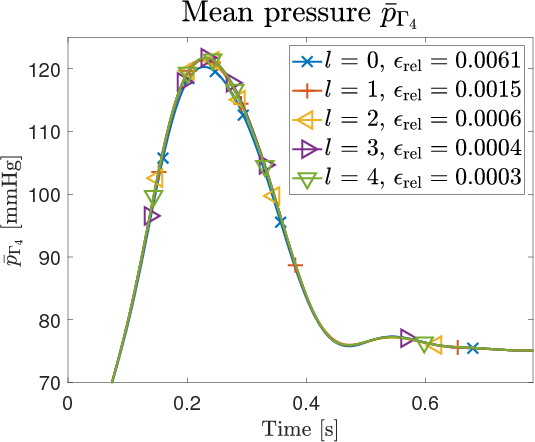}
        \end{overpic}
    }
    \\
    \vspace{-0.275cm}
    {
        \includegraphics[height=4.5cm,draft=\draftSim]{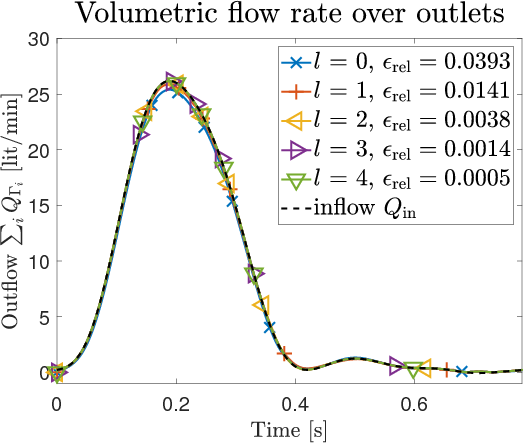}
    }
    \hspace{0.02\linewidth}
    {
        \includegraphics[height=4.5cm,draft=\draftSim]{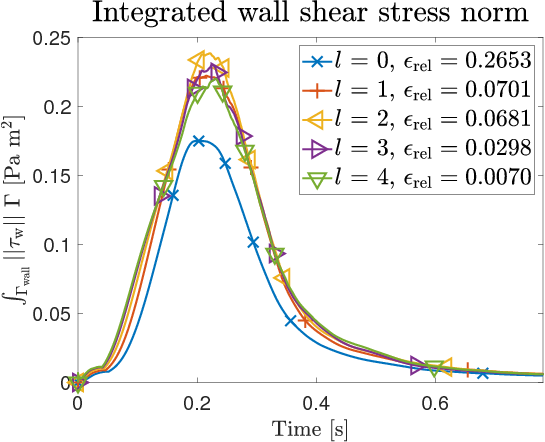}
    }
    \end{center}
    \caption{QoIs for the third cardiac cycle and mesh levels $l=0,\dots,4$ with relative error $\epsilon_\mathrm{rel}$ to the reference solution.}
    \label{fig:convergence_results}
\end{figure}

Within this setup, the construction of nested mesh hierarchies via~\cite{Bosnjak_2023a} is vital to obtain a suitable geometric coarsening sequence within the $hp$-multigrid preconditioner, which enables sufficiently rich sampling sets. Therefore, the coarsest mesh meeting engineering accuracy requirements is determined by solving the flow problem on successively refined meshes, see Fig.~\ref{fig:MeshPerturbation_Input} and Table~\ref{tab:convergence_grids}. The relative error is defined for scalar $f(t)$ as $\epsilon_\mathrm{rel}(f):=\int{(f-f_\mathrm{ref})}~\mathrm{d}t / \int{f_\mathrm{ref}~\mathrm{d}t}$ with a reference value $f_\mathrm{ref}(t)$. We compute a reference solution using four uniform refinements ($l=4$) and fourth order polynomials to approximate the geometry, while the remaining simulations use isoparametric mappings. Figure~\ref{fig:convergence_results}, e.g., shows the flow rates $Q_{\Gamma_4}$ and spatial mean pressures $\bar{p}_{\Gamma_4}$~\eqref{eqn:definitions_Q_pmean}, the volumetric flow rate over outlets and the wall shear stress norms over the vessel wall $\Gamma_\mathrm{wall}$ for different refinement levels $l = 0,\dots,4$. Due to the favorable mass conservation properties of the scheme, even the coarsest meshes deliver less than $4\%$ error in the flow rates, mass balance and pressures at the outlets. 
However, the wall shear stress is a function of the unit outward normal vector and the velocity gradient has larger errors. The non-perturbed case yields relative errors of 1.4\% in the summed flow rate, 1.6\% in the spatial mean pressure on $\Gamma_4$ and an integrated wall shear stress norm of 7.3\% when comparing the refinement level $l=1$ with the reference.
\begin{table}
	\centering
	\caption{Meshes considered for the convergence study, resulting degrees of freedom and resulting time steps for 3 cardiac cycles $t\in[0,2.34]~\mathrm{s}$ and a upper CFL limit of 0.5.}
	\vspace{2mm}
	\label{tab:convergence_grids}
    \begin{tabular}{||l|>{\raggedleft\arraybackslash}p{0.1\linewidth}|>{\raggedleft\arraybackslash}p{0.1\linewidth}|>{\raggedleft\arraybackslash}p{0.1\linewidth}|>{\raggedleft\arraybackslash}p{0.1\linewidth}|>{\raggedleft\arraybackslash}p{0.1\linewidth}||}
        \hline
		&&&&&\\[-1.6ex]
        Uniform refinements $l$ & 0&1&2&3&4\\
		\hline\hline
		&&&&&\\[-1ex]
		  Elements                         &224    &1\,792  &14\,336    &114\,688    &917\,504  \\
        Degrees of freedom               &19\,936&159\,488&1\,275\,904&10\,207\,232&81\,657\,856\\ 
        Time steps ($\mathrm{CFL} = 0.5$)&6\,288 &17\,005 &38\,029    &87\,929     &208\,842
		\\[1.4ex]
        \hline
	\end{tabular}
\end{table}

\begin{figure}
    \centering
    \centering
    {
        \begin{overpic}[width=0.30\textwidth,draft=\draftSim]{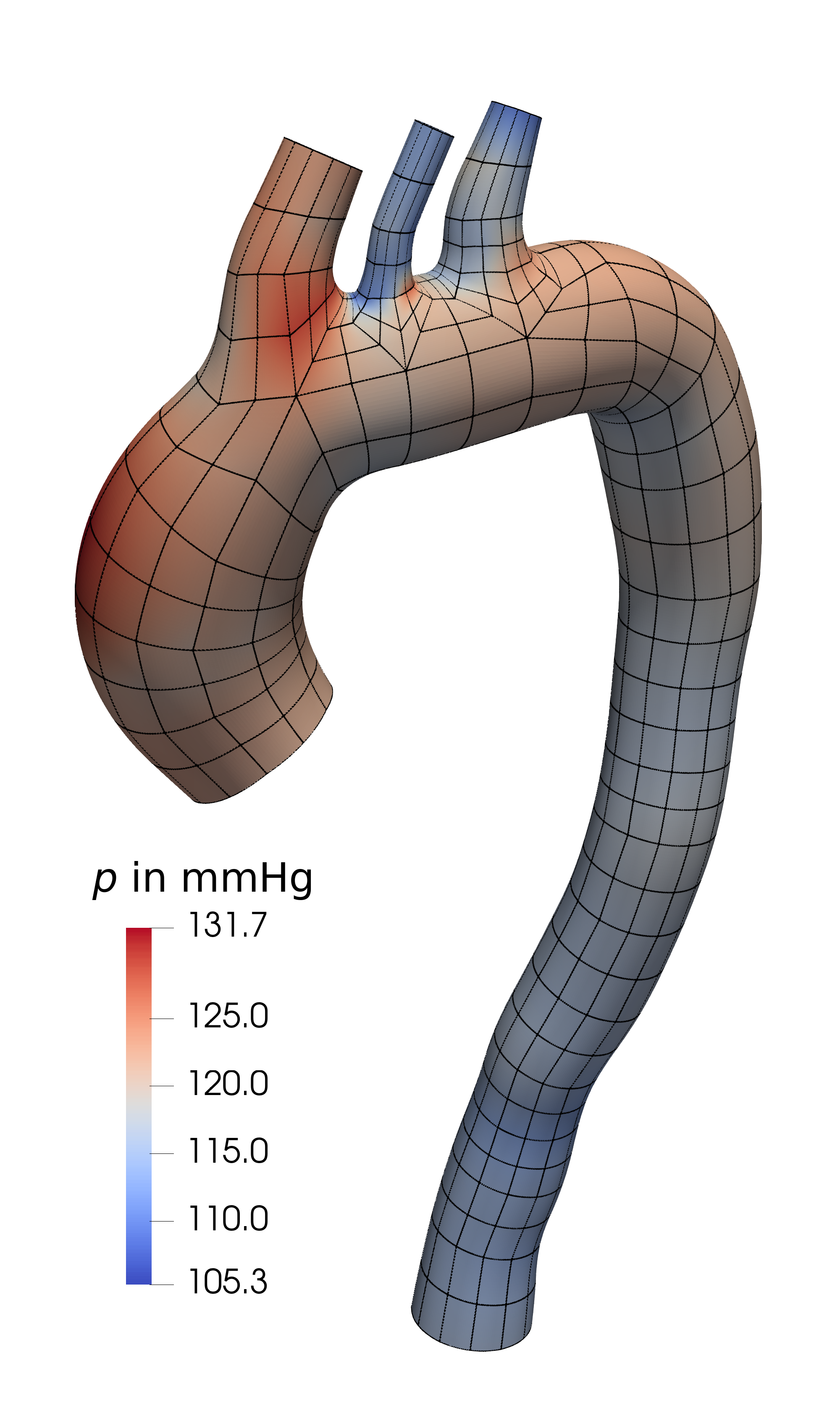}
            \put(55, 382.5){\small\colorbox{white}{$p$ [mmHg]}}
        \end{overpic}
    }
    \hspace{0.02\linewidth}
    {
        \begin{overpic}[width=0.30\textwidth,draft=\draftSim]{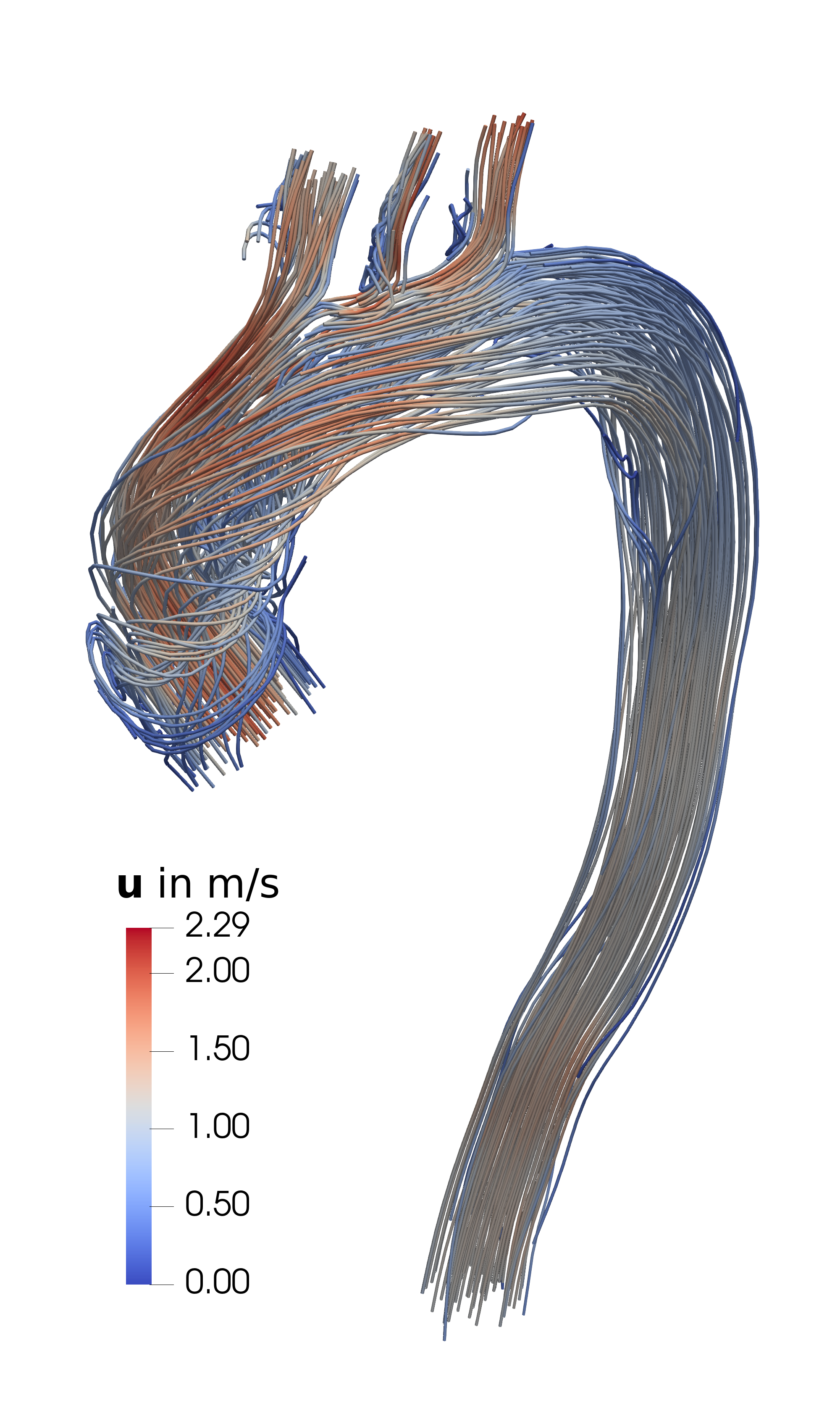}
            \put(55, 382.5){\small\colorbox{white}{$\ve{u}$ [m/s]}}
        \end{overpic}
    }
    \hspace{0.02\linewidth}
    {
        \begin{overpic}[width=0.30\textwidth,draft=\draftSim]{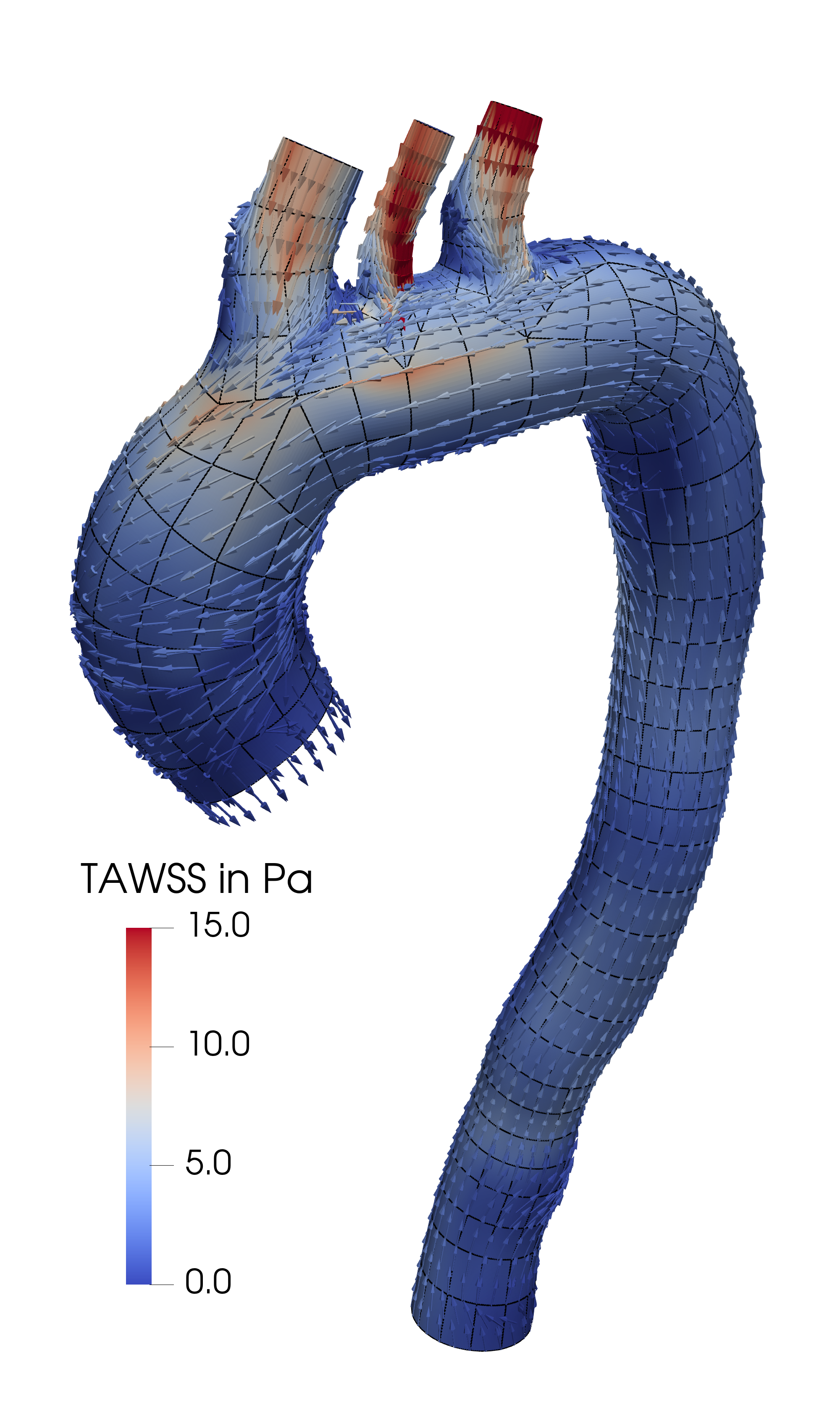}
            \put(15, 382.5){\small\colorbox{white}{TAWSS [Pa]}}        
        \end{overpic}
    }
    \caption{Solution using the $l=1$ mesh: pressure (left) and velocity streamlines (middle) at peak systole and time-averaged wall shear stress (right).}
    \label{fig:solution_case_0}
\end{figure}

Since the Reynolds number is constant under limited transformations of the domain, we consider the mesh $l=1$ shown in Fig.~\ref{fig:MeshPerturbation_Input}(b) consisting of 1\,792 $Q_2/Q_1$ Taylor--Hood elements for all perturbed configurations, noting that the errors increase for large perturbations of the domain. To further reduce computational costs, we consider the first cardiac cycle only. This is admissible since after $t\approx0.05~\mathrm{s}$, periodicity is already reached due to the deliberately chosen Windkessel parameters. That is, relative errors in flow rates, spatial mean pressures and integral mean WSS in the reference run ($A=0.0~\text{mm}$) is $<4$\% comparing values in the time interval $[0.05, \, 0.78]$ with the periodic solution. 
Example results using the $l=1$ mesh and no perturbation are shown in Fig.~\ref{fig:solution_case_0}. They show recirculations due to the inlet profile, Dean vortices due to the arc curvature and local pressure peaks on the branching vessels, and the TAWSS vector field with increased values on the branching vessel due to their smaller diameter.

Altogether, the main flow features are resolved, while full convergence cannot be demonstrated for all 300\,000 cases given a reasonable computational budget. Especially under large perturbations, the error achievable with the $l=1$ mesh might be significantly increased compared to the non-perturbed case. To exemplarily verify low errors in the observed norms, we consider a single randomly chosen realization of the mesh perturbation with a mean displacement from the undeformed geometry of $A=0.5$ and $1.0$~mm and perform a convergence study once again. As the relative errors in the QoIs as listed in 
Table~\ref{tab:convergence_grids_perturbed} show, the perturbed mesh yields slightly increased errors as expected, while the mesh level $l=1$ still delivers reasonable engineering accuracy. For the perturbed meshes ($A=0.5$ and $1.0$~mm cases), we hence still expect $\leq5\%$ error in QoIs related to the primary unknowns, and $\leq10\%$ error in the QoIs related to the velocity gradient. Note, however, that the case with $A=1.0~\text{mm}$ is already on the verge of being too inaccurate. Refining the mesh once leads to an increase of computational costs by a factor of \(2^4=16\) ($\times8$ due to the increased number of unknowns and $\times2$ due to the CFL condition). To limit computational costs, we hence nonetheless consider $l=1$.
\begin{table}
	\centering
	\caption{Relative errors $\epsilon_\mathrm{rel}$~[\%] in QoIs using $Q_2/Q_1$ interpolation on mesh levels $l$ constructed by uniform refinement, using $l=4$ and fourth order polynomials for geometry approximation as baseline: comparison of undeformed reference mesh ($A=0$~mm) and a sample perturbed mesh with $A=0.5$ and $1.0$~mm mean perturbation.}
	\vspace{2mm}
	\label{tab:convergence_grids_perturbed}
    \begin{tabular}{||c|c|c|c|c|c|c|c|c|c|c|c|c||}
        \hline
        &
        \multicolumn{3}{|c|}{} & \multicolumn{3}{|c|}{} & \multicolumn{3}{|c|}{} & \multicolumn{3}{|c||}{}
        \\[-1.6ex]
        & \multicolumn{3}{|c|}{$\epsilon_\mathrm{rel}\left(Q_{\Gamma_4}\right)$}
        &
        \multicolumn{3}{|c|}{$\epsilon_\mathrm{rel}\left(\bar{p}_{\Gamma_4}\right)$}
        &
        \multicolumn{3}{|c|}{$\epsilon_\mathrm{rel}\left(\sum_i\bar{Q}_{\Gamma_i}\right)$}
        &
        \multicolumn{3}{|c||}{$\epsilon_\mathrm{rel}\left(\int_{\Gamma_\mathrm{wall} ||\ve{\tau}_\mathrm{wall}||}\right)$}
        \\
        $l\downarrow$, $A\rightarrow$
        & 0.0 & 0.5 & 1.0 & 0.0 & 0.5 & 1.0 & 0.0 & 0.5 & 1.0 & 0.0 & 0.5 & 1.0
        \\
        \hline\hline
        &&&&&&&&&&&&\\[-1ex]
        0  &4.25&4.63&5.60&  0.61&0.65&0.75&  3.93&4.33&5.33&  26.53&30.50&33.84 \\
        1  &1.33&1.55&2.01&  0.15&0.20&0.25&  1.41&1.42&1.60&  7.01 &8.84 &10.73 \\
        2  &0.55&0.56&0.63&  0.06&0.06&0.06&  0.38&0.45&0.50&  6.81 &5.09 &4.04 \\
        3  &0.42&0.46&0.40&  0.04&0.05&0.04&  0.14&0.18&0.18&  2.98 &2.93 &2.55 \\
        4  &0.23&0.22&0.21&  0.03&0.02&0.02&  0.05&0.08&0.07&  0.70 &0.66 &0.61
        \\[1.4ex]
        \hline
    \end{tabular}
\end{table}


\subsection{Uncertainty Quantification}
\label{sec:numerical_results-UQ-analysis}
We consider uncertainty quantification of the QoIs under increasing mean perturbations $A=0.25, 0.5$ and $1.0$~mm. That is, the spatial mean pressure and flow rate per outlet and the (spatial integral) mean of the wall shear stress (see Sec.~\ref{sec:QoIs} regarding the respective definitions) are recorded, sampling \(100\,000\) random fields per mean perturbation level $A$ and solving the flow problem for the resulting geometry. Up to \(650\) geometries per perturbation case yielded invalid meshes, and were discarded. Figs.~\ref{fig:resultsUQ:flow_rates} and \ref{fig:resultsUQ:mean_pressure} show the mean $\mu$ and the uncertainty $\Delta$ in the flow rate $Q_i$~\eqref{eqn:definitions_Q_pmean} and mean outlet pressure $\bar{p}_{\Gamma_i}$ on outlets $\Gamma_i$ $i=1,\dots,4$. In the Figs.~\ref{fig:resultsUQ:flow_rates} and \ref{fig:resultsUQ:mean_pressure}, the first considered cardiac cycle is shown, and the uncertainty band is scaled by a factor of \(5\) to ease comparison. Note that the absolute uncertainty $\Delta$, given in the bottom rows of the respective figures, encodes the same information as the uncertainty band. Since the flow rates and pressure are inherently coupled through the Windkessel models, the same observations hold:
\begin{enumerate}
    \item[(i)] The respective means are independent of the mean perturbation $A$.
    \item[(ii)] The \textit{relative} uncertainty resulting from geometric variations is $<2$\% for the flow rates and $<1$\% for the spatial mean pressures \textit{during systole}, even for the most complex case $A=1.0~\text{mm}$.
    \item[(iii)] The uncertainty increases with the mean perturbation $A$. 
    \item[(iv)] The uncertainty is highest in the deceleration phase. 
\end{enumerate}
Furthermore, we note that the increase in uncertainty with increasing $A$ is sub-linear and dependent on the outlet. Doubling the mean perturbation $A=0.25~\text{mm}$ does not lead to a doubled uncertainty in the flow rate and spatial mean pressure, a similar relation holds for $A=0.5~\text{mm}$. The effects of geometric variations on the flow field are strongest in the deceleration phase, where the laminar, unidirectional flow pattern observed during acceleration and peak flow rate in systole breaks to a recirculatory, less organized flow.

In the spatial mean pressure per outlet, a sharp peak in the uncertainty is seen at $t\approx 0.01~\text{s}$, which relates to a fluctuating flow rate and pressure due to the Windkessel model. These fluctuations are expected and lie outside of the credible (nearly periodic) time interval $[0.05,\,  0.78]~\text{s}$, chosen for efficiency reasons, and where the solution reaches the periodic state up to engineering accuracy.
\begin{figure}[!ht]
    \centering
    {\begin{overpic}[width=0.3\linewidth,draft=\draftUQresults]{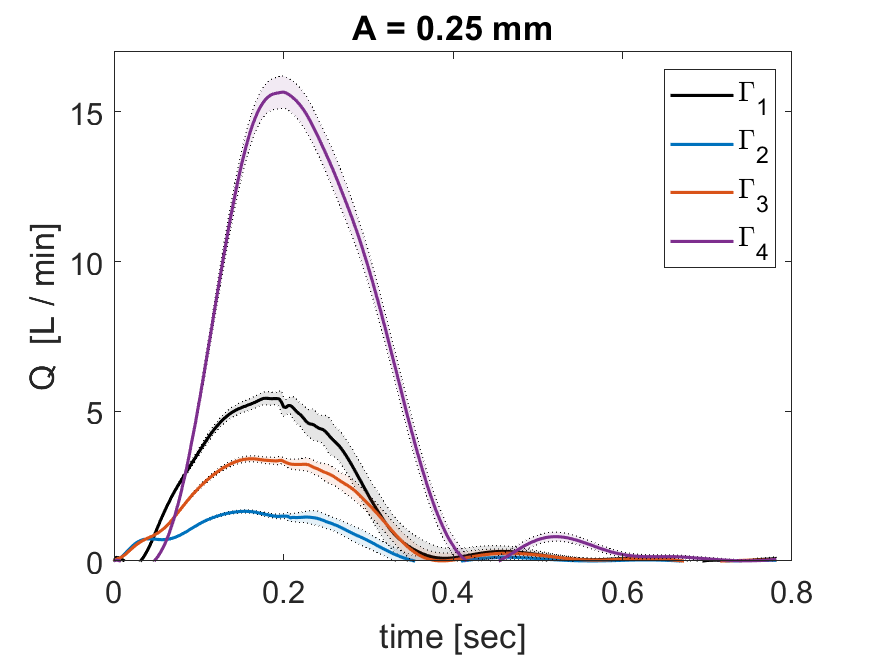}
        \put(-10, 382.5){\footnotesize\colorbox{white}{\rotatebox[origin=c]{90}{$Q_{\Gamma_i}$ [lit/min]}}}
        \put(375, 0){\footnotesize\colorbox{white}{\phantom{Time [s]}}}
    \end{overpic}}\hspace{0.02\linewidth}%
    {\begin{overpic}[width=0.3\linewidth,draft=\draftUQresults]{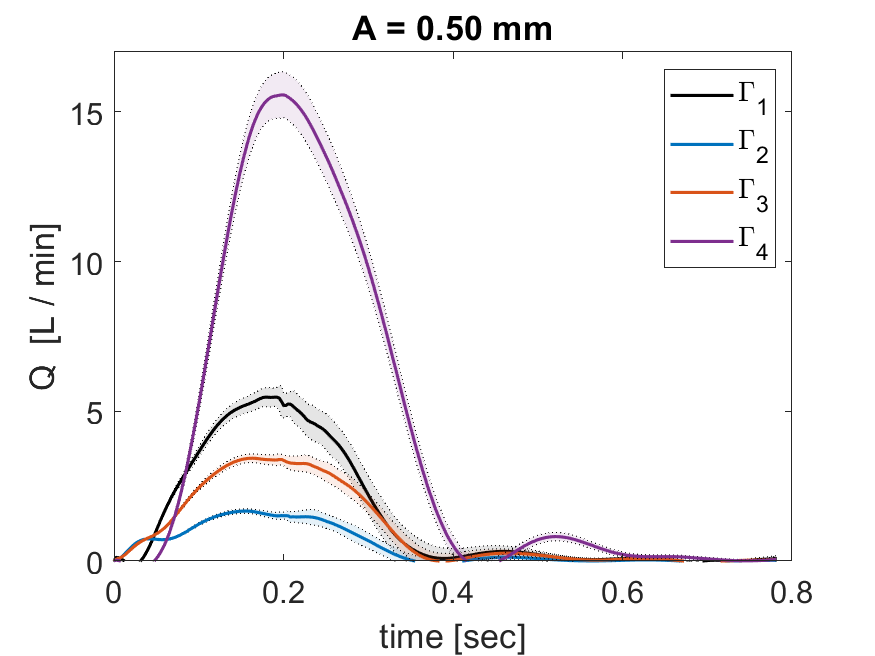}
        \put(-10, 382.5){\footnotesize\colorbox{white}{\rotatebox[origin=c]{90}{\phantom{$Q_{\Gamma_i}$ [lit/min]}}}}
        \put(720, 450){\colorbox{white}{\phantom{\scalebox{2.5}[4.5]{O}}}}
        \put(375, 0){\footnotesize\colorbox{white}{\phantom{Time [s]}}}
    \end{overpic}}\hspace{0.02\linewidth}%
    {\begin{overpic}[width=0.3\linewidth,draft=\draftUQresults]{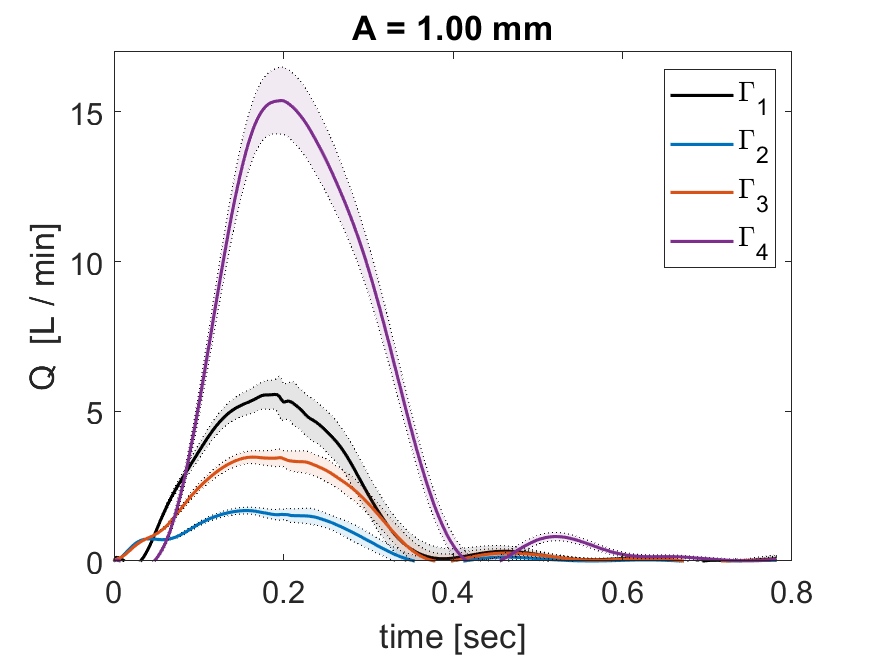}
        \put(-10, 382.5){\footnotesize\colorbox{white}{\rotatebox[origin=c]{90}{\phantom{$Q_{\Gamma_i}$ [lit/min]}}}}
        \put(720, 450){\colorbox{white}{\phantom{\scalebox{2.5}[4.5]{O}}}}
        \put(375, 0){\footnotesize\colorbox{white}{\phantom{Time [s]}}}
    \end{overpic}}
    \\
    {\begin{overpic}[width=0.3\linewidth,draft=\draftUQresults]{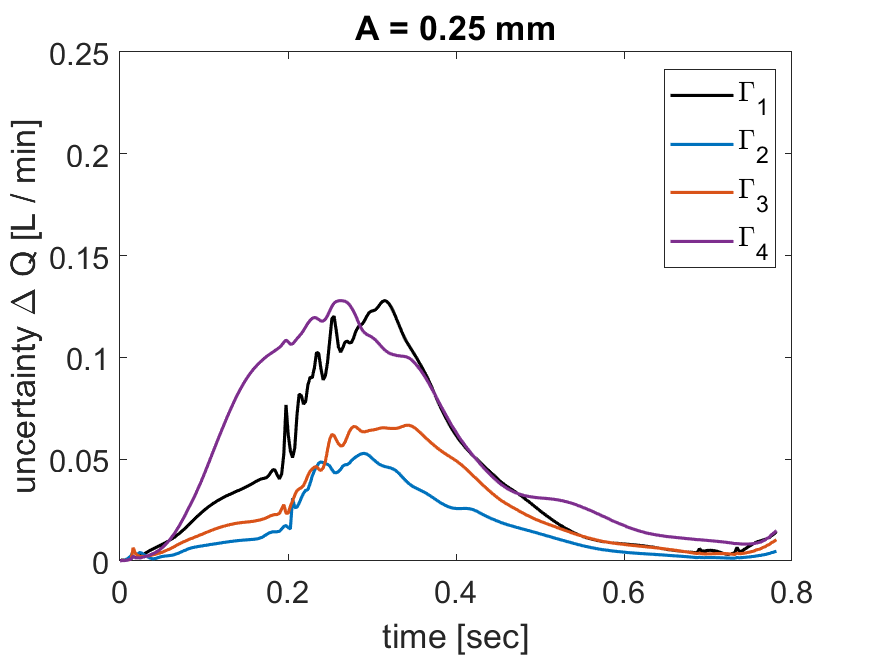}
        \put(-42.5, 382.5){\footnotesize\colorbox{white}{\rotatebox[origin=c]{90}{\phantom{Uncertainty $\Delta Q_{\Gamma_i}$ [lit/min]}}}}
        \put(-42.5, 382.5){\footnotesize\colorbox{white}{\rotatebox[origin=c]{90}{Uncertainty $\Delta$ [lit/min]}}}
            \put(720, 450){\colorbox{white}{\phantom{\scalebox{2.5}[4.5]{O}}}}
            \put(395, 0){\footnotesize\colorbox{white}{{Time [s]}}}
    \end{overpic}}\hspace{0.02\linewidth}%
    {\begin{overpic}[width=0.3\linewidth,draft=\draftUQresults]{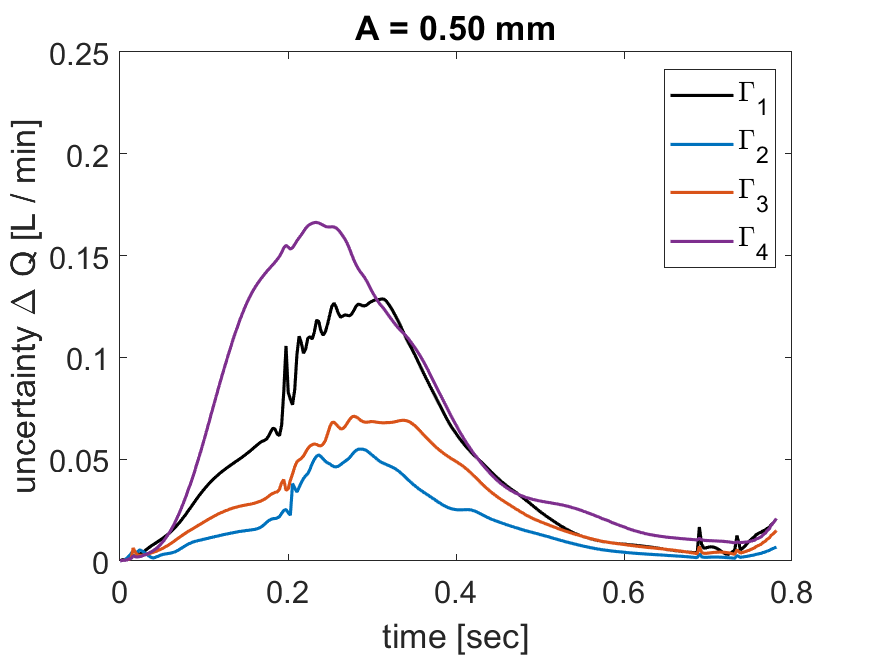}
        \put(-42.5, 382.5){\footnotesize\colorbox{white}{\rotatebox[origin=c]{90}{\phantom{Uncertainty $\Delta Q_{\Gamma_i}$ [lit/min]}}}}
        \put(720, 450){\colorbox{white}{\phantom{\scalebox{2.5}[4.5]{O}}}}
        \put(395, 0){\footnotesize\colorbox{white}{{Time [s]}}}
    \end{overpic}}\hspace{0.02\linewidth}%
    {\begin{overpic}[width=0.3\linewidth,draft=\draftUQresults]{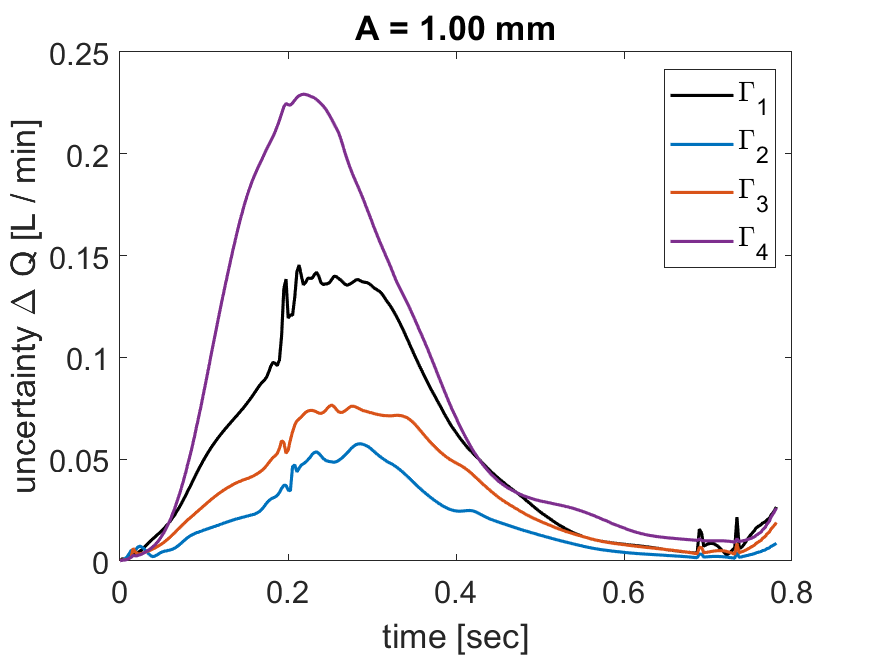}
        \put(-42.5, 382.5){\footnotesize\colorbox{white}{\rotatebox[origin=c]{90}{\phantom{Uncertainty $\Delta Q_{\Gamma_i}$ [lit/min]}}}}
        \put(720, 450){\colorbox{white}{\phantom{\scalebox{2.5}[4.5]{O}}}}
        \put(395, 0){\footnotesize\colorbox{white}{{Time [s]}}}
    \end{overpic}}
    \caption{Mean $\mu$ (top row) and uncertainty $\Delta$ (bottom row) of the outlet flow rates $Q_{\Gamma_i}$, $i=1,\dots,4$ (see Fig.~\ref{fig:wss_spheres}, middle) for varying perturbation levels $A=0.25, 0.5$ and $1.0$~mm. Note that the uncertainty bands are scaled by a factor of 5 for visualization purposes. The legend in the first diagram applies to all plots.}
    \label{fig:resultsUQ:flow_rates}
\end{figure}
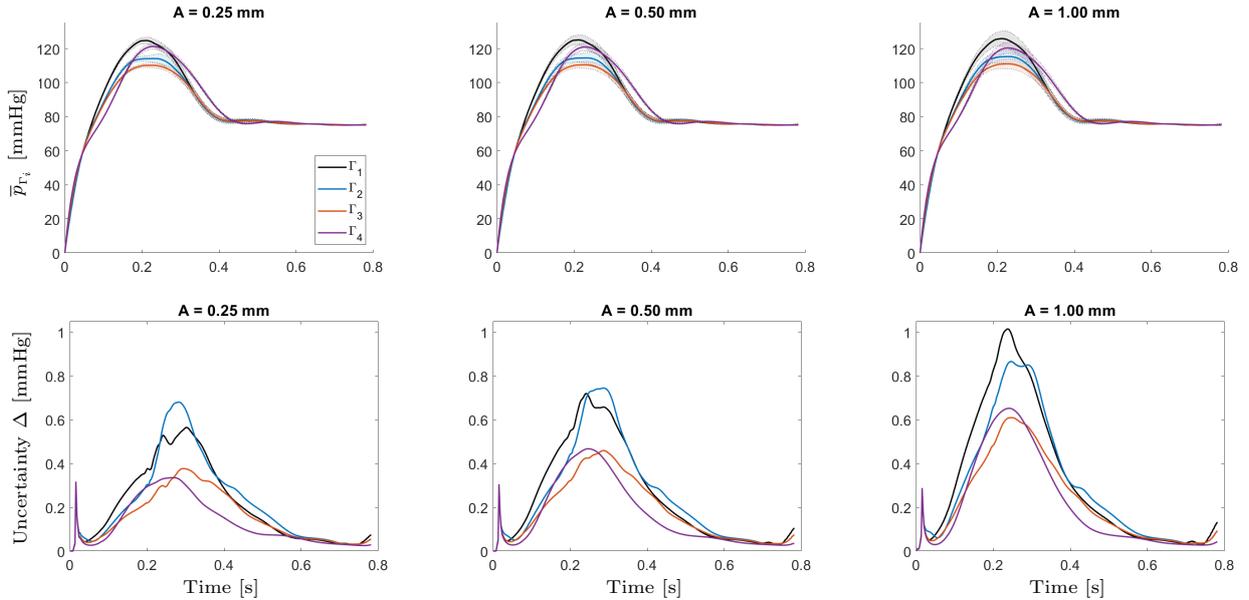
\begin{figure}[!ht]
    \centering
    {\begin{overpic}[width=0.3\linewidth,draft=\draftUQresults]{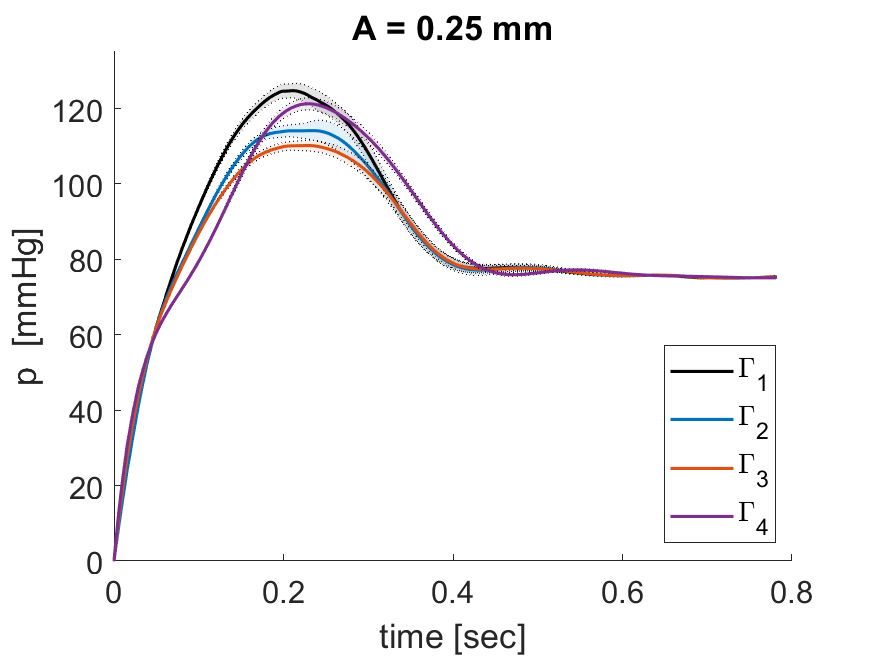}
        \put(-35, 382.5){\footnotesize\colorbox{white}{\rotatebox[origin=c]{90}{$\bar{p}_{\Gamma_i}$ [mmHg]}}}
        \put(395, 0){\footnotesize\colorbox{white}{\phantom{Time [s]}}}
    \end{overpic}
    }\hspace{0.02\linewidth}%
    {\begin{overpic}[width=0.3\linewidth,draft=\draftUQresults]{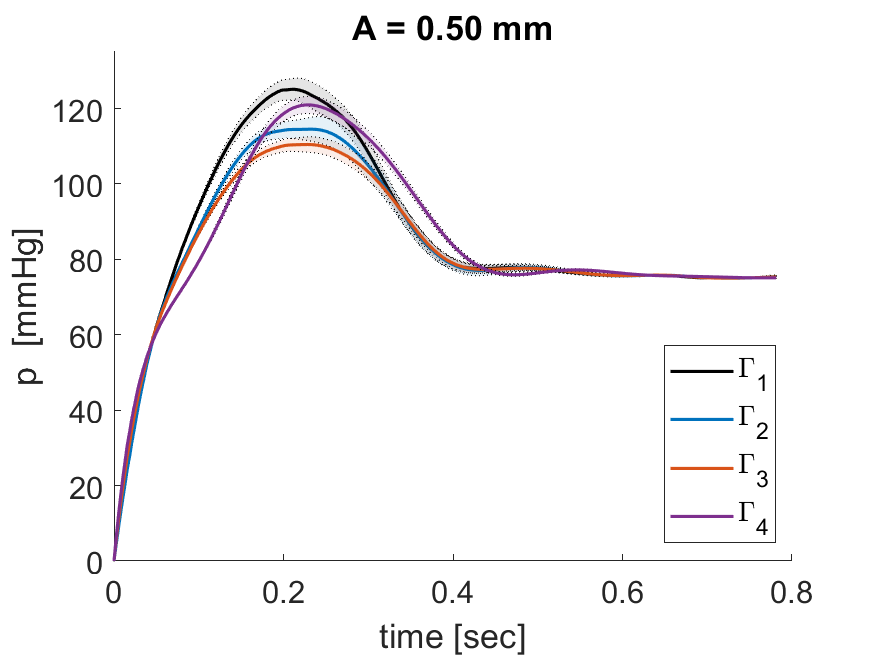}
        \put(-35, 382.5){\footnotesize\colorbox{white}{\rotatebox[origin=c]{90}{\phantom{$\bar{p}_{\Gamma_i}$ [mmHg]}}}}
        \put(720, 145){\colorbox{white}{\phantom{\scalebox{2.5}[4.5]{O}}}}
        \put(395, 0){\footnotesize\colorbox{white}{\phantom{Time [s]}}}
    \end{overpic}}\hspace{0.02\linewidth}%
    {\begin{overpic}[width=0.3\linewidth,draft=\draftUQresults]{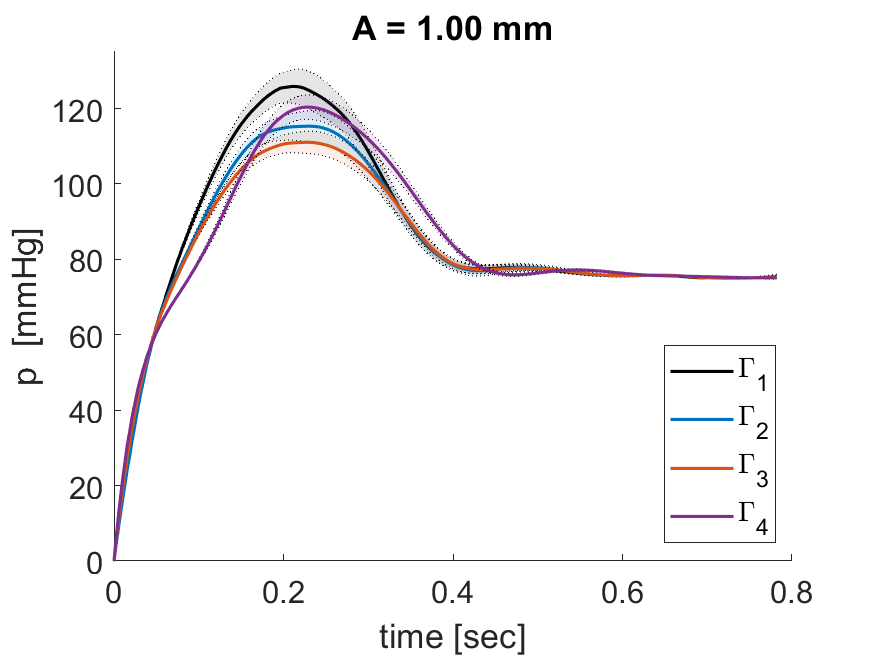}
        \put(-35, 382.5){\footnotesize\colorbox{white}{\rotatebox[origin=c]{90}{\phantom{$\bar{p}_{\Gamma_i}$ [mmHg]}}}}
        \put(720, 145){\colorbox{white}{\phantom{\scalebox{2.5}[4.5]{O}}}}
        \put(395, 0){\footnotesize\colorbox{white}{\phantom{Time [s]}}}
    \end{overpic}}
    \\
    {\begin{overpic}[width=0.3\linewidth,draft=\draftUQresults
    ]{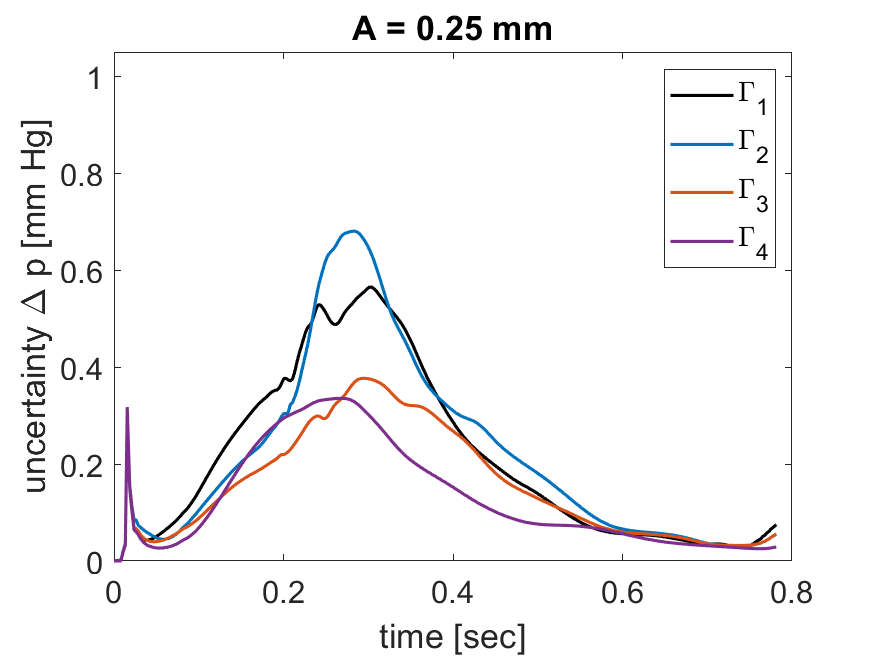}
        \put(-35, 382.5){\footnotesize\colorbox{white}{\rotatebox[origin=c]{90}{\phantom{Uncertainty $\Delta\bar{p}_{\Gamma_i}$ [mmHg]}}}} 
        \put(-35, 382.5){\footnotesize\colorbox{white}{\rotatebox[origin=c]{90}{{Uncertainty $\Delta$ [mmHg]}}}}
        \put(720, 450){\colorbox{white}{\phantom{\scalebox{2.5}[4.5]{O}}}}
        \put(395, 0){\footnotesize\colorbox{white}{{Time [s]}}}
    \end{overpic}}\hspace{0.02\linewidth}%
    {\begin{overpic}[width=0.3\linewidth,draft=\draftUQresults
    ]{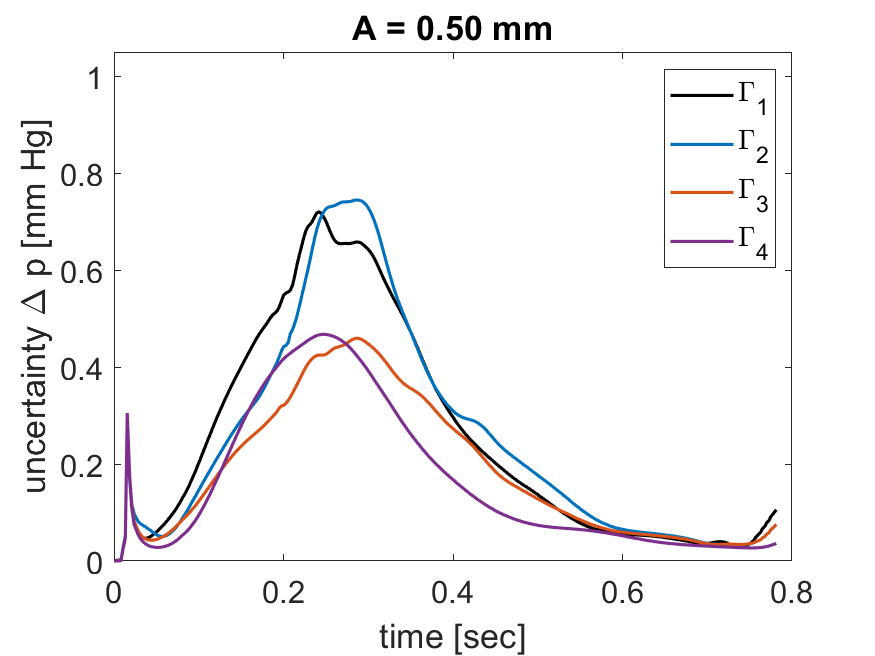}
        \put(-35, 382.5){\footnotesize\colorbox{white}{\rotatebox[origin=c]{90}{\phantom{Uncertainty $\Delta\bar{p}_{\Gamma_i}$ [mmHg]}}}}
        \put(720, 450){\colorbox{white}{\phantom{\scalebox{2.5}[4.5]{O}}}}
        \put(395, 0){\footnotesize\colorbox{white}{{Time [s]}}}
    \end{overpic}}\hspace{0.02\linewidth}%
    {\begin{overpic}[width=0.3\linewidth,draft=\draftUQresults
    ]{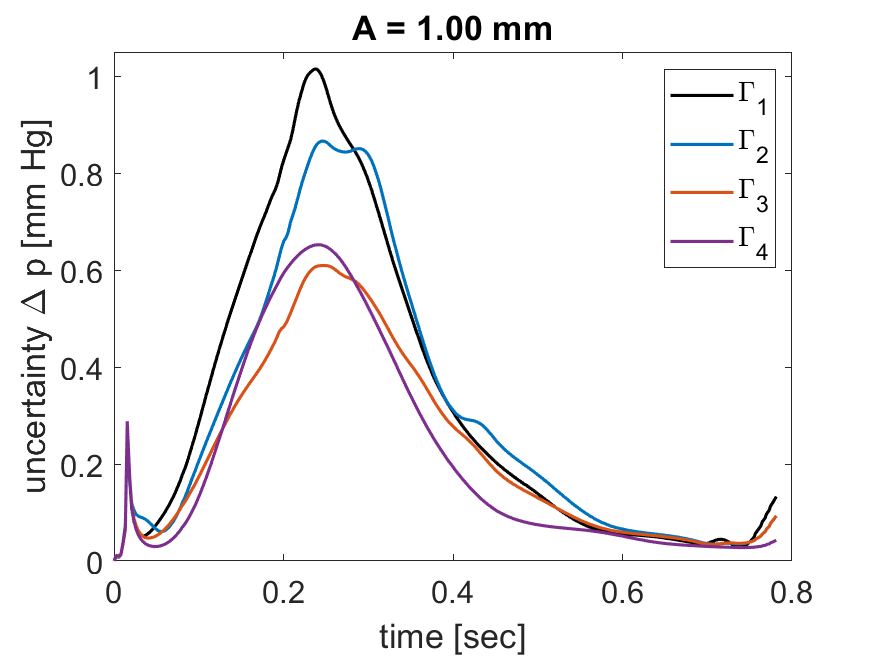}
        \put(-35, 382.5){\footnotesize\colorbox{white}{\rotatebox[origin=c]{90}{\phantom{Uncertainty $\bar{p}_{\Gamma_i}$ [mmHg]}}}}
        \put(720, 450){\colorbox{white}{\phantom{\scalebox{2.5}[4.5]{O}}}}
        \put(395, 0){\footnotesize\colorbox{white}{{Time [s]}}}
    \end{overpic}}
    \caption{Mean $\mu$ (top row) and uncertainty $\Delta$ (bottom row) in the spatial mean pressure $\bar{p}_{\Gamma_i}$, $i=1,\dots,4$ (see Fig.~\ref{fig:wss_spheres}, middle) for varying perturbation levels $A=0.25, 0.5$ and $1.0$~mm. Note that the uncertainty bands are scaled by a factor of 5 for visualization purposes. The legend in the first diagram applies to all plots.}
    \label{fig:resultsUQ:mean_pressure}
\end{figure}

\FloatBarrier

\begin{figure}[!ht]
    \centering
    {\begin{overpic}[width=0.3\linewidth,draft=\draftUQresults]{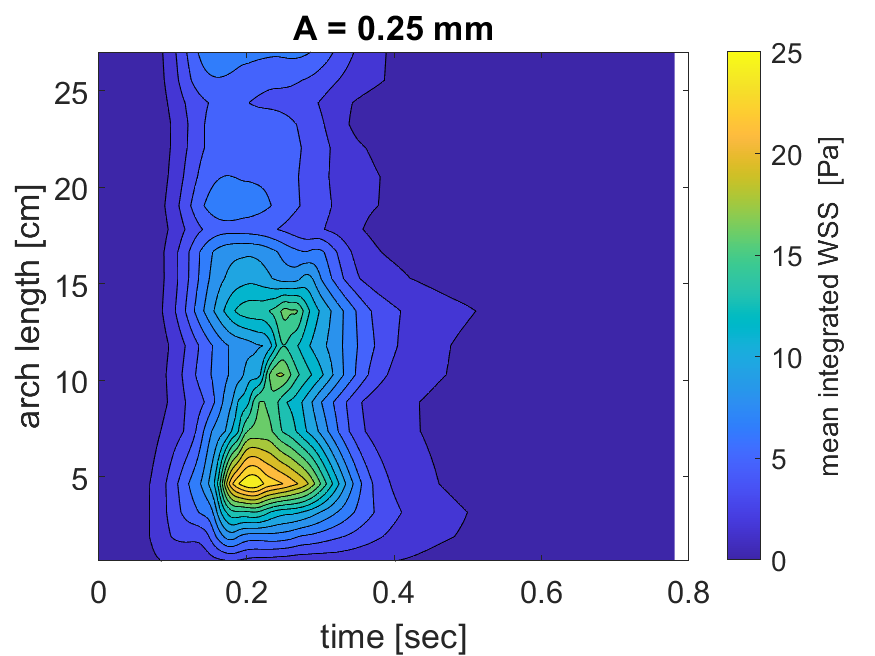}
        \put(-32.5, 382.5){\footnotesize\colorbox{white}{\rotatebox[origin=c]{90}{{Arc length [cm]}}}}
        \put(360, 0){\footnotesize\colorbox{white}{\phantom{Time [s]}}}
        \put(920, 382.5){\footnotesize\colorbox{white}{\rotatebox[origin=c]{90}{\phantom{Int. mean WSS $w_{\Gamma_\mathrm{wall}}$ [Pa]}}}} 
        \put(920, 382.5){\footnotesize\colorbox{white}{\rotatebox[origin=c]{90}{{Mean $\mu$ [Pa]}}}}
    \end{overpic}
    }\hspace{0.02\linewidth}%
    {\begin{overpic}[width=0.3\linewidth,draft=\draftUQresults]{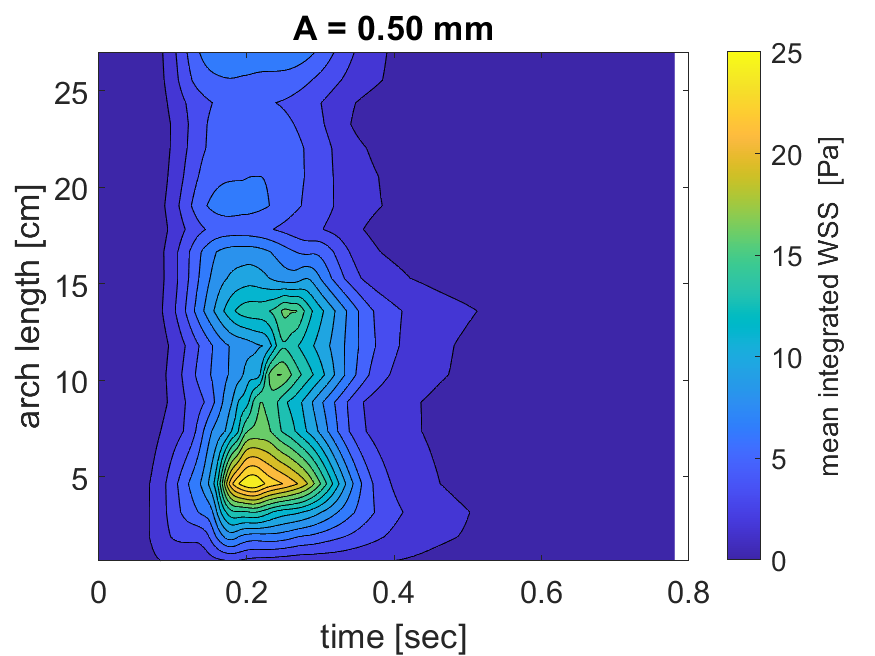}
        \put(-32.5, 382.5){\footnotesize\colorbox{white}{\rotatebox[origin=c]{90}{\phantom{Arc length [cm]}}}}
        \put(360, 0){\footnotesize\colorbox{white}{\phantom{Time [s]}}}
        \put(920, 382.5){\footnotesize\colorbox{white}{\rotatebox[origin=c]{90}{\phantom{Int. mean WSS $w_{\Gamma_\mathrm{wall}}$ [Pa]}}}}
    \end{overpic}
    }\hspace{0.02\linewidth}%
    {\begin{overpic}[width=0.3\linewidth,draft=\draftUQresults]{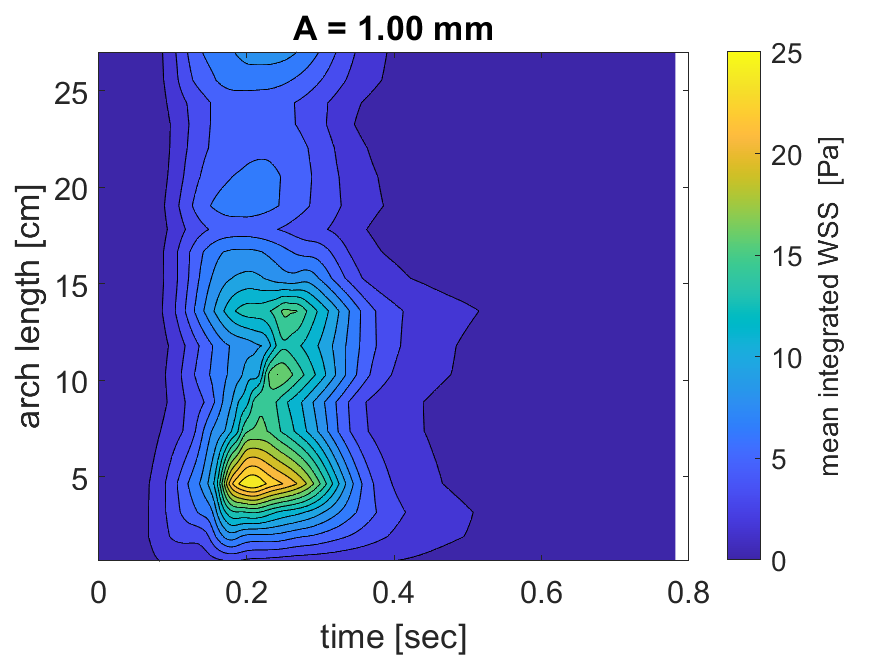}
        \put(-32.5, 382.5){\footnotesize\colorbox{white}{\rotatebox[origin=c]{90}{\phantom{Arc length [cm]}}}}
        \put(360, 0){\footnotesize\colorbox{white}{\phantom{Time [s]}}}
        \put(920, 382.5){\footnotesize\colorbox{white}{\rotatebox[origin=c]{90}{\phantom{Int. mean WSS $w_{\Gamma_\mathrm{wall}}$ [Pa]}}}}
    \end{overpic}
    }
    \\
    {\begin{overpic}[width=0.3\linewidth,draft=\draftUQresults]{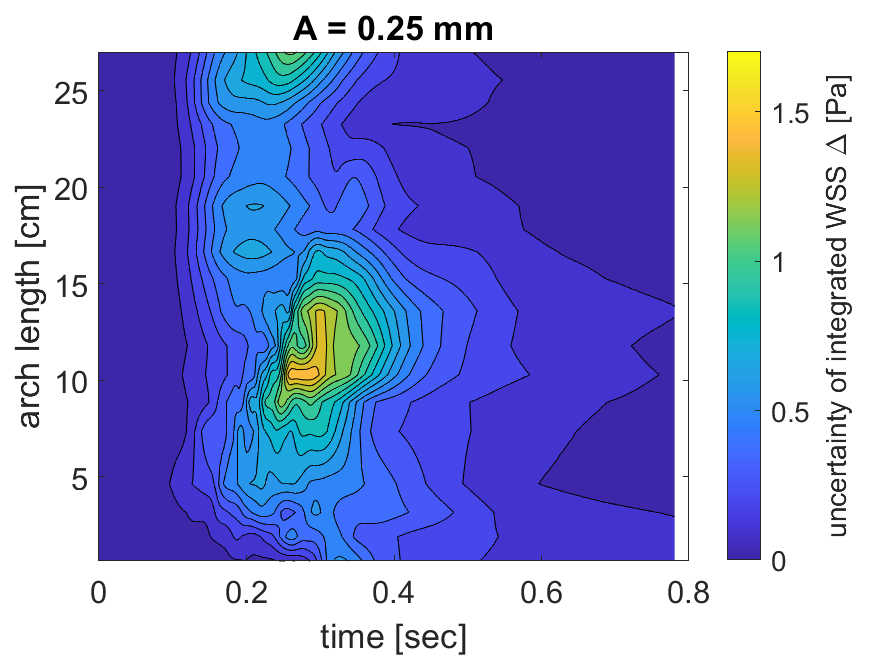}
        \put(-32.5, 382.5){\footnotesize\colorbox{white}{\rotatebox[origin=c]{90}{{Arc length [cm]}}}}
        \put(360, 0){\footnotesize\colorbox{white}{\phantom{Time [s]}}}
        \put(925, 382.5){\footnotesize\colorbox{white}{\rotatebox[origin=c]{90}{\phantom{Uncertainty $\Delta w_{\Gamma_{\mathrm{wall}}}$ [Pa]}}}} 
        \put(925, 382.5){\footnotesize\colorbox{white}{\rotatebox[origin=c]{90}{{Uncertainty $\Delta$ [Pa]}}}}
    \end{overpic}
    }\hspace{0.02\linewidth}%
    {\begin{overpic}[width=0.3\linewidth,draft=\draftUQresults]{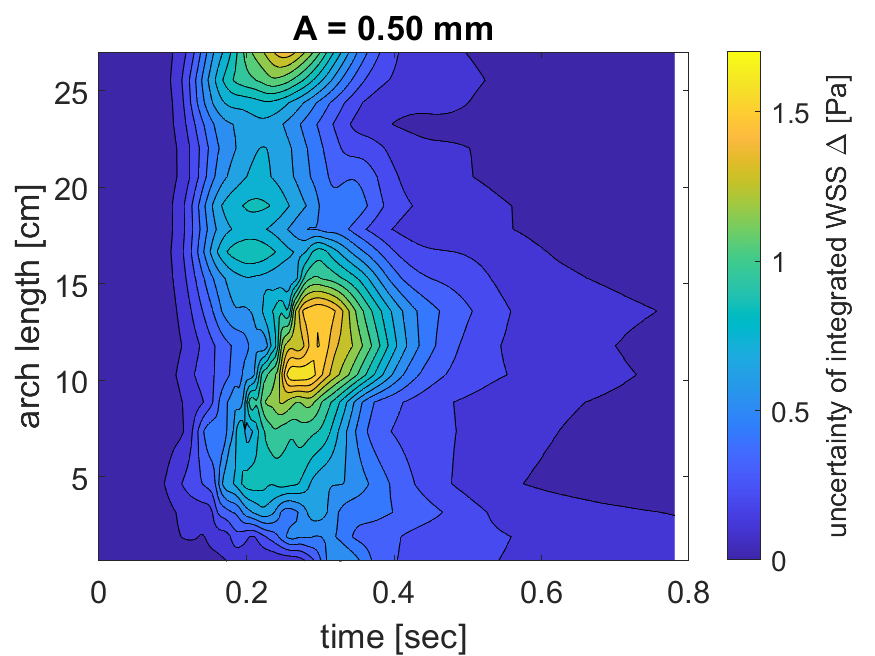}
        \put(-32.5, 382.5){\footnotesize\colorbox{white}{\rotatebox[origin=c]{90}{\phantom{Arc length [cm]}}}}
        \put(360, 0){\footnotesize\colorbox{white}{\phantom{Time [s]}}}
        \put(925, 382.5){\footnotesize\colorbox{white}{\rotatebox[origin=c]{90}{\phantom{Uncertainty $\Delta w_{\Gamma_{\mathrm{wall}}}$ [Pa]}}}}
    \end{overpic}
    }\hspace{0.02\linewidth}%
    {\begin{overpic}[width=0.3\linewidth,draft=\draftUQresults]{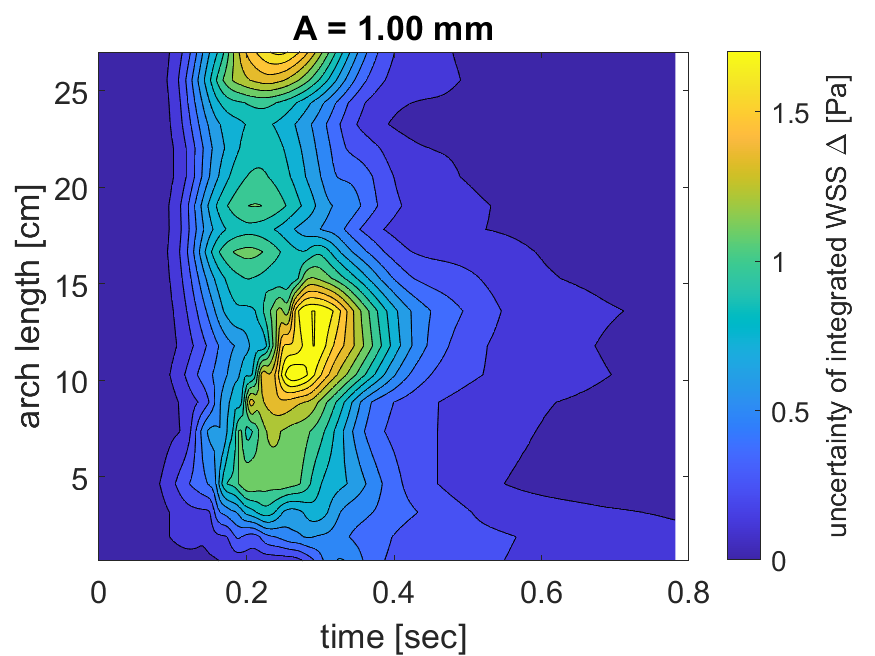}
        \put(-32.5, 382.5){\footnotesize\colorbox{white}{\rotatebox[origin=c]{90}{\phantom{Arc length [cm]}}}}
        \put(360, 0){\footnotesize\colorbox{white}{\phantom{Time [s]}}}
        \put(925, 382.5){\footnotesize\colorbox{white}{\rotatebox[origin=c]{90}{\phantom{Uncertainty $\Delta w_{\Gamma_{\mathrm{wall}}}$ [Pa]}}}}
    \end{overpic}
    }
    \\
    {\begin{overpic}[width=0.3\linewidth,draft=\draftUQresults]{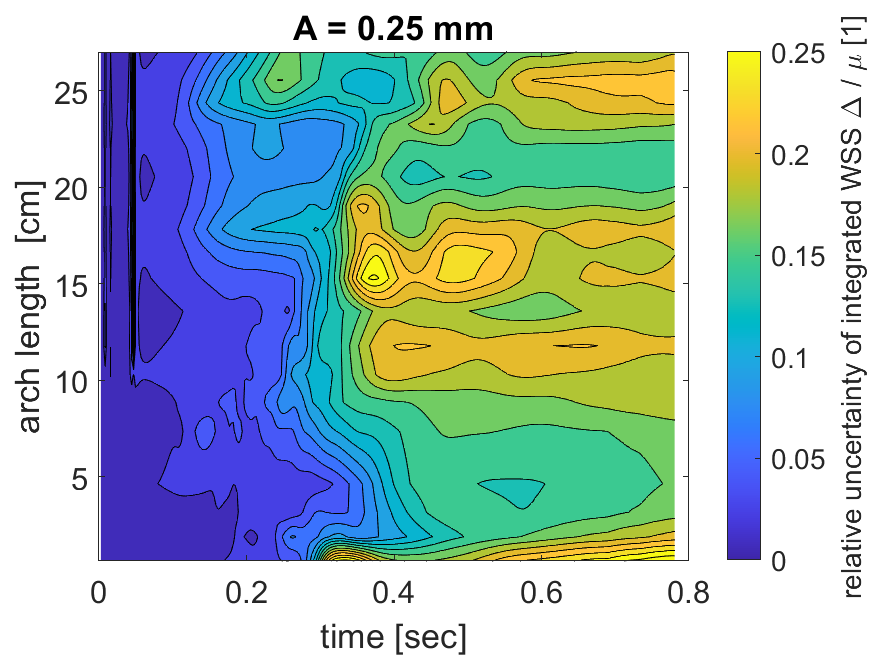}
        \put(-32.5, 382.5){\footnotesize\colorbox{white}{\rotatebox[origin=c]{90}{{Arc length [cm]}}}}
        \put(360, 0){\footnotesize\colorbox{white}{{Time [s]}}}
        \put(950, 382.5){\footnotesize\colorbox{white}{\rotatebox[origin=c]{90}{\phantom{Rel. uncertainty in $w_{\Gamma_{\mathrm{wall}}}$ [-]}}}} 
        \put(950, 382.5){\footnotesize\colorbox{white}{\rotatebox[origin=c]{90}{{Rel. uncertainty $\Delta/\mu$ [-]}}}}
    \end{overpic}
    }\hspace{0.02\linewidth}%
    {\begin{overpic}[width=0.3\linewidth,draft=\draftUQresults]{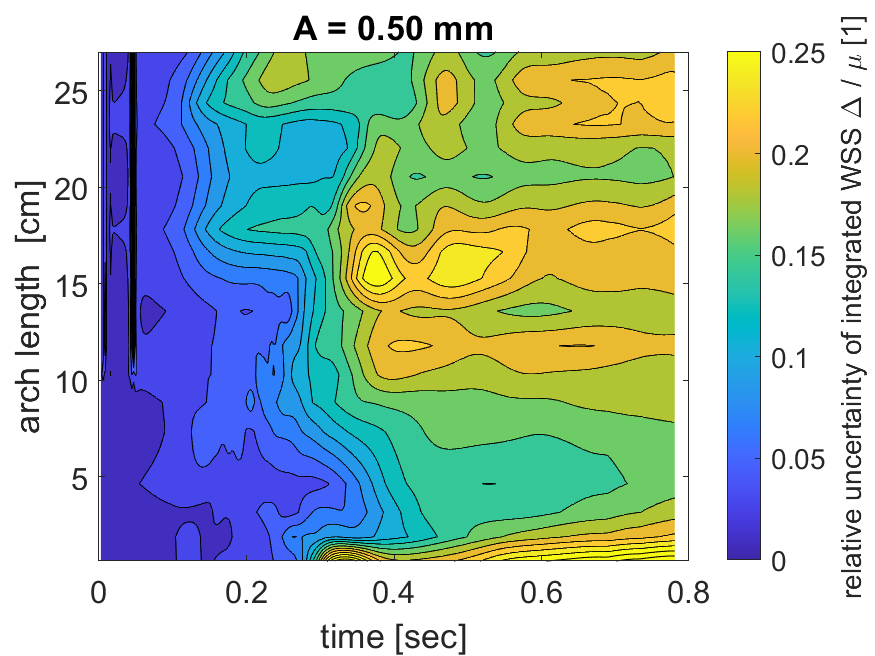}
        \put(-32.5, 382.5){\footnotesize\colorbox{white}{\rotatebox[origin=c]{90}{\phantom{Arc length [cm]}}}}
        \put(360, 0){\footnotesize\colorbox{white}{{Time [s]}}}
        \put(950, 382.5){\footnotesize\colorbox{white}{\rotatebox[origin=c]{90}{\phantom{Rel. uncertainty in $w_{\Gamma_{\mathrm{wall}}}$ [-]}}}}
    \end{overpic}
    }\hspace{0.02\linewidth}%
    {\begin{overpic}[width=0.3\linewidth,draft=\draftUQresults]{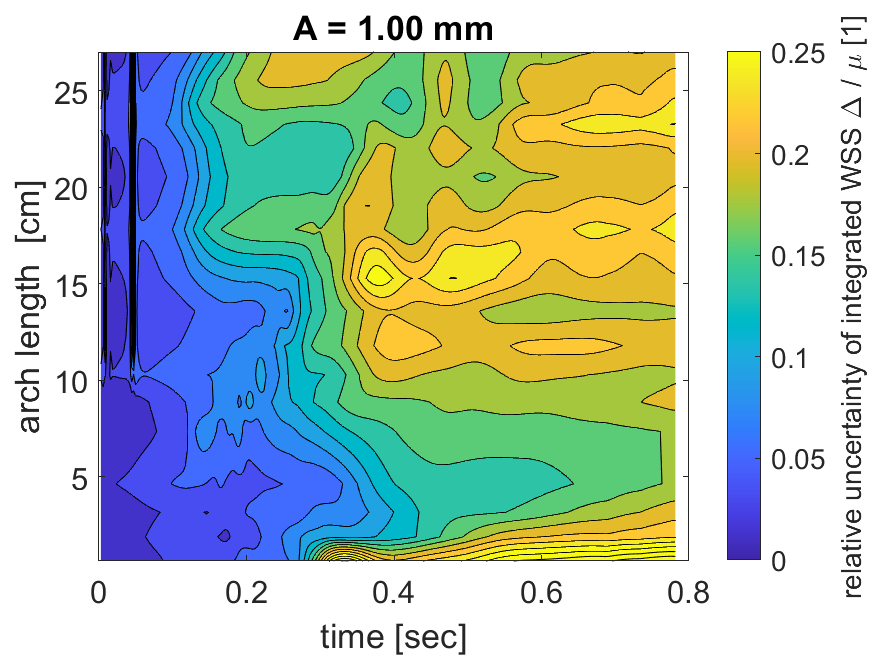}
        \put(-32.5, 382.5){\footnotesize\colorbox{white}{\rotatebox[origin=c]{90}{\phantom{Arc length [cm]}}}}
        \put(360, 0){\footnotesize\colorbox{white}{{Time [s]}}}
        \put(950, 382.5){\footnotesize\colorbox{white}{\rotatebox[origin=c]{90}{\phantom{Rel. uncertainty in $w_{\Gamma_{\mathrm{wall}}}$ [-]}}}}
    \end{overpic}}
    \caption{Mean $\mu$ (top row), uncertainty $\Delta$ (middle row), and relative uncertainty $\Delta/\mu$ (bottom row) in the spatial integral mean wall shear stress per subregion over arc length (see Fig.~\ref{fig:wss_spheres}, middle) for varying perturbation levels $A=0.25, 0.5$ and $1.0$~mm.}
    \label{fig:resultsUQ:smallS_WSS}
\end{figure}

Next, we summarize the mean and uncertainty in the integral mean WSS $w_\Gamma$~\eqref{eqn:wss_integral_avg} over subregions defined via spheres, as detailed in Sec.~\ref{sec:QoIs}, see Fig.~\ref{fig:wss_spheres} and Table~\ref{tab:spheres}. The data are analyzed incorporating the arc length along the aorta, such that in contrast to the previous results \textit{per outlet}, spatial patterns can be investigated. Again, the mean of the integral mean WSS values agree well for all perturbation levels in terms of localization and magnitude, as can be seen from Fig.~\ref{fig:resultsUQ:smallS_WSS} (top row). Increased WSSs are observed in the entire arc region from $s=2$ to $15$~cm, with distinct peak values around $s=5$~cm, that is, approximately at the brachiocephalic trunk. The observed values lie around \(20\)~Pa at their maximum, and are significantly lower during the diastolic phase. These values are within the physiological range, although we note that the TAWSS, the \textit{time-averaged} WSS, is often reported in the literature, which provides values that are well below the peak values reported here, but can be directly derived from the present results.

The observed position of the peak WSS can be easily explained by the curvature of the vessel, which is highest in the arc and leads to flow redirection. This in turn results in large shear stresses acting on the apex of the aortic arch. Also remember that the perturbations are only set to zero above $X_{0,3} =333$~mm, which corresponds to the lowest point of the subclavian arteries. Therefore, the perturbations are in fact nonzero in most regions with elevated WSSs. The uncertainty in the integral average WSS is $<1.75~\text{Pa}$, that is $<7$\% of the peak values $\approx25~\text{Pa}$ for all perturbation levels and for all sub-regions along the vessel's centerline.

Interestingly, the integral mean WSS in the abdominal aorta, especially at the main outlet and close to it (arc length $> 22.5~\text{cm}$), shows increased sensitivity with regards to geometric variations. Boundary effects of random field generation could be ruled out as the cause.
The uncertainty noticeably increases with $A$, but remain proportional to values observed in the remaining vessel. The Windkessel parameters are tuned such that $\approx60$\% of the volumetric flow exit via the computational domain via the main outlet $\Gamma_4$ in the abdominal aorta in the reference unperturbed geometry, i.e., $A=0.0~\text{mm}$. Geometric variations directly impact the cross sectional area, narrowing/widening the hydraulic radius, while the Windkessel parameters remain unchanged. The uncertainty in $Q_4$ is highest in absolute terms (see Fig.~\ref{fig:resultsUQ:flow_rates}), and hence, can lead to significantly increased integral mean WSS for increasing mean perturbations $A$. The increasing uncertainty close to the outlets is linked to the sensitivity of the Windkessel models. Their influence on the flow field in the interior domain slightly decreased with distance from the outlets due to dissipation.

Figure~\ref{fig:resultsUQ:smallS_WSS} (bottom row) depicts the \textit{relative} uncertainties in the integral mean WSS per sphere over the arc length. Especially at peak systole, i.e., around $t=0.2~\text{s}$, one observes relative uncertainties increasing from $5$\% to $<15$\% along the arc, while in diastole, \textit{relative} uncertainty is high (up to 25\%), while \textit{absolute} uncertainty is low (less than $\approx2.5~\text{Pa}$). The increase in the mean geometric perturbation also directly translates to a visible increase in the uncertainties as expected. 

\begin{figure} [!ht]
    \centering
    {\begin{overpic}[width=0.3\linewidth,draft=\draftUQresults]{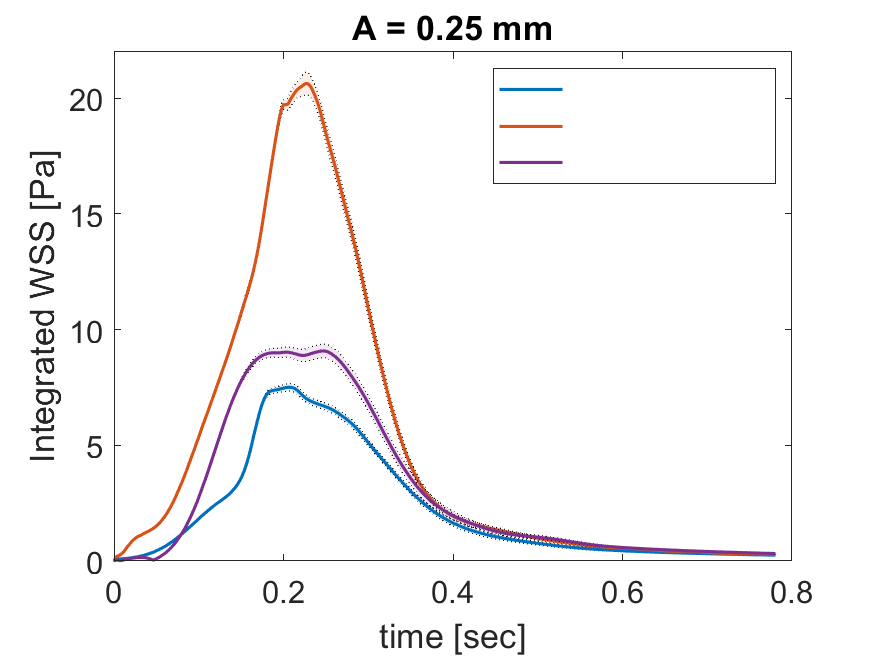}
        \put(-17.5, 382.5){\footnotesize\colorbox{white}{\rotatebox[origin=c]{90}{\phantom{Int. mean WSS $w_{\Gamma_\mathrm{wall}}$ [Pa]}}}} 
        \put(-17.5, 382.5){\footnotesize\colorbox{white}{\rotatebox[origin=c]{90}{{Mean $\mu$ [Pa]}}}}
        \put(375, 0){\footnotesize\colorbox{white}{\phantom{Time [s]}}}
        \put(650, 637){\tiny{{ascending ao.}}}
        \put(650, 596){\tiny{{aortic arc}}}
        \put(650, 555){\tiny{{abdominal ao.}}}
    \end{overpic}}\hspace{0.02\linewidth}%
    {\begin{overpic}[width=0.3\linewidth,draft=\draftUQresults]{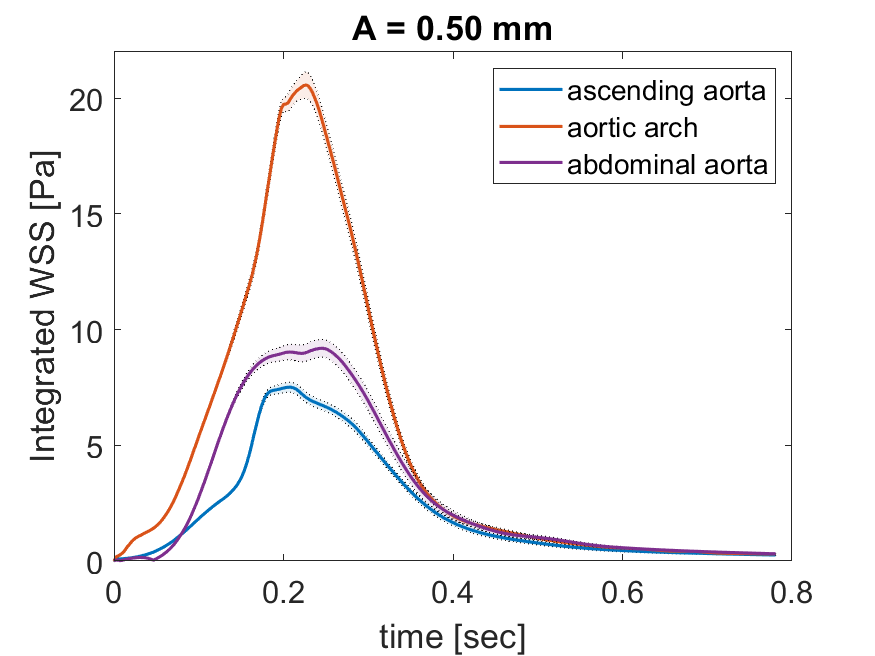}
        \put(-17.5, 382.5){\footnotesize\colorbox{white}{\rotatebox[origin=c]{90}{\phantom{Int. mean WSS $w_{\Gamma_\mathrm{wall}}$ [Pa]}}}}
        \put(375, 0){\footnotesize\colorbox{white}{\phantom{Time [s]}}}
        \put(510, 540){\colorbox{white}{\phantom{\scalebox{6.5}[2.5]{O}}}}
    \end{overpic}}\hspace{0.02\linewidth}%
    {\begin{overpic}[width=0.3\linewidth,draft=\draftUQresults]{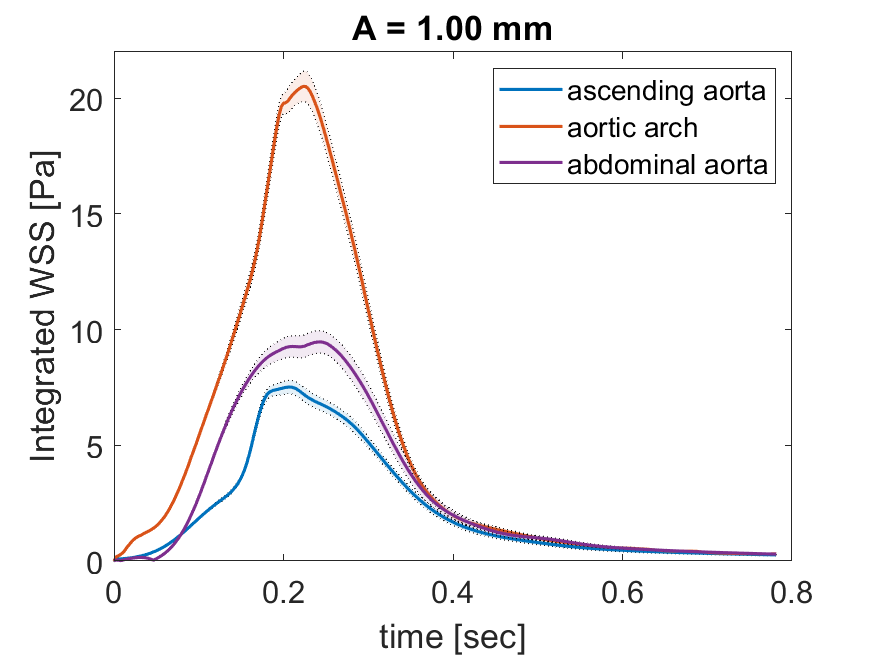}
        \put(-17.5, 382.5){\footnotesize\colorbox{white}{\rotatebox[origin=c]{90}{\phantom{Int. mean WSS $w_{\Gamma_\mathrm{wall}}$ [Pa]}}}}
        \put(375, 0){\footnotesize\colorbox{white}{\phantom{Time [s]}}}
        \put(510, 540){\colorbox{white}{\phantom{\scalebox{6.5}[2.5]{O}}}}
    \end{overpic}}
    \\
    {\begin{overpic}[width=0.3\linewidth,draft=\draftUQresults]{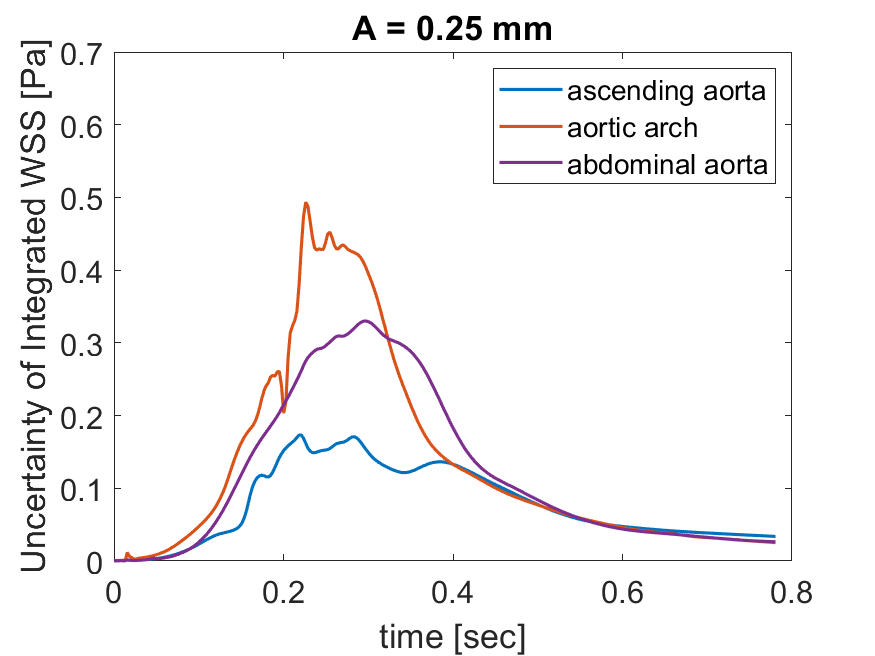}
        \put(-25, 382.5){\footnotesize\colorbox{white}{\rotatebox[origin=c]{90}{\phantom{Uncertainty $\Delta w_{\Gamma_\mathrm{wall}}$ [Pa]}}}} 
        \put(-25, 382.5){\footnotesize\colorbox{white}{\rotatebox[origin=c]{90}{{Uncertainty $\Delta$ [Pa]}}}}
        \put(375, 0){\footnotesize\colorbox{white}{\phantom{Time [s]}}}
        \put(510, 540){\colorbox{white}{\phantom{\scalebox{6.5}[2.5]{O}}}}
    \end{overpic}}\hspace{0.02\linewidth}%
    {\begin{overpic}[width=0.3\linewidth,draft=\draftUQresults]{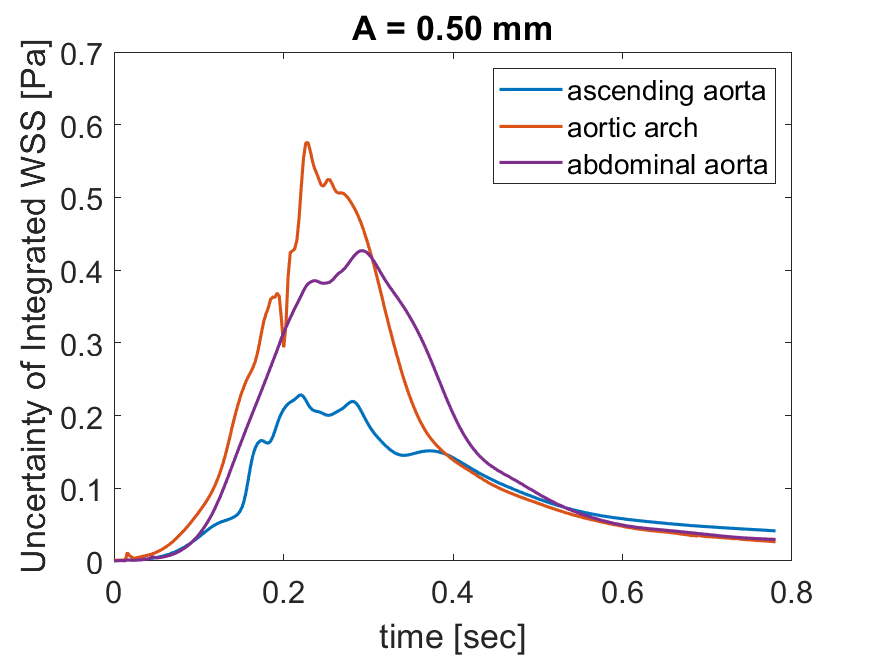}
        \put(-25, 382.5){\footnotesize\colorbox{white}{\rotatebox[origin=c]{90}{\phantom{Uncertainty $\Delta w_{\Gamma_\mathrm{wall}}$ [Pa]}}}}
        \put(375, 0){\footnotesize\colorbox{white}{\phantom{Time [s]}}}
        \put(510, 540){\colorbox{white}{\phantom{\scalebox{6.5}[2.5]{O}}}}
    \end{overpic}}\hspace{0.02\linewidth}%
    {\begin{overpic}[width=0.3\linewidth,draft=\draftUQresults]{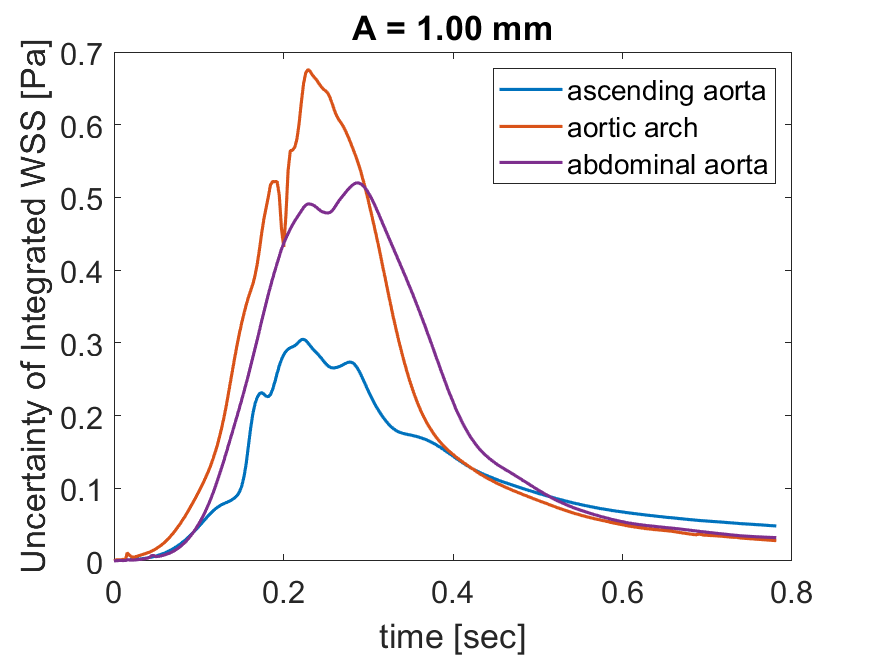}
        \put(-25, 382.5){\footnotesize\colorbox{white}{\rotatebox[origin=c]{90}{\phantom{Uncertainty $\Delta w_{\Gamma_\mathrm{wall}}$ [Pa]}}}}
        \put(375, 0){\footnotesize\colorbox{white}{\phantom{Time [s]}}}
        \put(510, 540){\colorbox{white}{\phantom{\scalebox{6.5}[2.5]{O}}}}
    \end{overpic}}
    \\
    {\begin{overpic}[width=0.3\linewidth,draft=\draftUQresults]{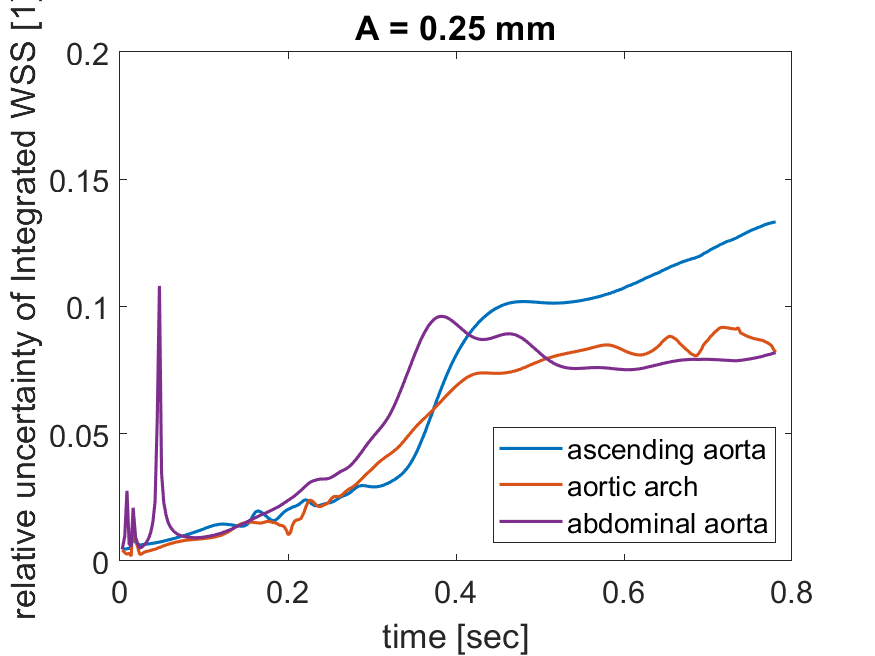}
        \put(-35, 382.5){\footnotesize\colorbox{white}{\rotatebox[origin=c]{90}{\phantom{Rel. uncertainty in $w_{\Gamma_\mathrm{wall}}$ [-]}}}} 
        \put(-35, 382.5){\footnotesize\colorbox{white}{\rotatebox[origin=c]{90}{{Rel. uncertainty $\Delta/\mu$ [-]}}}}
        \put(375, 0){\footnotesize\colorbox{white}{{Time [s]}}}
        \put(510, 145){\colorbox{white}{\phantom{\scalebox{6.5}[2.5]{O}}}}
    \end{overpic}}\hspace{0.02\linewidth}%
    {\begin{overpic}[width=0.3\linewidth,draft=\draftUQresults]{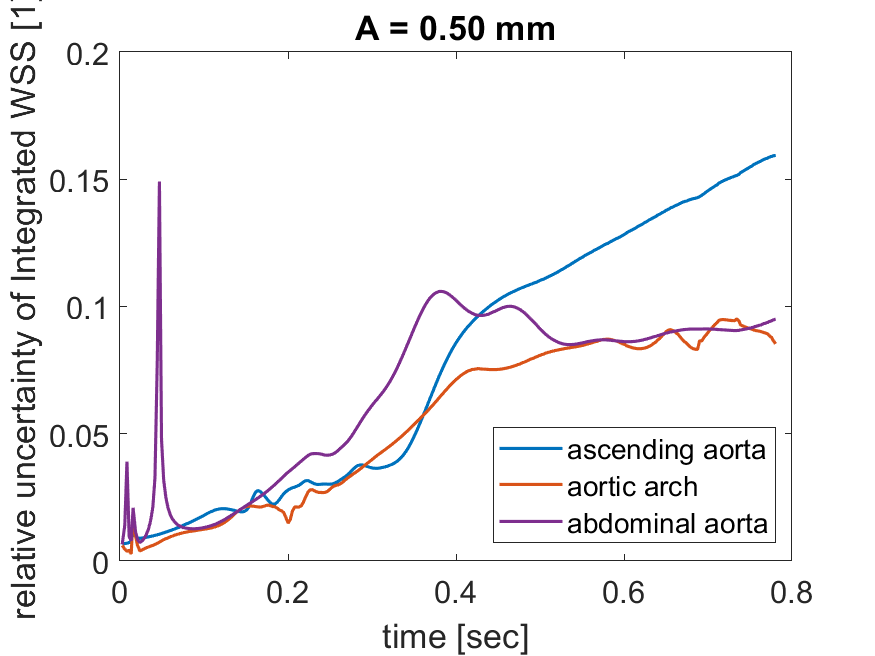}
        \put(-35, 382.5){\footnotesize\colorbox{white}{\rotatebox[origin=c]{90}{\phantom{Rel. uncertainty in $w_{\Gamma_\mathrm{wall}}$ [-]}}}}
        \put(375, 0){\footnotesize\colorbox{white}{{Time [s]}}}
        \put(510, 145){\colorbox{white}{\phantom{\scalebox{6.5}[2.5]{O}}}}
    \end{overpic}}\hspace{0.02\linewidth}%
    {\begin{overpic}[width=0.3\linewidth,draft=\draftUQresults]{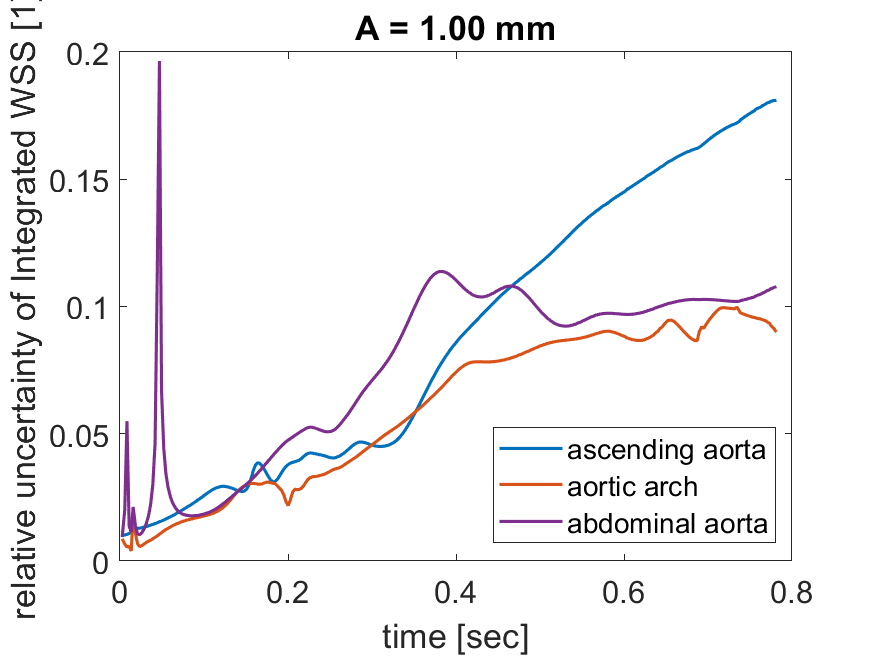}
        \put(-35, 382.5){\footnotesize\colorbox{white}{\rotatebox[origin=c]{90}{\phantom{Rel. uncertainty in $w_{\Gamma_\mathrm{wall}}$ [-]}}}}
        \put(375, 0){\footnotesize\colorbox{white}{{Time [s]}}}     
        \put(510, 145){\colorbox{white}{\phantom{\scalebox{6.5}[2.5]{O}}}}
    \end{overpic}}
    \caption{Mean $\mu$ (top row), uncertainty $\Delta$ (middle row), and relative uncertainty $\Delta/\mu$ (bottom row) of the spatial integral mean wall shear stress per subregion (see Fig.~\ref{fig:wss_spheres}, right) for varying perturbation levels $A=0.25, 0.5$ and $1.0$~mm. Note that the uncertainty bands are scaled by a factor of 5 for visualization purposes.  The legend in the first diagram applies to all plots.}
    \label{fig:resultsUQ:bigS_WSS}
\end{figure}
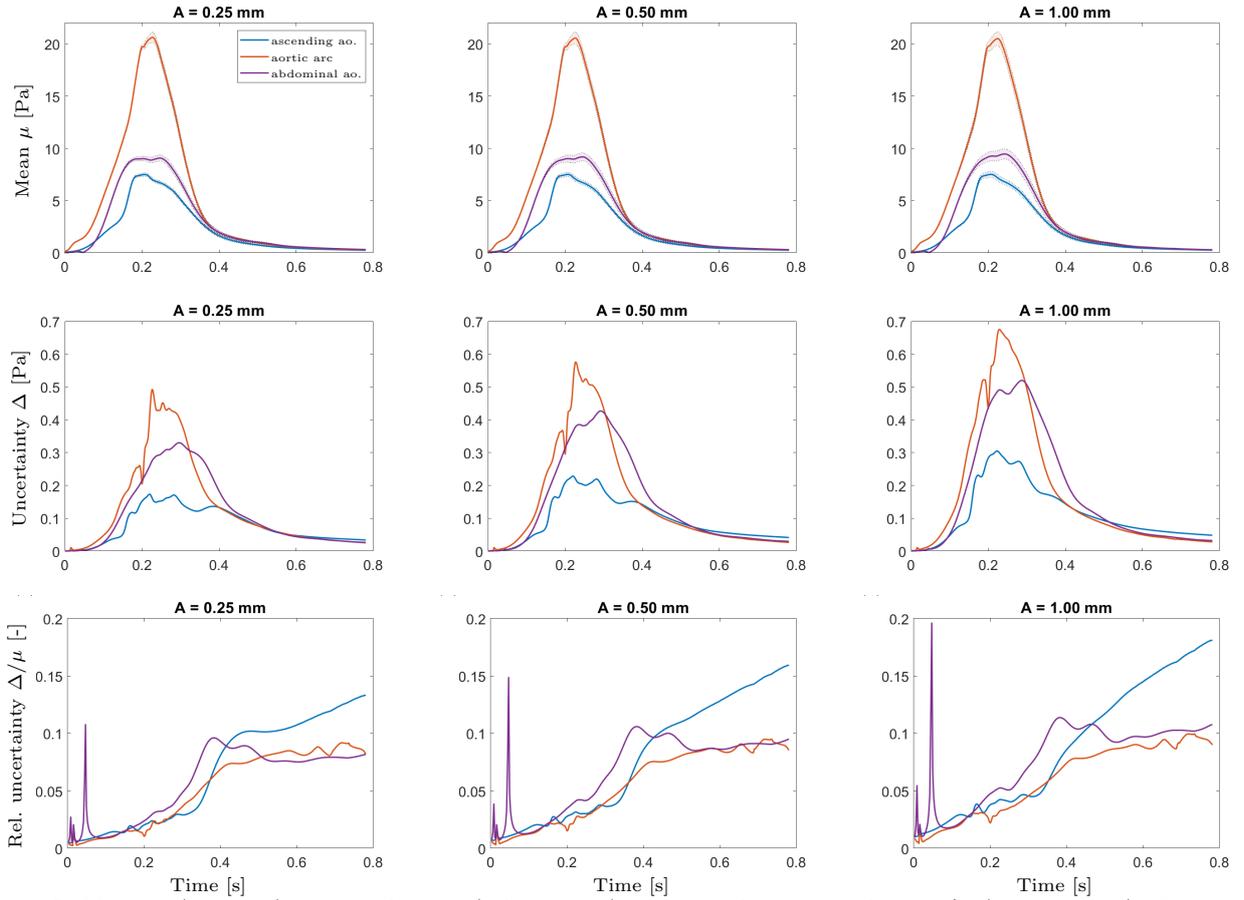

Results differentiating the regions of the aorta, namely ascending aorta, aortic arc and abdominal aorta (Fig.~\ref{fig:wss_spheres}, right), are given in Fig.~\ref{fig:resultsUQ:bigS_WSS}. The uncertainty in the integral average WSS in all regions is $<0.7~\text{Pa}$, that is $<3.5$\% of the peak values $\approx 20~\text{Pa}$ for all perturbation levels. Altogether, same trends for regional integral mean WSS, integral mean WSS over the arc length \textit{and} the flow rate and spatial mean pressure per outlet are seen, summarized as follows:
\begin{enumerate}
    \item[(i)] The respective means are independent of the mean perturbation $A$.
    \item[(ii)] The uncertainty increases with the mean perturbation $A$. 
    \item[(iii)] The uncertainty is highest in late systole. 
\end{enumerate}
Again, we observe a sharp peak in the uncertainty at $t\approx 0.01~\text{s}$, which relates to fluctuations in the pressure due to the Windkessel model. These fluctuations are expected and lie outside of the credible (nearly periodic) time interval $[0.05,\, 0.78]~\text{s}$ (chosen for efficiency; periodic solution reached up to engineering accuracy). The increase in uncertainty with increasing arc length is expressed here through an increased slope of the relative uncertainty (see Fig.~\ref{fig:resultsUQ:bigS_WSS}), while the absolute uncertainties in the diastolic phase remain low.

\section{Discussion}\label{sec:Discussion}

As mentioned in the introduction, the order of magnitude of the CT scan error of existing machines is no more than \(0.5\)~mm, with modern developments seeking to reduce it to errors smaller than \(0.2\)~mm. Thus, the imposed random fields in the cases of \(A=0.5\)~mm and \(A=1.0\)~mm certainly account for more than the CT scan error.

Regarding the meshing error, note that for any mesh refinement level, all of the mesh surface points are mapped to the surface of the actual domain. Therefore, a more refined (or higher-order) mesh herein also implies a higher geometrical representation accuracy. Thus, it is reasonable to assume, that for a relatively smooth domain like the aorta, a sufficiently refined mesh has a negligible error. To quantify the representation error of the mesh used in the examples (refinement level \(1\), order \(4\)), we compare it to the overkill mesh (refinement level \(4\), order \(4\)) in terms of volume and surface area. The absolute volume and surface area errors are \(21.87\)~mm$^3$ and \(2.13\)~mm\(^2\), yielding relative errors are \(0.02\%\) and \(0.01\%\), respectively. Thus, we conclude that the errors coming from the volumetric meshing are dominated by the CT errors described earlier.

Using physiological model parameters and sufficiently accurate numerical discretizations, the geometric variations as introduced in detail in Sec.~\ref{subsec:BoundaryRandomField}, related to medical imaging, segmentation and mesh construction errors, only mildly affect simulation results, if their magnitude can be kept below a reasonable threshold. The presented results show that increasing geometric perturbations increase uncertainty in clinically relevant QoIs merely in a sub-linear fashion, and absolute uncertainty lies within the order of magnitude of the approximation error (or lower). With increasing arc length, the relative uncertainty stemming from geometric perturbations builds up, since downstream areas are affected by upstream variations in geometry. Additionally, \textit{relative} uncertainty in the diastolic phase is higher due to lower absolute values, while absolute uncertainty is in general low. The peak spatial mean pressure, flow rates and WSSs show little relative uncertainty, which lies below or at the expected approximation error for the QoIs being functions of the primary variables (velocity and pressure) \textit{and/or} the velocity gradient. Altogether, the impact on uncertainty in computed QoIs is acceptable considering engineering error tolerances (e.g., $\leq10$\% in all QoIs in systole) under moderate mean geometric perturbations ($\leq1.0$~mm). The errors stemming from mesh construction, which \textit{can} be verified against a ground truth, certainly lie below this threshold as described above. 

Geometric errors introduced by segmentation might be higher, but are also less straight-forward to verify, since the ground truth is often not available, and can only be approximated on a best effort basis from assisted or fully manual segmentation~\cite{Pepe2020, Pepe2024}. In this regard, fully automated segmentation strategies are promising, potentially achieving more accurate results than trained individuals or domain experts. However, segmentation errors are to be accepted up to a certain degree, which has to be discussed in connection with the assumptions entering the random fields' construction. The geometric variation is assumed to be a Gaussian random field with specified correlation hyperparameter. Hence, the correlation length chosen and kept fixed herein is a reasonable choice, but far from being the only possible one. This specific choice is motivated by the assumption of scarcely placed segmentation points for a cross section and a limited number of cross sections segmented along the aorta's centerline, assumptions which might impact the geometric representation fed into mesh construction differently depending on the segmentation algorithm.

\section{Conclusion and future work}
\label{sec:Conclusion}

This contribution discusses the impact of local perturbations in the geometric representation of patient-specific blood vessels on the aortic hemodynamics. Local perturbations refer to correlated distortions of the vessel wall, widening or shrinking the lumen \textit{locally} in an anisotropic fashion. These perturbations are constructed as random fields, describing the displacement normal to the vessel wall, with mean perturbation $A=0.25, 0.5$ and $1.0$~mm, and the correlation length $\ell = 5$~mm. Based on this random field defined on a reference geometry, a mapping is constructed to the final geometry 
(with \lq bumps'), which then serves as the computational domain for blood flow simulations. These simulations are used to recover QoIs such as flow rates, pressures and wall shear stresses. The errors in these QoIs are first observed under spatiotemporal refinement to identify the most efficient numerical settings that still lead to the engineering accuracy of the QoI.

The investigations are based on \(100\, 000\) blood flow simulations \textit{per perturbation level} to achieve statistical significance. The major takeaways from the numerical results are:
\begin{itemize}
    \item[(i)] the respective means of the QoIs are not effected (\lq net zero effect'),
    \item[(ii)] uncertainty increases with the perturbation mean,
    \item[(iii)] downstream regions are affected by upstream perturbations,
    \item[(iv)] \textit{absolute} uncertainty is highest ($<2$\% for flow rates, $<1$\% for spatial mean pressures, $\approx 10$\% for WSS) in late systole, where most QoIs feature maximal values, while \textit{relative} uncertainty is highest in diastole (up to 25\%, but with nearly zero QoI).
\end{itemize}
Hence, we conclude that moderate geometric variations ($\leq1.0$~mm) investigated here can be accepted when aiming for engineering accuracy in the primary variables and the related QoIs. QoIs relating to the velocity gradient such as the wall shear stress are more sensitive due to their inherent dependence on the surface normal vectors, such that point-wise differences might be drastic. However, aiming for relative errors in the single-digit percentage range in the relevant QoIs, all perturbation levels considered here lead to acceptable errors in the systolic phase. The absolute errors in the diastolic phase remained acceptable as well. Naturally, the requirements with respect to absence of geometric variations are more stringent, if increased accuracy is needed. Within the present study, rather loose error tolerances in the simulation tools were accepted to achieve a larger number of samples per perturbation level, whereas these criteria might be more strict in practice. Therefore, depending on the target errors in the QoIs within a specific application, even geometric variations with mean $A=0.25~\text{mm}$ might be considered too inaccurate. The relations between geometric perturbation and QoIs quantified within this work hence reassure the community's understanding that accurate geometric modeling is an important aspect for vascular simulations. However, our results also indicate that lengthy segmentation arguments whether to label single, particular voxels as inside/outside the aorta, be it in manual or automatic segmentation, pose a level of detail that does not matter for CFD simulations of aortic hemodynamics observed within the scope of this study.
Nevertheless, our observations serve as strong motivation for further development of (i) segmentation and meshing algorithms, to increase representation accuracy, speed/convenience, and mesh quality, and (ii) numerical methods for solving the underlying partial differential equations within the clinical timeframe. These ongoing developments will further boost significance of simulation tools used in clinical support, where limited geometric perturbations do not restrict significance of in silico modeling.

Future developments using similar methodologies could be centered around uncertainties in QoIs for structural biomechanics problems and vessel geometries, given geometric and material uncertainties. Furthermore, geometric parametrization allows simulation and subsequent uncertainty quantification of aneurysmatic or stenosed vessels, where local hemodynamics are significantly altered, and biomarkers may be subject to vast uncertainties stemming from inter-patient variability.


\section*{Acknowledgements}
The authors gratefully acknowledge Graz University of Technology, Austria for the financial support of the LEAD-project: Mechanics, Modelling and Simulation of Aortic Dissection.
\section*{Declarations}
\emph{Competing interests:} the authors declare that they have no known competing financial interests or personal relationships that could have appeared to influence the work reported in this paper.


\begin{appendices}

\section{Spheres defining wall regions}
The integration domains for measuring the wall shear stress, see Eq.~\eqref{eqn:wss_integral_avg}, are only completely defined if the center point, radii and corresponding arc length positions of the spheres are also specified, as given in Table~\ref{tab:spheres}. An illustration can be found in Fig.~\ref{fig:wss_spheres}.
\begin{table}[h!]
	\centering
    \caption{Spheres employed to define the domains of integration for the wall shear stress measure $w_\Gamma$~\eqref{eqn:wss_integral_avg} defined by their centers $[x,y,z]$ and radii $r$. The centers lie at arc length $s$ on the centerline of the vessel, see Fig.~\ref{fig:wss_spheres}. The last three entries in the table correspond to the larger spheres with a center not lying on the centerline of the vessel to differentiate between the ascending aorta, aortic arc and abdominal aorta regions.}
	\vspace{2mm}
	\label{tab:spheres}
	\begin{tabular}{||c|>{\centering\arraybackslash}p{0.125\linewidth}|>{\centering\arraybackslash}p{0.125\linewidth}|>{\centering\arraybackslash}p{0.125\linewidth}|>{\centering\arraybackslash}p{0.125\linewidth}||}
        \hline
		&&&&\\[-1.6ex]
		$s$~[cm]& $x$~[cm]& $y$~[cm]&$z$~[cm] & $r$~[cm]\\[0.5ex]
		\hline\hline
		&&&&\\[-1ex]
        \phantom{0}0.63 & 15.15 & 16.73 & 28.35 & 1.24\\
        \phantom{0}1.89 & 14.87 & 16.34 & 29.47 & 1.42\\
        \phantom{0}3.16 & 15.09 & 16.30 & 30.67 & 1.39\\
        \phantom{0}4.63 & 15.70 & 16.69 & 31.94 & 1.15\\
        \phantom{0}7.35 & 16.97 & 18.62 & 32.92 & 1.10\\
        \phantom{0}8.86 & 17.48 & 20.02 & 33.09 & 1.08\\
        10.28 & 17.71 & 21.35 & 32.85 & 1.03\\
        11.78 & 17.79 & 22.28 & 31.80 & 1.03\\
        13.58 & 17.86 & 22.77 & 30.08 & 1.02\\
        15.27 & 17.86 & 22.88 & 28.41 & 0.92\\
        16.64 & 17.83 & 22.74 & 27.05 & 0.89\\
        17.81 & 17.84 & 22.56 & 25.89 & 0.90\\
        19.03 & 17.69 & 22.36 & 24.70 & 0.89\\
        20.54 & 17.29 & 22.16 & 23.27 & 0.90\\
        22.02 & 16.81 & 21.82 & 21.91 & 0.88\\
        23.26 & 16.44 & 21.58 & 20.77 & 0.83\\
        24.36 & 16.19 & 21.48 & 19.70 & 0.87\\
        25.54 & 15.98 & 21.25 & 18.57 & 0.89\\
        27.00 & 15.82 & 21.00 & 17.15 & 0.91\\[0.5ex]
		\hline\hline
		&&&&\\[-1ex]
        Ascending aorta
        & 15.00 & 15.00 & 29.00 & 4.00 \\
        Aortic arc
        & 17.50 & 18.00 & 35.00 & 4.00 \\
        Abdominal aorta
        & 21.00 & 31.00 & 22.00 & 15.00 \\[1.4ex]
        \hline
	\end{tabular}
\end{table}

\end{appendices}

\clearpage
\bibliography{BosnjakRefsBibTeX}

\end{document}